\documentclass[aps,prd,showpacs,twocolumn,superscriptaddress,floatfix]{revtex4}  
\usepackage{graphicx}  
\usepackage{dcolumn}   
\usepackage{bm}        
\usepackage{amssymb}   

\usepackage{psfrag}
\usepackage[T1]{fontenc}

\newcommand{\MET}{\mbox{$\not\!\!E_T$}}

\begin{document}

\begin{minipage}[height=5cm]{0.45\textwidth}
\vskip 0.25cm
\end{minipage}
\hfill
\begin{minipage}[height=5cm]{0.45\textwidth}
\raggedleft
\end{minipage}
\hfill

\hspace{5.2in}\mbox{FERMILAB-PUB-12/062-E}


\title{\boldmath  
\vspace{-0.1cm}
~\\
Search for $WH$ associated production in $p \bar{p}$ collisions at $\sqrt{s}=1.96\,{\rm TeV}$ \\
~\\}
\affiliation{LAFEX, Centro Brasileiro de Pesquisas F\'{i}sicas, Rio de Janeiro, Brazil}
\affiliation{Universidade do Estado do Rio de Janeiro, Rio de Janeiro, Brazil}
\affiliation{Universidade Federal do ABC, Santo Andr\'e, Brazil}
\affiliation{University of Science and Technology of China, Hefei, People's Republic of China}
\affiliation{Universidad de los Andes, Bogot\'a, Colombia}
\affiliation{Charles University, Faculty of Mathematics and Physics, Center for Particle Physics, Prague, Czech Republic}
\affiliation{Czech Technical University in Prague, Prague, Czech Republic}
\affiliation{Center for Particle Physics, Institute of Physics, Academy of Sciences of the Czech Republic, Prague, Czech Republic}
\affiliation{Universidad San Francisco de Quito, Quito, Ecuador}
\affiliation{LPC, Universit\'e Blaise Pascal, CNRS/IN2P3, Clermont, France}
\affiliation{LPSC, Universit\'e Joseph Fourier Grenoble 1, CNRS/IN2P3, Institut National Polytechnique de Grenoble, Grenoble, France}
\affiliation{CPPM, Aix-Marseille Universit\'e, CNRS/IN2P3, Marseille, France}
\affiliation{LAL, Universit\'e Paris-Sud, CNRS/IN2P3, Orsay, France}
\affiliation{LPNHE, Universit\'es Paris VI and VII, CNRS/IN2P3, Paris, France}
\affiliation{CEA, Irfu, SPP, Saclay, France}
\affiliation{IPHC, Universit\'e de Strasbourg, CNRS/IN2P3, Strasbourg, France}
\affiliation{IPNL, Universit\'e Lyon 1, CNRS/IN2P3, Villeurbanne, France and Universit\'e de Lyon, Lyon, France}
\affiliation{III. Physikalisches Institut A, RWTH Aachen University, Aachen, Germany}
\affiliation{Physikalisches Institut, Universit\"at Freiburg, Freiburg, Germany}
\affiliation{II. Physikalisches Institut, Georg-August-Universit\"at G\"ottingen, G\"ottingen, Germany}
\affiliation{Institut f\"ur Physik, Universit\"at Mainz, Mainz, Germany}
\affiliation{Ludwig-Maximilians-Universit\"at M\"unchen, M\"unchen, Germany}
\affiliation{Fachbereich Physik, Bergische Universit\"at Wuppertal, Wuppertal, Germany}
\affiliation{Panjab University, Chandigarh, India}
\affiliation{Delhi University, Delhi, India}
\affiliation{Tata Institute of Fundamental Research, Mumbai, India}
\affiliation{University College Dublin, Dublin, Ireland}
\affiliation{Korea Detector Laboratory, Korea University, Seoul, Korea}
\affiliation{CINVESTAV, Mexico City, Mexico}
\affiliation{Nikhef, Science Park, Amsterdam, the Netherlands}
\affiliation{Radboud University Nijmegen, Nijmegen, the Netherlands}
\affiliation{Joint Institute for Nuclear Research, Dubna, Russia}
\affiliation{Institute for Theoretical and Experimental Physics, Moscow, Russia}
\affiliation{Moscow State University, Moscow, Russia}
\affiliation{Institute for High Energy Physics, Protvino, Russia}
\affiliation{Petersburg Nuclear Physics Institute, St. Petersburg, Russia}
\affiliation{Instituci\'{o} Catalana de Recerca i Estudis Avan\c{c}ats (ICREA) and Institut de F\'{i}sica d'Altes Energies (IFAE), Barcelona, Spain}
\affiliation{Stockholm University, Stockholm and Uppsala University, Uppsala, Sweden}
\affiliation{Lancaster University, Lancaster LA1 4YB, United Kingdom}
\affiliation{Imperial College London, London SW7 2AZ, United Kingdom}
\affiliation{The University of Manchester, Manchester M13 9PL, United Kingdom}
\affiliation{University of Arizona, Tucson, Arizona 85721, USA}
\affiliation{University of California Riverside, Riverside, California 92521, USA}
\affiliation{Florida State University, Tallahassee, Florida 32306, USA}
\affiliation{Fermi National Accelerator Laboratory, Batavia, Illinois 60510, USA}
\affiliation{University of Illinois at Chicago, Chicago, Illinois 60607, USA}
\affiliation{Northern Illinois University, DeKalb, Illinois 60115, USA}
\affiliation{Northwestern University, Evanston, Illinois 60208, USA}
\affiliation{Indiana University, Bloomington, Indiana 47405, USA}
\affiliation{Purdue University Calumet, Hammond, Indiana 46323, USA}
\affiliation{University of Notre Dame, Notre Dame, Indiana 46556, USA}
\affiliation{Iowa State University, Ames, Iowa 50011, USA}
\affiliation{University of Kansas, Lawrence, Kansas 66045, USA}
\affiliation{Kansas State University, Manhattan, Kansas 66506, USA}
\affiliation{Louisiana Tech University, Ruston, Louisiana 71272, USA}
\affiliation{Boston University, Boston, Massachusetts 02215, USA}
\affiliation{Northeastern University, Boston, Massachusetts 02115, USA}
\affiliation{University of Michigan, Ann Arbor, Michigan 48109, USA}
\affiliation{Michigan State University, East Lansing, Michigan 48824, USA}
\affiliation{University of Mississippi, University, Mississippi 38677, USA}
\affiliation{University of Nebraska, Lincoln, Nebraska 68588, USA}
\affiliation{Rutgers University, Piscataway, New Jersey 08855, USA}
\affiliation{Princeton University, Princeton, New Jersey 08544, USA}
\affiliation{State University of New York, Buffalo, New York 14260, USA}
\affiliation{Columbia University, New York, New York 10027, USA}
\affiliation{University of Rochester, Rochester, New York 14627, USA}
\affiliation{State University of New York, Stony Brook, New York 11794, USA}
\affiliation{Brookhaven National Laboratory, Upton, New York 11973, USA}
\affiliation{Langston University, Langston, Oklahoma 73050, USA}
\affiliation{University of Oklahoma, Norman, Oklahoma 73019, USA}
\affiliation{Oklahoma State University, Stillwater, Oklahoma 74078, USA}
\affiliation{Brown University, Providence, Rhode Island 02912, USA}
\affiliation{University of Texas, Arlington, Texas 76019, USA}
\affiliation{Southern Methodist University, Dallas, Texas 75275, USA}
\affiliation{Rice University, Houston, Texas 77005, USA}
\affiliation{University of Virginia, Charlottesville, Virginia 22901, USA}
\affiliation{University of Washington, Seattle, Washington 98195, USA}
\author{V.M.~Abazov} \affiliation{Joint Institute for Nuclear Research, Dubna, Russia}
\author{B.~Abbott} \affiliation{University of Oklahoma, Norman, Oklahoma 73019, USA}
\author{B.S.~Acharya} \affiliation{Tata Institute of Fundamental Research, Mumbai, India}
\author{M.~Adams} \affiliation{University of Illinois at Chicago, Chicago, Illinois 60607, USA}
\author{T.~Adams} \affiliation{Florida State University, Tallahassee, Florida 32306, USA}
\author{G.D.~Alexeev} \affiliation{Joint Institute for Nuclear Research, Dubna, Russia}
\author{G.~Alkhazov} \affiliation{Petersburg Nuclear Physics Institute, St. Petersburg, Russia}
\author{A.~Alton$^{a}$} \affiliation{University of Michigan, Ann Arbor, Michigan 48109, USA}
\author{G.~Alverson} \affiliation{Northeastern University, Boston, Massachusetts 02115, USA}
\author{M.~Aoki} \affiliation{Fermi National Accelerator Laboratory, Batavia, Illinois 60510, USA}
% ...
%
%
\collaboration{The D0 Collaboration\footnote{with visitors from
$^{a}$Augustana College, Sioux Falls, SD, USA,
$^{b}$The University of Liverpool, Liverpool, UK,
$^{c}$UPIITA-IPN, Mexico City, Mexico,
$^{d}$DESY, Hamburg, Germany,
,
$^{e}$SLAC, Menlo Park, CA, USA,
$^{f}$University College London, London, UK,
$^{g}$Centro de Investigacion en Computacion - IPN, Mexico City, Mexico,
$^{h}$ECFM, Universidad Autonoma de Sinaloa, Culiac\'an, Mexico
and
$^{i}$Universidade Estadual Paulista, S\~ao Paulo, Brazil.
$^{\ddag}$Deceased.
}} \noaffiliation
\vskip 0.25cm
           
\date{March 5, 2012}

\begin{abstract}
\vskip 0.4cm
This report describes a search for associated production of $W$ and Higgs bosons 
based on data corresponding to an integrated luminosity of $\cal{L}$$~ \approx 5.3 ~\rm fb^{-1}$ collected 
with the D0 detector at the Fermilab Tevatron $p\bar{p}$ Collider. Events containing a $W\rightarrow \ell \nu$ candidate 
(with $\ell$ corresponding to $e$ or $\mu$) are selected in association with two 
or three reconstructed jets. One or two of the jets are required to be consistent 
with having evolved from a $b$ quark. 
A multivariate discriminant technique is used to improve the 
separation of signal and backgrounds. Expected and observed upper limits are obtained for the 
product of the $WH$ production cross section and branching ratios and reported
in terms of ratios relative to the prediction of the standard model as a function of the mass of the Higgs boson ($M_{H}$). 
The observed and expected 95\% C.L. upper limits obtained for an assumed $M_{H}=115~\rm GeV$ are, respectively, 
factors of 4.5 and 4.8 larger than the value predicted by the standard model.

\end{abstract}
\maketitle

\section{Introduction}

In the standard model (SM) of particle physics, the  
masses of the weakly interacting $W$ and $Z$ gauge bosons are accommodated through the process of electroweak symmetry breaking, and the 
masses of fermions through their Yukawa couplings to the Higgs field.
The search for the Higgs boson, whose mass $M_{H}$ is not predicted by the SM, is a test of this hypothesis and is a major component of the
experimental programs at particle colliders. At the Fermilab Tevatron  $p\bar{p}$ 
Collider, this search is carried out using  multiple statistically independent search samples, 
each sensitive to different Higgs boson production processes and decay channels, providing increased sensitivity 
in the search for direct evidence for this SM mechanism \cite{combTEV,combD0}.  

This paper presents an extended description of the previously reported search \cite{WH} for SM Higgs boson production 
through the process $p\bar{p}\rightarrow WH$, in which a Higgs boson is produced in association with a $W$ boson. The search is based on 
data corresponding to an integrated luminosity $\cal{L}$ $\approx 5.3 ~\rm fb^{-1}$ collected with the D0 detector at the Fermilab Tevatron 
$p\bar{p}$ Collider with a center-of-mass energy $\sqrt{s}=1.96~\rm TeV$. 
The events are required to contain a $W\rightarrow e \nu$ or $W\rightarrow \mu \nu$ candidate, thereby suppressing background 
from inclusive $b$-jet production processes, and enhancing sensitivity to signal by several orders of magnitude.  The event selection is 
also sensitive to $W\rightarrow \tau \nu$ events with $\tau$ decay into electrons or muons. The events are required to contain a 
$H\rightarrow b\bar{b}$ candidate because of large branching 
fraction for this decay in the $M_{H}$ region considered here ($100 < M_{H} < 150 ~ \rm GeV$). 
The experimental signature is therefore a single 
isolated lepton, an imbalance in the measured transverse energy ($\MET$), and either two or three jets, at 
least one of which is consistent with having been initiated by a $b$ quark.
The three-jet sample is included to provide additional sensitivity for $WH$ events
containing gluon radiation from the initial or final-state particles of the hard collision. The data are examined in separate search samples 
of different sensitivity and a multivariate random forest technique \cite{RF1,RF2} 
is applied to each sample, further enhancing sensitivity to signal.

Direct searches for the process $e^{+}e^{-}\rightarrow ZH$ at the CERN $e^+e^-$ Collider (LEP) experiments set the 
SM Higgs mass to $M_{H} > 114.4 ~\rm GeV$ \cite{CERN}. In addition, a fit to electroweak precision measurements of the 
masses of the $W$ boson and the top quark from both Tevatron and LEP experiments leads to an upper limit of $M_{H} < 161 ~\rm GeV$ 
for SM Higgs production at the 95\% C.L. \cite{EWFIT}. 
Both the CDF and D0 Collaborations have extensively investigated the $WH$ associated production mechanism 
\cite{D0WH1,D0WH2,D0WH3,CDFWH1,CDFWH2,CDFWH3}, and a region at larger $ 156 < M_{H} < 177  ~\rm GeV$ has also been excluded at 95\% C.L. by 
direct searches for $H\rightarrow W^{+}W^{-}$ decays \cite{Exclude}. Results from the CERN Large Hadron Collider (LHC) Collaborations 
\cite{ref:ATLAS,ref:CMS} also exclude regions at higher $M_{H} > 127  ~\rm GeV$ and indicate that 
the most interesting region for the search for the SM Higgs boson is the one where 
the sensitivity of the search discussed in this article is maximal. The analysis presented here is expected to be a highly sensitive channel 
in the mass range $100 \lesssim M_{H} \lesssim 135 ~\rm GeV$, and complements searches at the LHC
 which rely primarily on different SM Higgs production and decay mechanisms
 in this mass range.

\section{The D0 Detector}

The main components of the D0 detector used in this investigation
are the tracking detectors, calorimeters, muon detectors, and 
the luminosity system. Protons and antiprotons 
interact close to the origin of the D0 detector
coordinate system, which is at the center of the detector. A right-handed Cartesian 
coordinate system is used with the positive $z$ axis pointing along the nominal direction of 
the incoming proton beam (the positive $y$ axis points toward the top of the detector) and the pseudorapidity variable, 
defined as $\eta = - \ln \tan \frac{\theta}{2}$, where $\theta$ is the polar angle in the corresponding spherical 
polar coordinate system. The kinematic properties of particles and jets are measured with respect to the reconstructed
$p\bar{p}$ collision vertex.  More details on D0 construction and component design are available in Refs.\
\cite{d0det_run2,d0det_run1}. Upgrades to the tracking and trigger systems 
were installed during the summer of 2006 and the data samples collected 
prior to and after this upgrade are referred to as pre- and post-upgrade 
samples in the following.

\subsection{Tracking Detectors}

The D0 tracking system surrounds the interaction point and consists of an inner silicon
microstrip tracker (SMT) followed by an outer central scintillating 
fiber tracker (CFT). Both the SMT and CFT are situated within a 2 T 
magnetic field provided by a superconducting solenoidal coil surrounding the entire
tracking system.

The silicon microstrip tracker is used for tracking charged particles and reconstructing interaction
and decay vertices. In the central region there are six barrel sections each comprising 
four detector layers. The barrel sections are interspersed and capped with disks composed of 
12 double-sided silicon wedge detectors. The first and 
second detector layers of each barrel contain 12 silicon modules and 
24 modules are installed in the third and fourth detector layers. An additional inner
layer was added to the silicon tracker system in 2006 \cite{smt_layer}.
In the high  $|\eta|$ region on either side of the three disk-barrel assemblies there are three further radial disk sections (F-disks), 
and in the far-forward region, large-diameter 
disks (H-disks) provide tracking at larger $|\eta|$. The tracks of particles with $|\eta| < 1.7 $ 
are measured using the CFT and the barrel and F-disk sections of the SMT, whereas tracks for 
particles at larger $|\eta|$  are reconstructed using the the F- and H-disks.

The CFT comprises scintillating fibers ($835 ~\rm \mu m$ in diameter) mounted on  
eight concentric support cylinders. The cylinders occupy the radial space from 20 to 52 cm 
from the center of the beam pipe. The two innermost cylinders are $1.66 ~\rm m$ long whereas 
the outer six cylinders are $2.52 ~\rm m$ long. The outer cylinder provides tracking coverage 
extending to $|\eta| = 1.7$.

\subsection{Calorimeters}

The D0 calorimeter system is used to measure energies as well as to identify electrons, 
photons, and jets. The calorimeter also helps to identify muons and provides a measure of the $\MET$ in events. The central calorimeter 
(CC) and the two end calorimeters (EC) are
contained within three individual cryostats located outside of the superconducting solenoid.
The central calorimeter covers detector pseudorapidities $|\eta| \lesssim 1.1$ 
and the end calorimeters extend the range to $|\eta| = 4.2$. 
The active material in each calorimeter section is liquid argon. Extending radially outwards
from the detector center, the calorimeters are subdivided into electromagnetic (EM), fine hadronic, and coarse hadronic (CH) sections. 
The absorber material of the EM sections is uranium, whereas for the fine hadronic sections a uranium-niobium alloy is used. The CH absorbers are made of copper in the CC region and stainless steel in the EC region.
To improve measurements in the  intercryostat regions, plastic-scintillator detectors and ``massless gap'' detectors are used to 
sample showers between cryostats, enhancing calorimeter coverage in the region $0.8<|\eta|<1.4$.

\subsection{Muon Detectors}

The muon detector system \cite{d0muo_run2} consists of a central muon detector system 
covering the range $|\eta|<1$ and a forward muon system that covers
the region $1<|\eta|<2$. The central muon system comprises
aluminum proportional drift chambers whereas aluminum mini drift tubes are used in the forward system. Scintillation counters are included 
for triggering purposes,
and 1.8 T toroidal magnets make it possible to determine muon momenta and 
perform tracking measurements based on the muon system alone.

The proportional drift chambers are arranged
in three layers, one of which (A layer) is located within the toroid,
with the remaining two (B and C) layers located beyond the toroid, with the
C layer radially furthest from the interaction point.  In the central muon system, 
the B and C layers have three planes of drift cells. The A layer
has four planes, except  at the support structure at the bottom of the detector, 
where the A layer has three planes of cells. In the forward region, mini drift tubes 
are arranged in 
eight octants with four planes in the A layer while the B and C layers 
each have three planes.

\subsection{Luminosity System}

The D0 luminosity system is used to determine the instantaneous luminosity and also to measure beam-halo 
rates. The system is composed of two disks 
of scintillating tile detectors that are positioned in front of the ECs on both sides 
of the D0 detector at $z = \pm 140~\rm cm$. Each of the disks 
consists of 24 plastic scintillation counters that cover pseudorapidity 
regions $2.7<|\eta|<4.4$. The total integrated luminosity ($\cal{L}$) is determined via 
the average instantaneous number of observed inelastic collisions ($N_{\text{inel}}$), according to
$f N_{\text{inel}} / \sigma_{\text{inel}}$, where $f$ is the frequency of $p\bar{p}$
Tevatron bunch crossings, and $\sigma_{\text{inel}}$ is the effective inelastic production
cross section \cite{inelas} within the luminosity system acceptance, after taking into account beam-halo events 
and multiple collisions within a single beam crossing. In practice, $N_{\text{inel}}$
is calculated by inverting the Poisson probability of observing no hits in either of the two disks \cite{lumi_ID}.

\section{Triggering}

The D0 trigger system has three levels referred to as L1, L2, and L3. Each consecutive
level receives a lower rate of events for further examination.  
The L2 software-based algorithms refine the L1 information they receive and 
the L3 software-based algorithms then run simplified versions of offline identification algorithms based on the full detector readout.

The $W\rightarrow e \nu$ candidates of this search are collected using the 
logical OR \cite{Boole} of different triggers requiring a candidate electromagnetic 
object. The L1 electron triggers require calorimeter
energy signatures consistent with those of an electron. The logical OR also includes trigger 
algorithms requiring an electromagnetic object together with at least 
one jet, for which the L1 requirement includes a calorimeter energy deposition expected for 
jets at large transverse momenta $p_{T}$. The triggers have different 
minimum electron and jet $p_{T}$ thresholds, and each has a typical 
efficiency of (90--100)\% for the signal events satisfying the selection 
requirements discussed below, depending on the 
trigger type and sampled region of the detector.
The trigger efficiencies are determined using samples of  $Z/\gamma^{*} \rightarrow e^{+}e^{-}$ 
events and are modeled as functions of the $p_{T}$ and $\eta$ of the  leading (largest $p_{T}$) 
electromagnetic object in the event.  Event weights are used to 
apply the measured trigger efficiencies to the simulated signal and background
samples. Since the triggers undergo periodic
changes, these efficiencies depend on specific running periods. In particular,
an improved calorimeter trigger was added during the 2006 detector upgrade \cite{caltrig}.

$W\rightarrow \mu \nu$ candidates are triggered using the logical OR of the full set of 
available triggers and expected to be fully efficient for the selection 
criteria used. For muons, the selected pseudorapidity range of this analysis 
is $|\eta| < 1.6$, where the majority of the events are collected by 
triggers that require a large-$p_{T}$ muon at L1 (single-muon
triggers). The efficiency of the single-muon triggered component of the data is determined using
$Z/\gamma^{*} \rightarrow \mu^{+}\mu^{-}$ events, again separately for specific running periods. It is typically $\approx$ 70\% and is 
well modeled in simulation.  
The remainder of the events are collected primarily using jet 
triggers. The efficiency for these  triggers is determined separately
by taking the ratio of this component of the triggered data set to Monte Carlo (MC) 
simulation with triggering probabilities set to unity 
[after correcting the data for multijet (MJ) background as 
described separately in Sec.\ VII]. The ratio is 
parameterized as a function of the scalar sum ($H_{T}$) of the transverse momenta of the jets
in the event, and compared to the well-modeled 
single-muon triggered data set. The simulated 
probability for 
events to pass at least one of the single high-$p_{T}$ muon triggers 
is then scaled to the efficiency of the complete set of triggers used.  
The most recently collected data correspond to the highest 
instantaneous luminosities, and because different 
proportions of multijet, $\MET$+jet, and 
muon+jet triggered events are observed as a function of luminosity, the additional probability 
factor is computed separately for events collected before and after the 2006 D0 upgrade. The remaining triggers provide a gain in 
probability of $\approx$ 0.23 
before the 2006 upgrade and range from 0.23--0.33 following the upgrade. 

After additional detector status quality requirements, applied to ensure subdetector systems are operational, the total integrated luminosity 
is $\cal{L}~$$= 5.32 ~\rm fb^{-1}$ for the electron channel and $\cal{L}~$$= 5.36 ~\rm fb^{-1}$ for the muon channel. 
The contribution to the total integrated luminosity from the pre-2006 upgrade part of the data set is about $1 ~\rm fb^{-1}$
in each case.
The uncertainty on the experimentally measured integrated luminosity  
is $6.1$\% \cite{lumi_ID} and is dominated by the uncertainity in 
the effective inelastic production 
cross section \cite{inelas}.

\section{Identification of Leptons, ${\boldsymbol \MET}$, \newline and Jets}

Candidate events with $W$ bosons are selected by requiring a single reconstructed 
lepton together with large $\MET$ and the selected $W\rightarrow \ell \nu$ 
samples are also required to contain either two or three reconstructed jets.

Electrons of $p_{T}> 15 ~\rm GeV$ are reconstructed in the CC or EC calorimeters in the 
pseudorapidity regions $|\eta|<1.1$ and $1.5 < |\eta| < 2.5$, respectively. In the CC (EC),  
a shower is required to deposit 97\%(90\%) of its total energy [as measured in a cone of
radius $\Delta R = \sqrt{(\Delta \eta)^{2} + (\Delta \phi)^{2}} = 0.4$] within 
a cone of 
radius $\Delta R = 0.2$ in the electromagnetic layers. 
The showers must have transverse and longitudinal distributions that are consistent
with those expected from electrons. In the CC region, a 
reconstructed track, isolated from other tracks, is required to have a 
trajectory that extrapolates to the EM shower. The isolation criteria 
restricts the sum of the scalar $p_{T}$ of tracks of $p_{T}>0.5 ~\rm GeV$ 
within a hollow cone 
of radius $\Delta R = \sqrt{(\Delta \eta)^{2} + (\Delta \phi)^{2}} = [0.05-0.4]$ 
surrounding the electron candidate to $<2.5 ~\rm GeV$. Additional information
on the number and scalar $p_{T}$ sum of tracks in cone of radius $\Delta R = 0.4$ surrounding the candidate cluster, track to cluster matching 
probability, the ratio of the transverse energy of the cluster and 
the transverse momentum of the track associated with the shower,
the EM fraction, and lateral and longitudinal shower shape characteristics 
are used to construct CC and EC electron likelihood discriminants. 
The discriminants are trained using  $Z/\gamma^{*} \rightarrow e^{+}e^{-}$ 
events and are applied to ensure that 
the observed particle characteristics are consistent with electrons 
\cite{ttbar-prd}.

Muons of $p_{T} > 15 ~\rm GeV$ are selected in the region $|\eta| < 1.6$. 
Muons are required to have track segments in both the A and B or C layers of the muon 
detectors, with a spatial 
match to a corresponding track 
in the central tracker. The scalar sum of the $p_{T}$ of tracks 
with $\Delta R < 0.5$ around the muon candidate is required to be less than $2.5 ~\rm GeV$. Furthermore, transverse energy deposits in 
the calorimeter in a hollow cone of $\Delta R = [0.1-0.4]$ around the muon must be less than 
$2.5 ~\rm GeV$. To suppress MJ background events that originate from 
semileptonic decays of hadrons, muon candidate tracks are required to be spatially separated from jets by $\Delta R (\mu, j) > 0.5 $. 
To suppress cosmic-ray muons, scintillator timing information is used to require the hits to coincide with a beam crossing. 

In addition to the selection criteria listed above, electrons and muon
samples are also selected using much looser reconstruction criteria. 
 For the electron channel, less restrictive calorimeter isolation 
and EM energy fraction criteria are used and the likelihood discriminants
are not applied. For 
the muon channel, less restrictive energy isolation and track-momentum 
criteria are used.
These samples are used only for the determination of the 
MJ background contributions to the final selected samples as described in Sec.\ VII.

The $\MET$ is calculated 
from individual calorimeter cell energies in the 
electromagnetic and fine hadronic parts of the calorimeter and is required to be
$\MET > 20~\rm GeV$ for both the electron and muon channels.
It is corrected for the presence of any muons. 
All energy corrections to leptons and to jets (including energy from the coarse
hadronic layers associated with jets) are propagated to the $\MET$.

 Jets are reconstructed in 
the calorimeters in the region 
$|\eta| < 2.5$ using the D0 Run II iterative midpoint cone algorithm, from energy deposits 
within cones of size $\Delta R = 0.5 $ \cite{d0jet}.
To minimize the possibility that jets are caused by noise or spurious energy deposits, the fraction of the total jet energy contained in 
the EM layers of the calorimeter is required to be between 5\% and 95\%, and the energy fraction
in the CH sections is required to be less than 40\%. To suppress noise, different cell 
energy thresholds are also applied to clustered and to isolated cells.  The energy of the jets is scaled by applying 
a correction determined from $\gamma$+jet events using the same jet
finding algorithm. This scale correction accounts for additional energy (e.g., residual energy
from previous bunch crossings and energy from multiple $p\bar{p}$ 
interactions) that is sampled within the finite cone size,  the calorimeter response to particles produced 
within the jet cone, and energy flowing outside the cone or moving into the cone via detector
effects (e.g the deflection of particles by the magnetic field). Details of the D0 jet energy scale correction can be 
found in Ref.\ \cite{d0jes}. 

In addition to the previously mentioned jet energy scale correction,
derived using $\gamma$+jet events, residual 
calibration differences between simulated and data-selected jets 
are also studied using $Z(\rightarrow e^{+}e^{-})$+jet events.
An additional energy recalibration and an energy smearing are then determined 
to adjust the $p_{T}$ imbalance between the $Z$ boson and the 
recoiling jet in simulation to that observed in data. The correction
is applied in simulation to gluon-dominated jet production processes. 
Differences in reconstruction thresholds in simulation and data are also 
taken into account. 

The jet identification efficiency and jet resolution are adjusted in the simulation to match those measured in data.
Following the 2006 upgrade of the D0 detector to handle higher instantaneous 
luminosity, all jets are also required to satisfy additional criteria 
for originating from the primary $p\bar{p}$ vertex (``vertex confirmation'').
The criteria are that the jets have at least two tracks, each of which have 
$p_{T}>0.5 ~\rm GeV$, at least one hit in the SMT detector, and distances of 
closest approach (DCA) of $<0.5$ and $<1.0 ~\rm cm$ from the primary 
$p\bar{p}$ interaction vertex in the transverse plane and 
along the $z$ axis, respectively.

\section{Tagging of ${\boldsymbol{b}}$ quark jets}

The final sample of $WH$ candidate events is selected by requiring that at 
least one of the jets produced in association with the $W$ boson is 
consistent with having been initiated by a $b$ quark, using the neural network
(NN) $b$-tagging algorithm described in detail in Ref. \cite{d0btag}.

Jets considered by the $b$-tagging algorithm are required to pass a ``taggability'' requirement that utilizes charged-particle tracking 
and vertexing information. 
The efficiency of this requirement accounts for variations in detector acceptance and track 
reconstruction efficiencies at different $z$ position values of the primary vertex (PV) through the interaction region, prior to the 
application of the $b$-tagging algorithm,
and depends on the $z$ position of the PV and the $\eta$ of the jet. 
More details on the reconstruction and selection 
of the primary interaction vertex are available in Ref.\ \cite{d0btag}.
The taggability requirement is that a calorimeter jet be matched to a track-jet 
within an angular separation of $\Delta R < 0.5$. 
Track-jets are formed starting from seed tracks of $p_{T} > 1 ~\rm GeV$ with at least one hit 
in the SMT detector and DCA requirements of $< 0.15~\rm cm$ and $<0.4~\rm cm$ to the primary vertex in the transverse plane and along the 
$z$ axis, respectively. The other tracks used to form 
track-jets must have $p_{T} > 0.5 ~\rm GeV$.  To reduce the probability of misidentified secondary vertices, tracks consistent with the decay of a 
long lived particle (e.g. $K_{s}, \Lambda$) or a converted 
photon are also removed, before application of the $b$-tagging algorithm. 

The efficiency of the taggability requirement in the selected samples is studied in data and in simulation, 
in four $z$-vertex intervals, as a function of jet $\eta$ and $p_{T}$.
Correction factors are determined and applied to the MC separately for the pre- and 
post-upgrade parts of the simulated samples. The corrections, 
which are of order 1\%, are applied as a function of jet $\eta$ (post-upgrade) and 
jet $p_{T}$ (pre-upgrade). The jet taggability efficiency is largest (80\%--90\%) around the center of
  the interaction region. More details on jet 
taggability and its efficiency can be found in Ref.\ \cite{d0btag}.

The NN $b$-tagging algorithm uses seven input variables, five of which make use of secondary decay vertex information.
These are the invariant masses (calculated from all contributing tracks assuming the pion mass) of secondary vertices, the number of tracks 
used to reconstruct the secondary vertex, the $\chi^{2}$ of the secondary vertex fit, the decay length significance 
of the secondary vertex with respect to the primary vertex in the transverse plane, and the number of secondary vertices reconstructed in the 
jet. Two further impact-parameter-based variables are also used. The first is a 
discrete signed impact parameter significance variable, which is a combination of four quantities related to the 
number and the quality of tracks within  a cone of radius $\Delta R = 0.5$ 
centered on the calorimeter jet. The second is a continuous jet-lifetime
variable, which is used to assign a total probability that tracks within a jet 
are consistent with the primary vertex position. The variable is calculated 
using the product of individual track probabilities, which indicate the likelihood that each track is consistent 
with the primary vertex position. The individual probabilities are based on the
impact parameter resolution of the tracks. The track impact parameters are given the same sign as the scalar product of the 
track DCA in the transverse plane and the jet $p_{T}$. The negative signed region is used
to calibrate the impact parameter resolution whereas tracks with positive values are used to calculate the total 
lifetime probability.

\section{Monte Carlo Simulation}

At each step of the selection, the data are compared to predictions obtained 
by combining the MC simulation of SM backgrounds with a data-based estimation 
of the instrumental background from MJ events containing 
misidentified leptons (discussed separately in Sec.\ VII).
All generated samples are passed through a detailed, {{\sc geant}}-based
simulation \cite{geant} of the D0 detector and the same reconstruction
algorithms used for data. Separate simulations are applied
for conditions prior to and after the 2006 detector upgrade. 
The SM predictions are used to set the relative normalizations for 
all of the generated samples, and additional reweighting factors are then 
applied to normalize samples generated using the leading order (LO) {{\sc alpgen}} \cite{alpgen} MC event generator to data. 
These factors are determined prior to the application of $b$-tagging 
(see Sec.\ V), where the signal 
contribution is expected to be negligible, and these are determined 
simultaneously with the MJ background, which is also obtained from data. 
The impact of multiple $p\bar{p}$  interactions and detector
noise is accounted for by adding data events recorded during
random beam crossings to the simulated events before they are
reconstructed. The instantaneous luminosity profile of these events
is matched, prior to the application of $b$-tagging, to that observed in the 
selected data samples. For all the MC samples the effects of beam remnants 
and of multiple partonic interactions (underlying event) are modeled using 
the {\sc{pythia}} parameters obtained from data in Ref.\ \cite{TuneA}.\\

{\bf $\bullet$ {\boldmath $WH$} production:} The $WH$ associated production process, 
with subsequent decay of the Higgs boson to a $b\bar{b}$ quark-antiquark pair, is 
modeled using the {{\sc pythia}} \cite{pythia63} MC event generator, according to the 
prescription of Refs.\ \cite{signal1,signal2,signal3,signal4,signal5} 
and the LH2003 Working Group \cite{signal}.
The events are generated using the CTEQ6L LO parton distribution 
functions \cite{CTEQ} with the renormalization and factorization scales set to the Higgs boson mass $M_{H}$. 

Eleven samples in total are generated, for  $M_{H}$ values spanning the range 
$M_{H} = 100-150 ~\rm GeV$ in intervals of $5 ~\rm GeV$.  Similarly, a set of 
11 $q\bar{q}\rightarrow ZH$  signal samples is also 
generated with {{\sc pythia}} to model the small contribution 
of signal events from $ZH$ associated production that passes all selections.  
These events are selected if one of the leptons from the decay 
$Z\rightarrow \ell^{+}\ell^{-}$ is
either not reconstructed or is produced outside of the detector acceptance.
The $WH$ and $ZH$ samples are referred to collectively as $WH$ in the 
figures and the remainder of the text.\\

{\bf $\bullet$ {\boldmath $W$}+light partons:} The SM background processes $W (\rightarrow 
\ell \nu) q\bar{q}$, where $q$ represents light quarks ($u$, $d$, $s$) and gluons are generated using the LO MC matrix element event generator 
{{\sc alpgen}} according to the parton-level cross section calculations of 
Ref.\ \cite{CTEQ}. Separate samples are generated for light parton 
multiplicities $0, 1, 2, 3, 4,$ and $\ge 5 $ with each case generated
for each of the final-state decay lepton flavors $\ell=e,\mu$, and $\tau$.
The {{\sc pythia}} generator is used to account for the subsequent 
hadronization and development of partonic showers. The MLM 
factorization (``matching'') scheme \cite{alw} is used to avoid
the possibility of overestimating the probability of further partonic 
emissions produced in {\sc pythia}. 
The samples are then normalized to data as described in Sec.\ VI A.
To avoid double counting of heavy quarks, $Wb\bar{b}$ and $Wc\bar{c}$ events, which are generated separately as described below, are removed.\\

{\bf $\bullet$ {\boldmath$Z/\gamma^{*}$}+light partons:} A corresponding set of  $Z/\gamma^{*} (\rightarrow \ell\ell) q\bar{q}$ samples are 
generated for light-parton multiplicities 0, 1, 2, and $\ge 3$. These samples also 
include each of the lepton flavors $\ell=e,\mu,$ and $\tau$. The $Z/\gamma^{*}$ contributions
are generated over the  $Z/\gamma^{*}$ mass region $M_{\ell\ell}=15-250~\rm GeV$ 
for $\ell=e,\mu$, and $M_{\ell\ell}=75-250~\rm GeV$ for $\tau$ decays.
The combined $W (\rightarrow 
\ell \nu) q\bar{q}$  and $Z/\gamma^{*} (\rightarrow \ell\ell) q\bar{q}$ samples
are referred to as $W$+light in the figures and the remainder of the text.\\

$\bullet$ {\boldmath $Wb\bar{b}$, $Wc\bar{c}$:} The channel $W (\rightarrow \ell\nu ) b\bar{b}$ and also  the channel
$W (\rightarrow \ell\nu) c\bar{c}$ (referred to collectively as 
$Wb\bar{b}$) are generated using {{\sc alpgen}} also according to the 
initial prescription of Ref.\ \cite{alpgen_xs}. The {{\sc pythia}} generator is again used to
account for subsequent shower development and the MLM matching scheme is again 
used for the treatment of further partonic emissions. 
Four separate samples are generated for  $0, 1, 2,$ and $\ge 3$ 
additional light partons. To avoid 
double counting,  $Wc\bar{c}$ states are removed from the $W (\rightarrow \ell\nu) b\bar{b} $ samples, and no events are removed 
from the $W (\rightarrow \ell \nu) c\bar{c}$ samples.\\

$\bullet$ {\boldmath $Z/\gamma^{*}b\bar{b}$, $Z/\gamma^{*}c\bar{c}$:} Corresponding samples of  $Z/\gamma^{*} (\rightarrow \ell\ell) b\bar{b}$ 
and $Z/\gamma^{*}(\rightarrow \ell\ell) c\bar{c}$ events are generated for each lepton flavor $\ell=e,\mu,\tau$ and for  0, 1, and $\ge 2$ 
additional light parton multiplicities. The combined $W (\rightarrow \ell\nu ) b\bar{b}$, $W (\rightarrow \ell\nu) c\bar{c}$,  
$Z/\gamma^{*} (\rightarrow \ell\ell) b\bar{b}$, and $Z/\gamma^{*}(\rightarrow \ell\ell) c\bar{c}$ samples are referred to as $Wb\bar{b}$ 
in the figures and the remainder of the text.\\

$\bullet$ {\boldmath $t\bar{t}$:} The background from $t\bar{t}$ 
interactions is generated using {{\sc alpgen}}, again interfaced with 
{{\sc pythia}}, and using the  MLM matching scheme. The cross section predictions contain the most important terms of 
the next-to-NLO (next-to-next-to-LO) corrections \cite{ttbar_xsecs}.  
The $t\bar{t}\rightarrow b\bar{b} + \ell^{+}\nu \ell^{'-} \bar{\nu}_{\ell^{'}}$ and 
$t\bar{t}\rightarrow b\bar{b} + 2j + \ell\nu$ final states are considered, 
including $0, 1,$ and $2$ additional light parton multiplicities, and 
all decay lepton flavors $\ell=e,\mu,\tau$.\\

{\bf $\bullet$ Single top quarks}: Background processes initiated by single top quark production 
are generated using {{\sc CompHep}} \cite{COMPHEP,COMPHEP1}. The cross sections \cite{stop_xsecs} are calculated at
NLO  and {{\sc pythia}} is again used for subsequent hadronization and 
partonic-shower development, along with the  MLM matching scheme.
The $s$-channel ($t\bar{b}\rightarrow \ell\nu b\bar{b}$) processes and $t$-channel ($tq\bar{b}\rightarrow \ell\nu bq\bar{b}$) processes are 
generated for the three lepton flavors 
$\ell= e, \mu,$ and $\tau$. The single top samples are referred to
collectively as $\rm{s-top}$ in the figures and the remainder of the text.\\

{\bf $\bullet$ Diboson}: Backgrounds from the hadronic production of diboson pairs ( 
$p\bar{p}\rightarrow V_{1}V_{2}$, where $V_{1},V_{2}=W^{\pm}$, or $Z/\gamma^{*}$) are simulated 
using {{\sc pythia}}. The cross sections are calculated at NLO according to the prescription of Ref.\ \cite{mcfm}, obtained using the {{\sc mcfm}} 
program, and incorporating spin correlations in partonic matrix elements. The diboson samples are generated 
inclusively for all boson decay leptonic flavors $\ell=e,\mu,\tau$ and are referred to collectively as $WZ$ in the figures
and remainder of the text.

\subsection{MC Reweighting}

Because of problems in the modeling of background processes in MC
 simulations, we apply corrections summarized in the following.
Distributions of the summed $W$+light and 
$W b\bar{b}$ simulated samples are compared to data, prior to the application of $b$-tagging, and corrections are developed to reweight shape 
discrepancies. 
The correction factors are calculated, prior to 
the determination of the {{\sc alpgen}} normalization factors. 
These corrections are motivated by previous comparisons of {{\sc alpgen}} with data \cite{alw_data} and with other event generators \cite{alw}.
The overall
event yields are preserved in the reweighting, and the same weight
functions are applied to all the $W$+jets {{\sc alpgen}} backgrounds, at 
reconstruction level in the MC. In this section, we describe the applied reweighting functions in detail.

The reweighting functions are determined from the ratio of the total
$W$+light and $W b\bar{b} $ distributions to the corresponding distributions
obtained from the high statistics selected $N^{\text{data}}_{W+\text{jets}}$ component of the data.  
The expected signal contribution is negligible in this sample.
$N^{\text{data}}_{W+\text{jets}}$ is obtained after correcting the total selected data sample $N_{\text{data}}$ for
MJ background ($N_{\text{MJ}}$) and the
total expected contributions from other SM background sources ($N^{\text{MC}}_{\text{\
SM}}$):

\begin{equation}
 N^{\text{data}}_{W+\text{jets}} = N_{\text{data}} - N_{\text{MJ}} - N^{\text{MC}}_{\text{SM}}.
\end{equation}

Prior to determining the {{\sc alpgen}} reweightings, calibration differences 
in data and MC for overlaid events in the post-2006 
upgrade samples are corrected for. Two reweighting constants are
applied which scale down the MC, for the leading and subleading jet $|\eta|$ 
distributions, thereby improving detector modeling in the
intercryostat pseudorapidity region $0.8<|\eta|<1.4$. Separate 
constants are used for the positive and negative pseudorapidity intervals, for each of the two leading jets.
The reweighting constants reduce the simulated contribution by 1\%--10\%.

The overall description of the $N^{\text{data}}_{W+\text{jets}}$ lepton $\eta$
distribution as well as the corresponding leading and subleading jet
$\eta$ distributions are then adjusted by applying a first-order polynomial
reweighting function in $\eta^{2}$ of the simulated lepton $\eta$
and second-order reweighting functions in $\eta^{2}$ to the $\eta$
of the leading and subleading jets. Only the two leading jets are
reweighted in the $W$+3 jet selections. These reweighting functions have
the primary effect of improving the {{\sc alpgen}} description of the $\eta$
distributions by increasing the MC component for $|\eta| \ge 1.5$.

Discrepancies observed in the correlation between the jet directions $\Delta R$($j_1$,$j_2$) and the $W$ boson $p_{T}$ 
are corrected through two reweighting functions in the
two-dimensional $\Delta R$$ - p_{T}^{W}$ plane. The functional form is a third-order polynomial in 
$\Delta R$($j_1$,$j_2$), increasing the {{\sc alpgen}} simulation by
$\approx 20\%$ at large $\Delta R$, times a constant plus Gaussian error function reweighting in 
$W$ boson $p_{T}$, applied to the $W$ boson {{\sc alpgen}} samples only, and which primarily increases the
simulation by $\approx 20\%$ for $p_{T}^{W}<20~\rm GeV$. The $p_{T}$ distribution
 for $Z/\gamma^{*}$ production is also adjusted to agree with the
observed distribution. The systematic uncertainties associated with these
reweightings are discussed in Sec.\ X.

\subsection{{\bf {\textsc{alpgen}}} Normalization Factors}

Two multiplicative scaling factors are used to normalize and to incorporate the effects of
higher-order terms in the {{\sc alpgen}} MC samples.
The first factor, $K^{W\text{+jets}}$, is applied to both the $W$ + light parton and $W b\bar{b}$
generated events, whereas
the second multiplicative factor, $S_{Wb\bar{b}}$, is applied only to the $Wb\bar{b}$ samples.

To determine $K^{W\text{+jets}}$, the number of selected
$W$ + light parton and  $W b\bar{b}$ events in {{\sc alpgen}} ($N^{\text{MC}}_{W\text{+jets}}$) is 
scaled to match the data ($N^{\text{data}}_{W\text{+jets}}$) contribution:

\begin{equation}
K^{W\text{+jets}} = \frac{ N^{\text{data}}_{W\text{+jets}}}{N^{\text{MC}}_{W\text{+jets}}} 
\end{equation}

The factors $K^{W\text{+jets}}$ are calculated separately for the electron and muon channel
samples and separately for both the $W$+2 jet and $W$+3 jet
selections. The obtained factors are found to be consistent within
their statistical and systematic uncertainties and are shown, after accounting for NLO corrections to the cross section \cite{mcfm} 
(already included in the generated samples), in Table \ref{SF}. The values are in the range
$K^{W\text{+jets}} \approx 1.0-1.16$ for the $W$+2 jet and $K^{W\text{+jets}} \approx 1.12-1.35$ for the
$W$+3 jet selected samples. The assigned systematic uncertainties are described in Sec.\ X.

\begin{table}[t]
\begin{center}
\caption{ \label{SF} 
The experimental $K^{W\text{+ jets}}$ factors (applied after taking into 
account the theoretical factor of 1.3) and the $S_{Wb\bar{b}}$ heavy flavor 
factors in a zero $b$-tagged sample (after accounting for the 
theoretical heavy flavor K factor of 1.47 for $W+$ jet). The errors shown
are statistical errors only. The systematic uncertainty on the $K^{W\text{+jets}}$ ($S_{Wb\bar{b}}$) factors 
is 6\% (20\%) as described in Sec.\ X.\\
}
\hspace*{-1ex}\scalebox{0.92}{
\begin{tabular}{llccc}
\hline
\hline
    &  & $K^{W\text{+2 jets}}$ & $K^{W\text{+3 jets}}$ & $S_{Wb\bar{b}}$\\ \hline
\hline
Pre-2006 & $e$        & 1.10 $\pm$ 0.01 & 1.21  $\pm$ 0.03  & 0.78 $\pm$ 0.09 \\
    & $\mu$            & 1.16 $\pm$ 0.01 & 1.35  $\pm$ 0.03 &  0.99 $\pm$ 0.11 \\ 
Post-2006   & $e$      & 1.05 $\pm$ 0.01 & 1.12  $\pm$ 0.01 &  1.14 $\pm$ 0.06\\
      & $\mu$          & 1.10 $\pm$ 0.01 & 1.21  $\pm$ 0.01 &  1.02 $\pm$ 0.06 \\    
\hline
\hline
\end{tabular}
}
\end{center}
\end{table}

As indicated above, the factor $S_{Wb\bar{b}}$ is applied additionally only
to the $W b\bar{b}$ heavy parton events:
\begin{equation}
N^{\text{MC}}_{W\text{+jets}} = N^{\text{MC}}_{W\text{+light}} + S_{Wb\bar{b}} N^{\text{MC}}_{Wb\bar{b}}
\end{equation}
[the same factor is used for the $W (\rightarrow l\nu) b\bar{b}$ and $W (\rightarrow l\nu) c\bar{c}$ generated 
samples and for the corresponding $Z/\gamma^{*}$ heavy flavor samples]. The heavy flavor factor $S_{Wb\bar{b}}$ 
is extracted by requiring either zero, one, or two $b$-tags (see Sec.\ V) to obtain samples
containing $N^{\text{tag,data}}_{W\text{+jets}}$ events, however it is dominated by the single $b$-tag samples that have 
the smallest expected signal contribution. The number of predicted $W$+jet events, $N^{\text{tag,MC}}_{W\text{+jets}}$, 
in the tagged samples, 
after application of the scaling factor $K^{W\text{+jets}}$, is given by 
$N^{\text{tag,data}}_{W\text{+jets}} = K^{W\text{+jets}} N^{\text{tag,MC}}_{W\text{+jets}} $, and the heavy flavor contribution can therefore 
be extracted from
\begin{equation}
 N^{\text{tag,data}}_{W\text{+jets}}  = \frac{ N_{\text{data}} - N_{\text{MJ}} - N^{\text{MC}}_{\text{SM}}}{N^{\text{MC}}_{W\text{+light}} 
+ S_{Wb\bar{b}} N^{\text{MC}}_{Wb\bar{b}}} N^{\text{tag,MC}}_{W\text{+jet}}.  
\end{equation}

The heavy flavor scale factors, determined using samples requiring
zero $b$-tagged jets are also shown in Table \ref{SF}. The factors are 
applied separately for the electron and muon channel samples and also for 
data before and after the D0 upgrade. The 
luminosity weighted average of the factors is found
to be consistent with the theoretically expected value \cite{mcfm}.

\section{Multijet Background}

\begin{figure}[t]
\vskip -0.1cm
\hspace{-1.0cm} \includegraphics[width=3.4in]{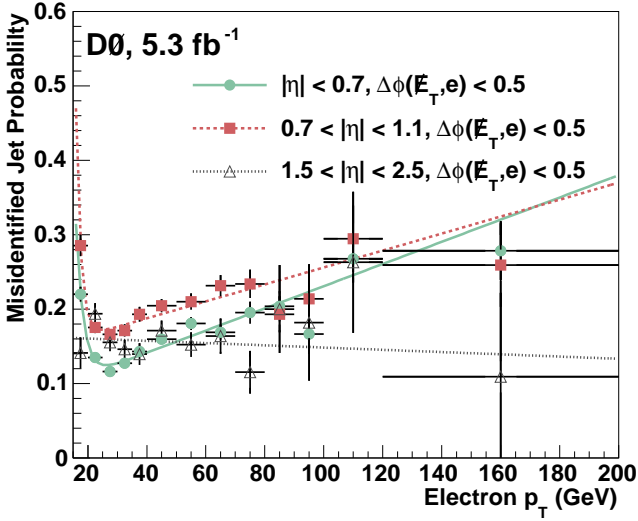}
\caption{[color online] The probability for MJ background events to enter the final $e+2$ jet (post-2006 upgrade) sample.
The solid, dashed, and dotted curves represent the result of fitted parameterizations in each interval.}
\label{fig:efakerate}
\end{figure}

The total MJ background contribution entering each of the final selected 
signal samples is determined from the data, prior to the application of $b$-tagging, using the prescription of Ref.\ \cite{ttbar-prd}. The MJ 
contributions are determined in conjunction with the previously described {{\sc alpgen}} 
normalization factors. Multijet 
templates are obtained from control samples in the data and normalized 
through a $\chi^{2}$ fit to the $W$ boson transverse mass distribution. For the
determination of the MJ contribution, the
{{\sc alpgen}} normalization factors described in 
Secs.\ VI A and VI B are varied in conjunction with the MJ normalization, such
that the total number of predicted MC and MJ events agrees with the 
total number of data events prior to the application of $b$-tagging.

For both the electron and muon selected events, additional data samples 
selected with the much looser lepton-identification criteria (see Sec.\ IV)
are used. Events entering the looser samples (L) are a combination of true leptonic events and
MJ background in which a jet is misidentified as a lepton. Upon application of the tighter (T) final selection criteria the 
remaining contributions of true leptonic and background events depend upon the 
(relative) efficiency $\epsilon^{\ell}_{\text{LT}}$ for true leptons to subsequently pass the final selection 
criteria, and the probability $P^{\text{MJ}}_{\text{LT}}$ that MJ background events 
in the looser sample subsequently 
enter the tighter, final signal samples. A weight $w_{i}$ is assigned to each 
event $i$ in the looser selected samples according to

\begin{equation}
w_{i} = \frac{P^{\text{MJ}}_{\text{LT},i}}{\epsilon^{\ell}_{\text{LT},i}-P^{\text{MJ}}_{\text{LT},i}} [\epsilon^{\ell}_{\text{LT},i} - \Theta^{i}]
\end{equation}
where $\Theta^{i} =  1$ if the event $i$ in the loose sample passes the tight selection 
requirements and is
zero otherwise. The total MJ background contributions in the final signal samples
are given by the sum of the event weights $w_{i}$ in the corresponding loose samples.
The efficiencies ${\epsilon^{\ell}_{\text{LT},i}}$ are functions of lepton $p_{T}$ and are determined from 
$Z/\gamma^{*} \rightarrow l^{+}l^{-}$ events.
The probabilities $P^{\text{MJ}}_{\text{LT},i}$ are determined from 
the measured ratio of the number of events 
in the final to loosely selected samples after correcting each 
sample for the expected MC contribution from the leptons in the specific kinematic interval. For both the final electron and muon samples, 
the probability for MJ events to enter the final selected samples is extracted in the region $5 <~ \MET < 15 ~\rm GeV$ [and without applying 
the additional requirement on $\MET$ in Eq.\ 7 of Sec.\ VIII].

\begin{figure}[t]
\vskip -0.1cm
\hspace{-0.5cm} \includegraphics[width=3.4in]{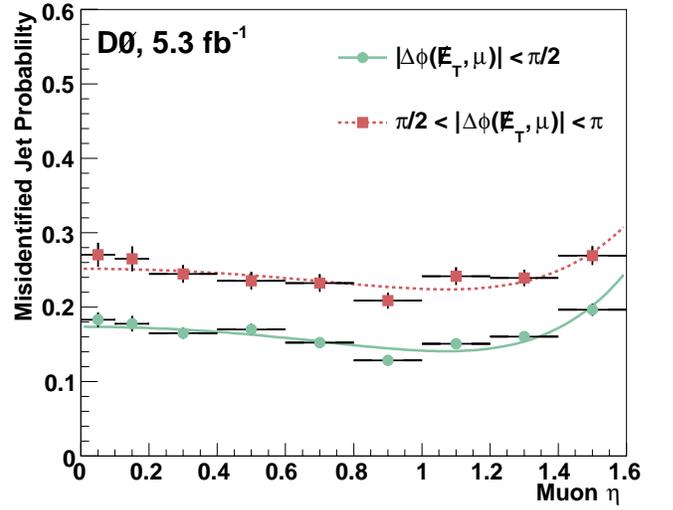}
\caption{[color online] The probability for MJ background events to enter the final $\mu+2$ jet (post-2006 upgrade) sample.
The solid and dashed curves represent the result of the fitted parameterizations in each interval.}
\label{fig:mufakerate}
\end{figure}

\subsection{Parameterization of the Misidentified Jet Probability}

\begin{figure*}[t]
\clearpage
\includegraphics[width=2.3in]{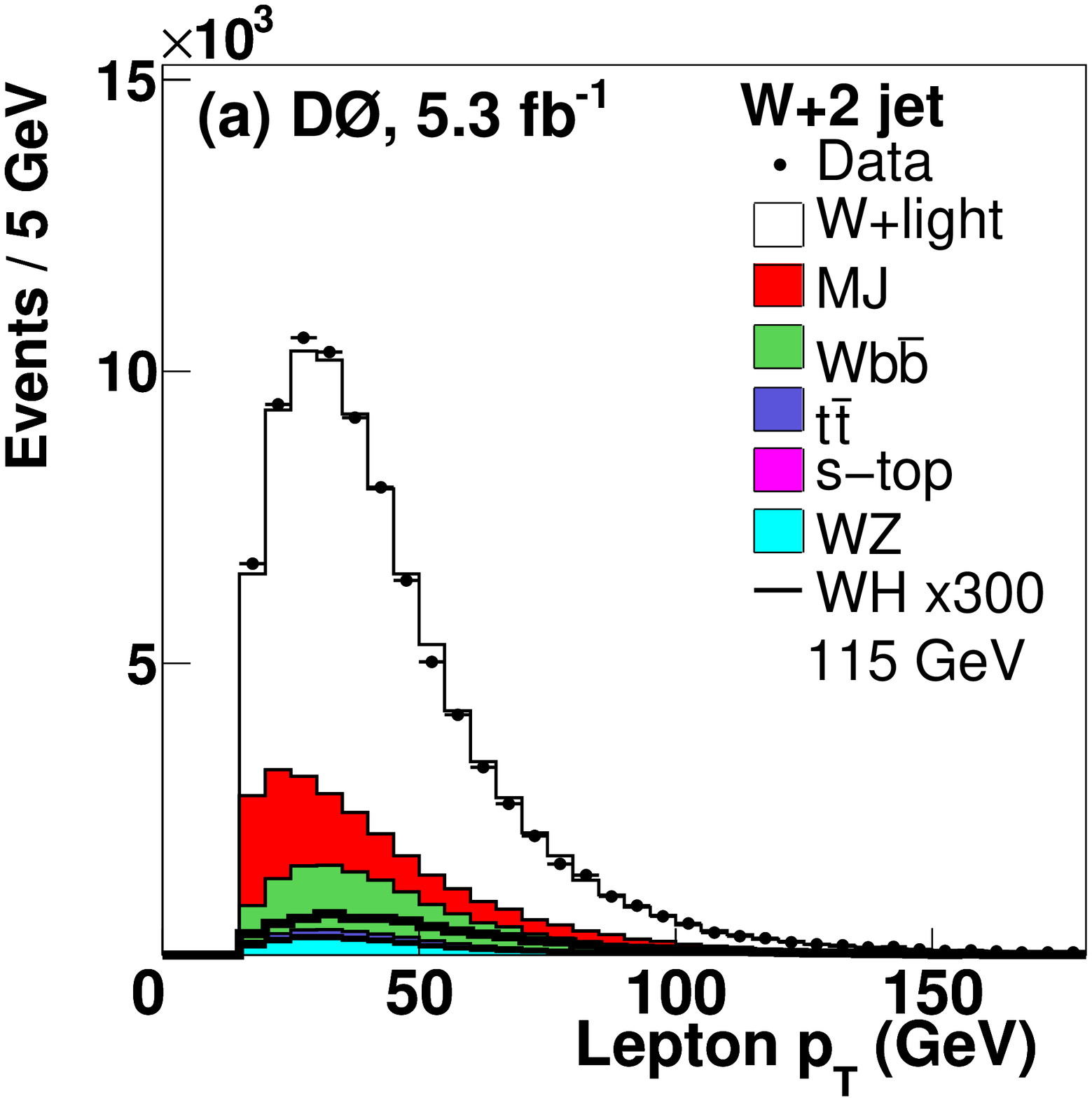}
\hskip 0.01cm \includegraphics[width=2.3in]{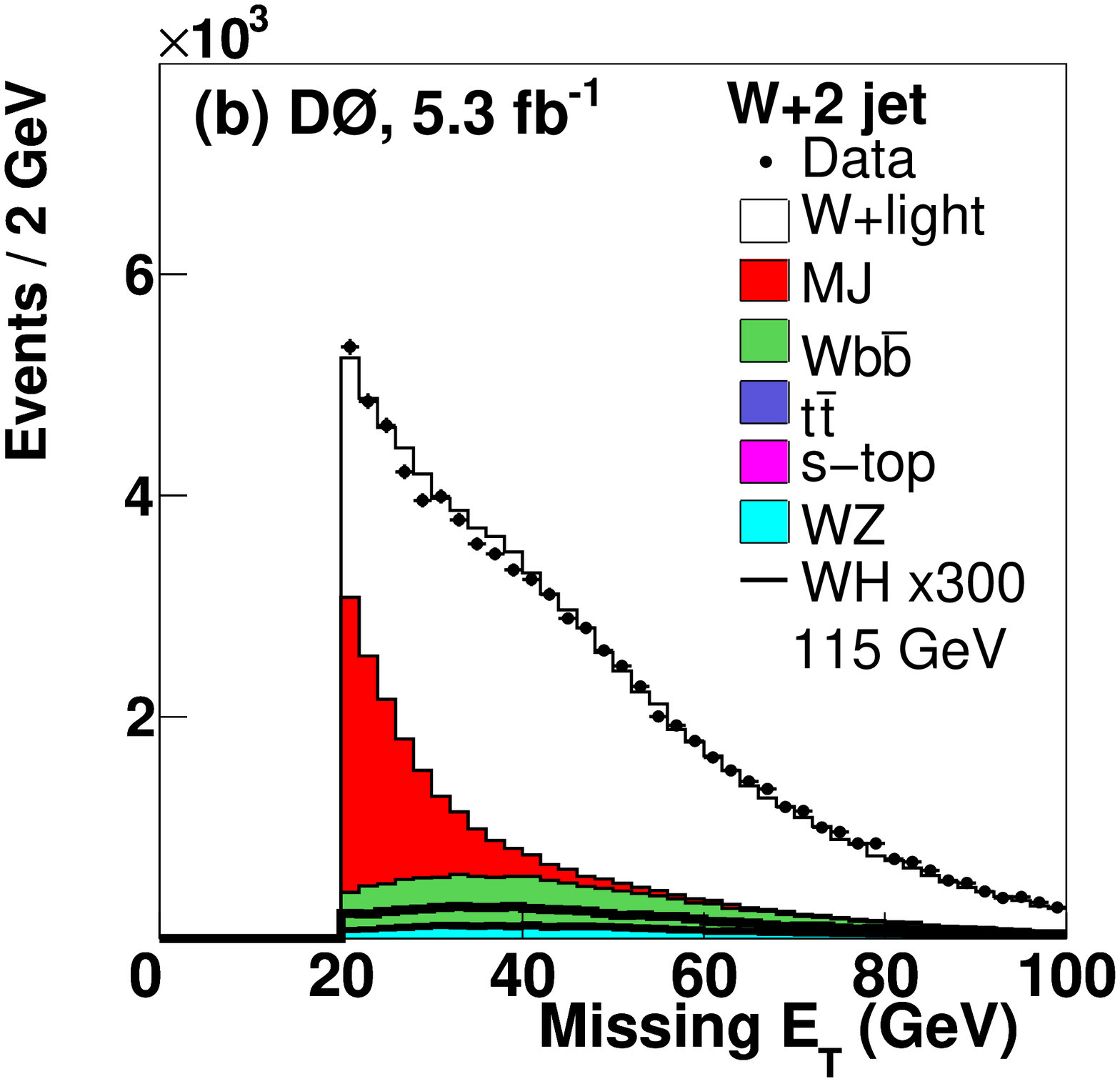}
\hskip 0.01cm \includegraphics[width=2.3in]{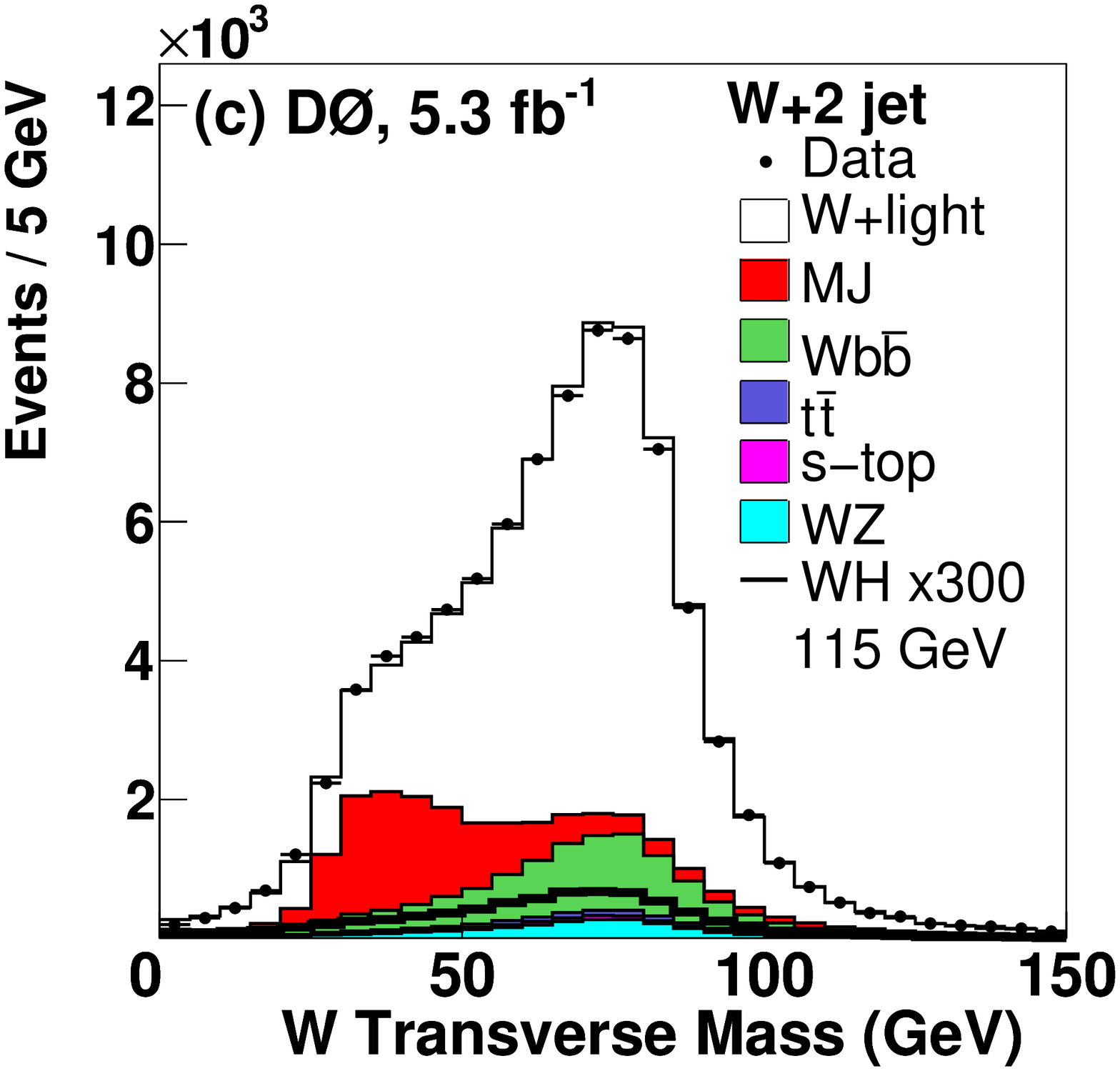}
\clearpage
\includegraphics[width=2.3in]{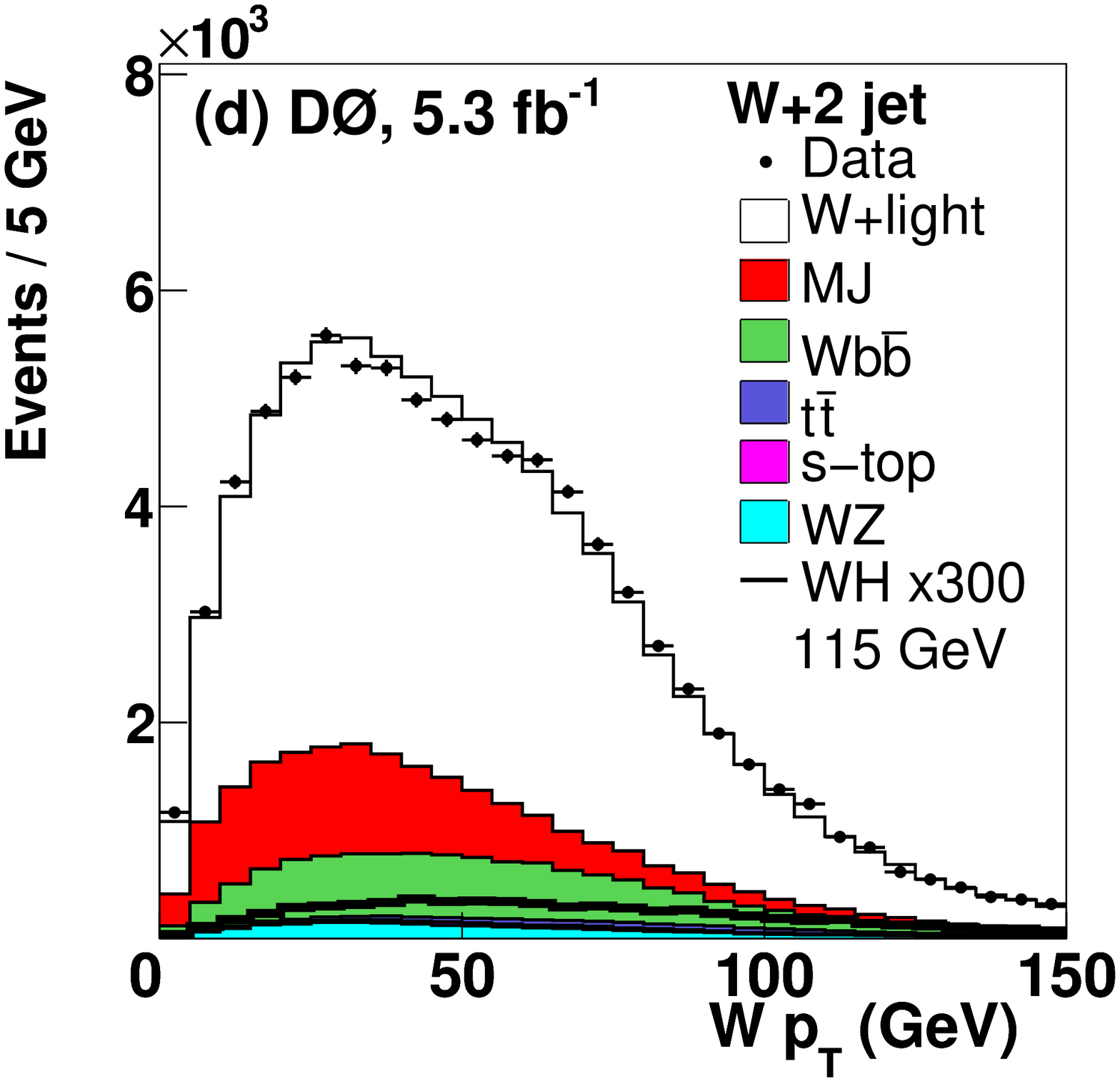}
\hskip 0.01cm \includegraphics[width=2.3in]{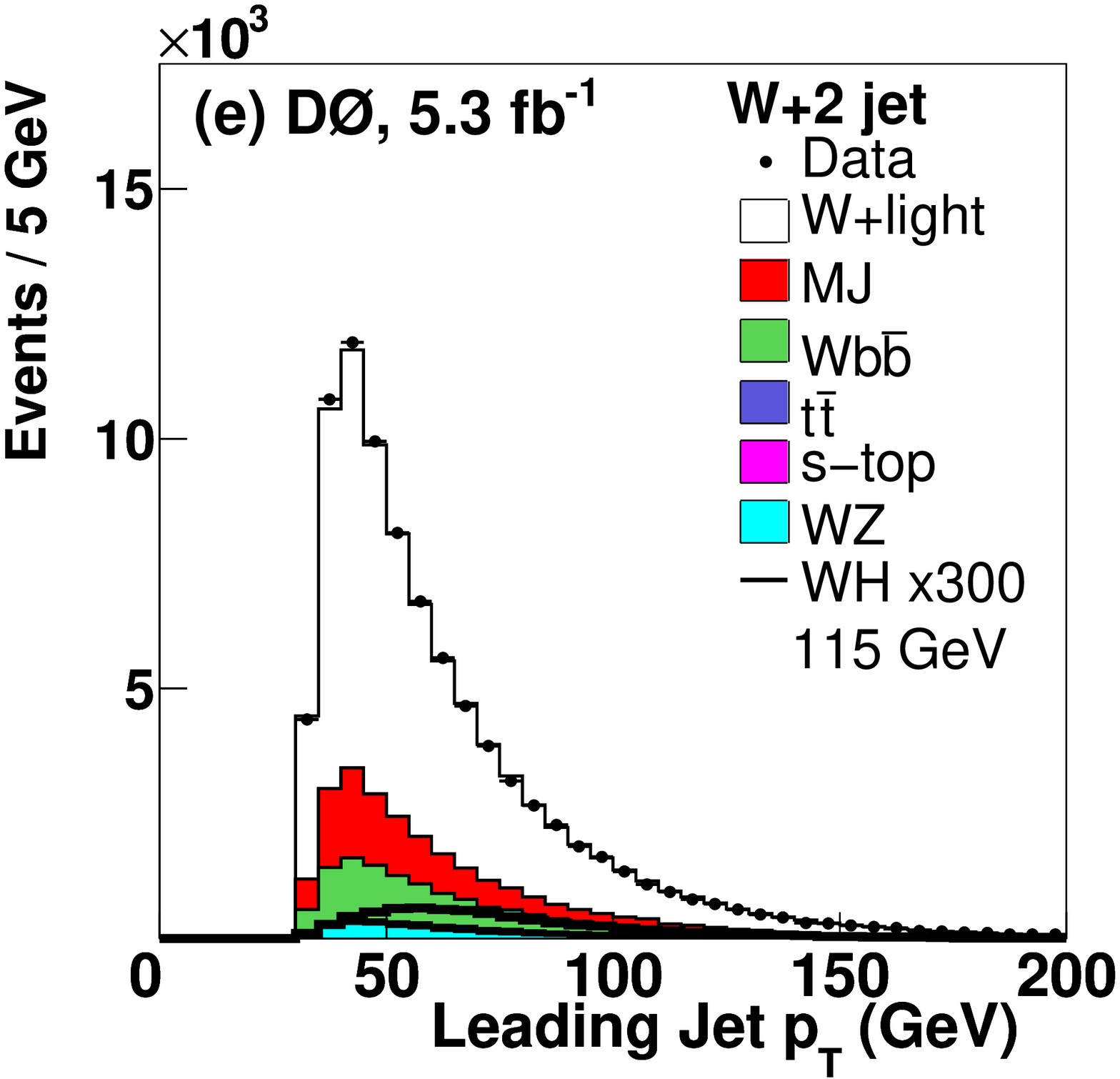}
\hskip 0.01cm \includegraphics[width=2.3in]{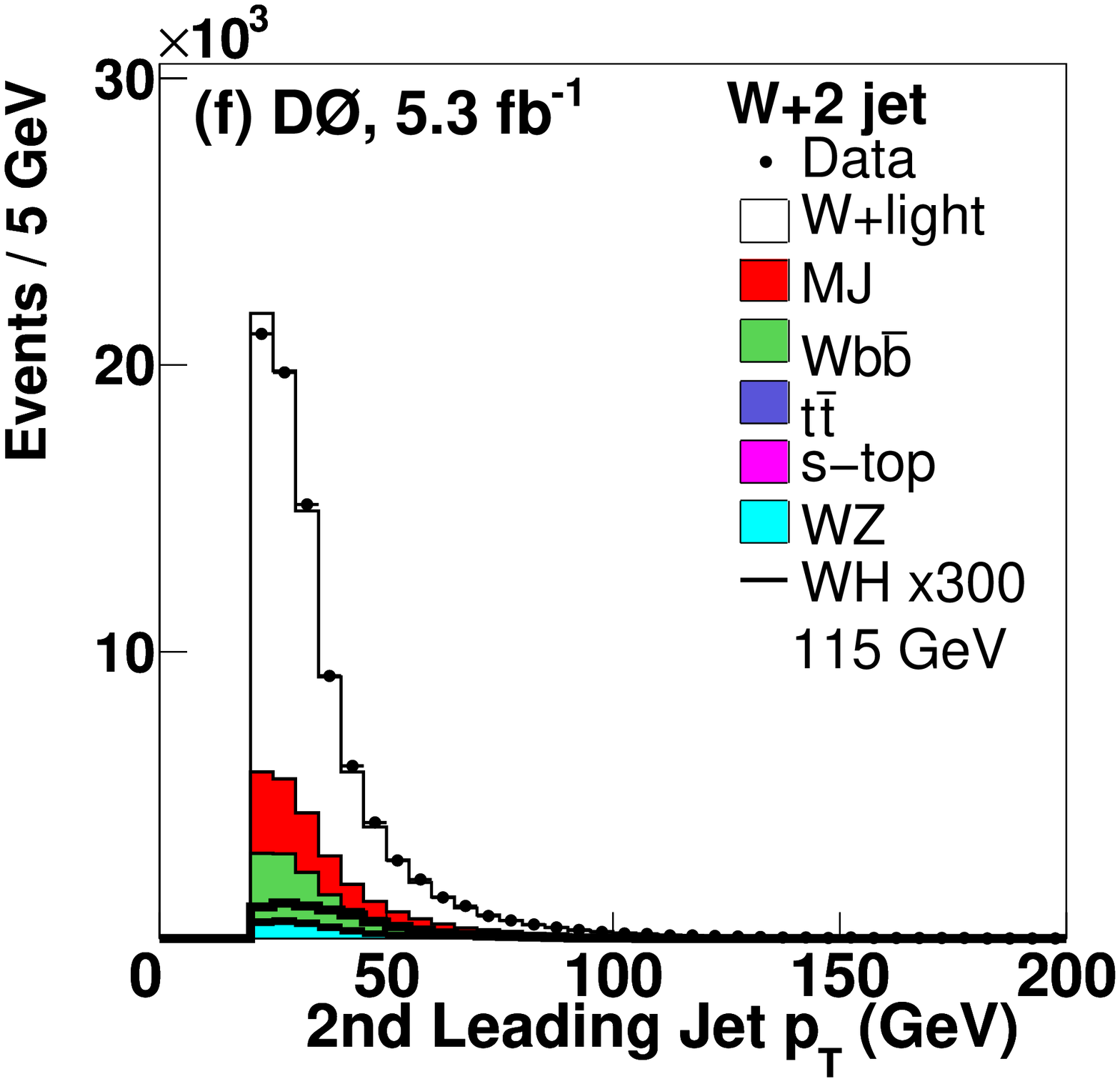}
\clearpage
\hskip -0.1cm \includegraphics[width=2.3in]{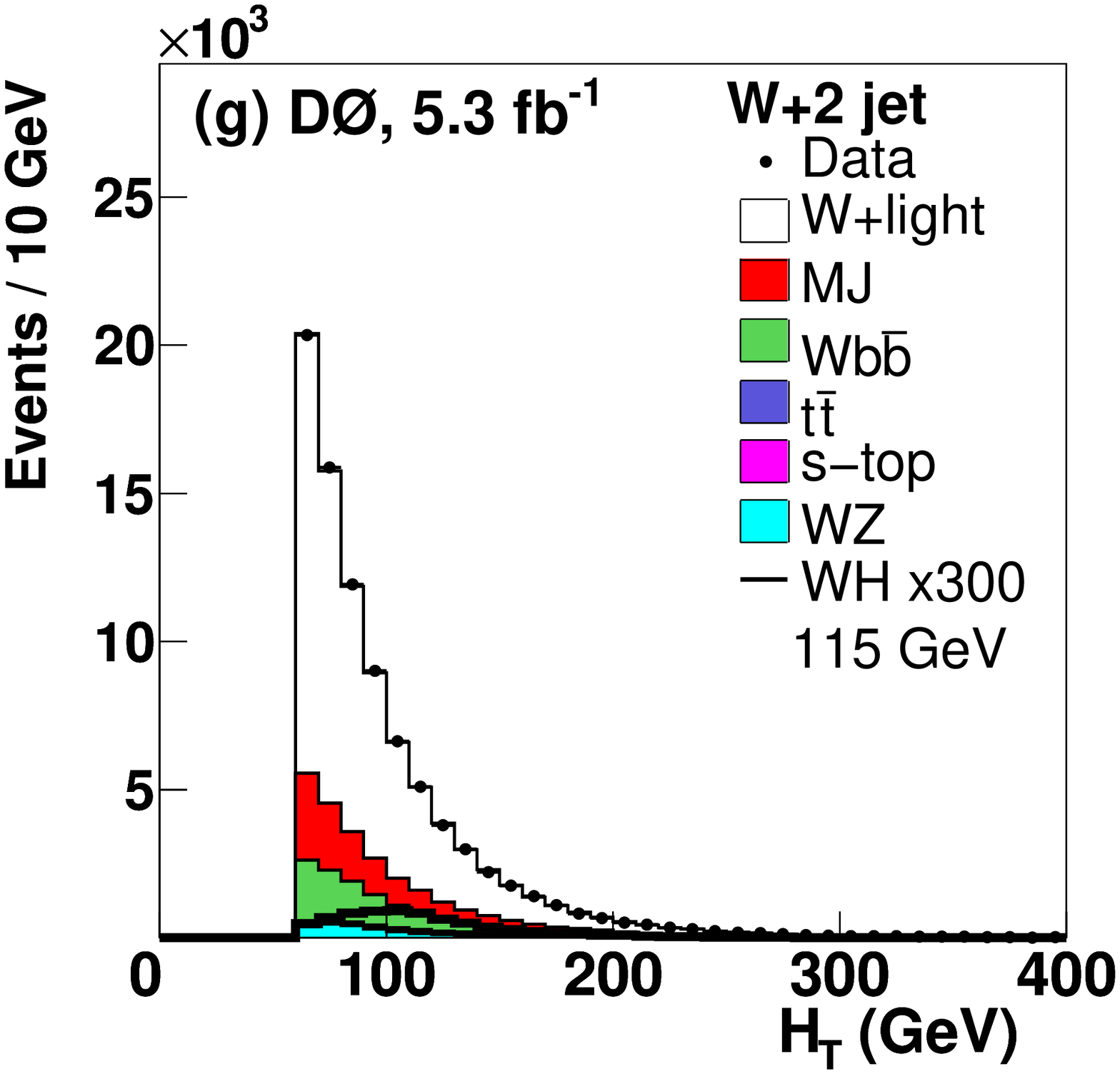}
\hskip 0.01cm \includegraphics[width=2.3in]{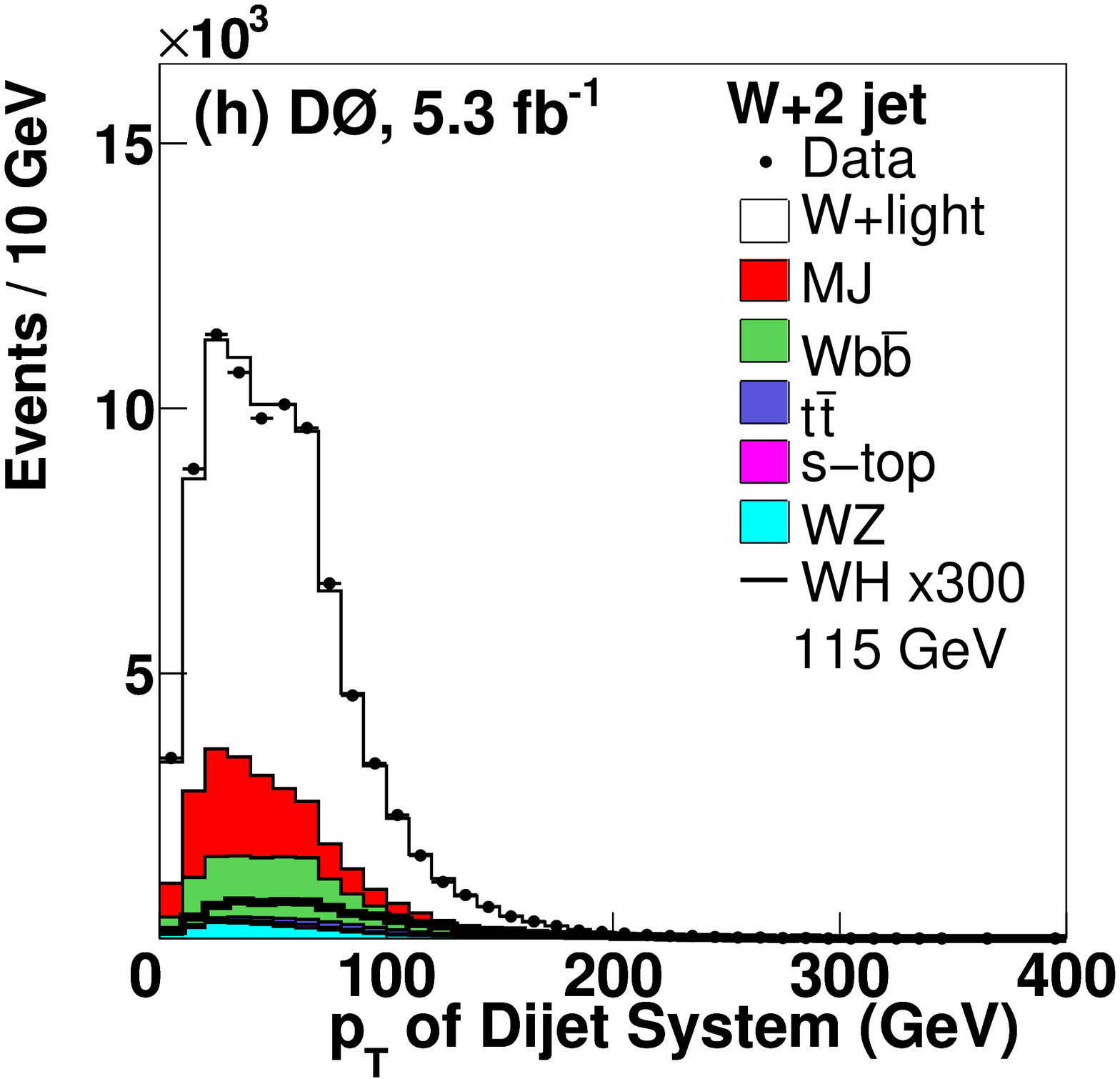}

\hskip -0.1cm \includegraphics[width=2.3in]{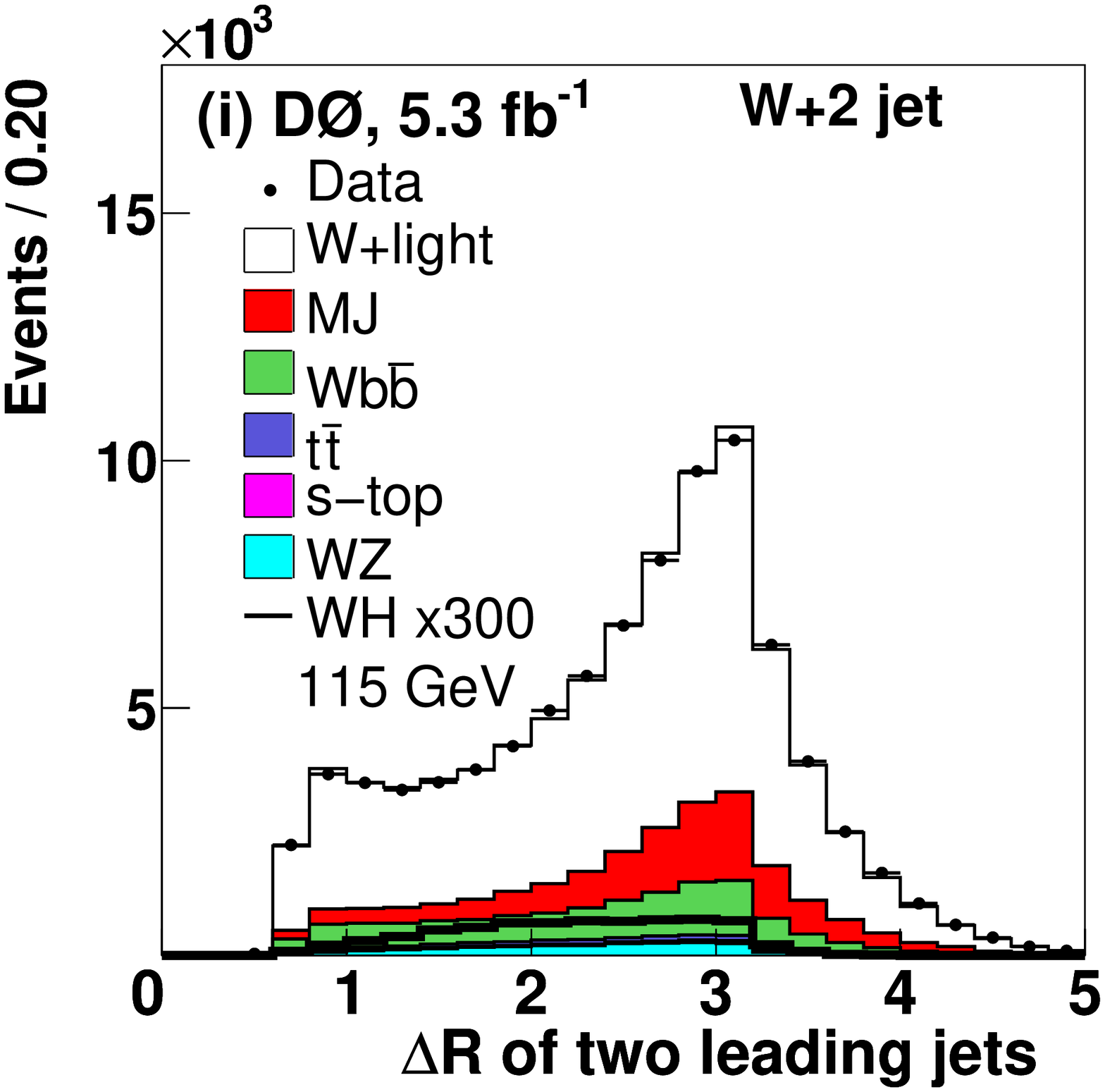}
\hskip 0.01cm \includegraphics[width=2.3in]{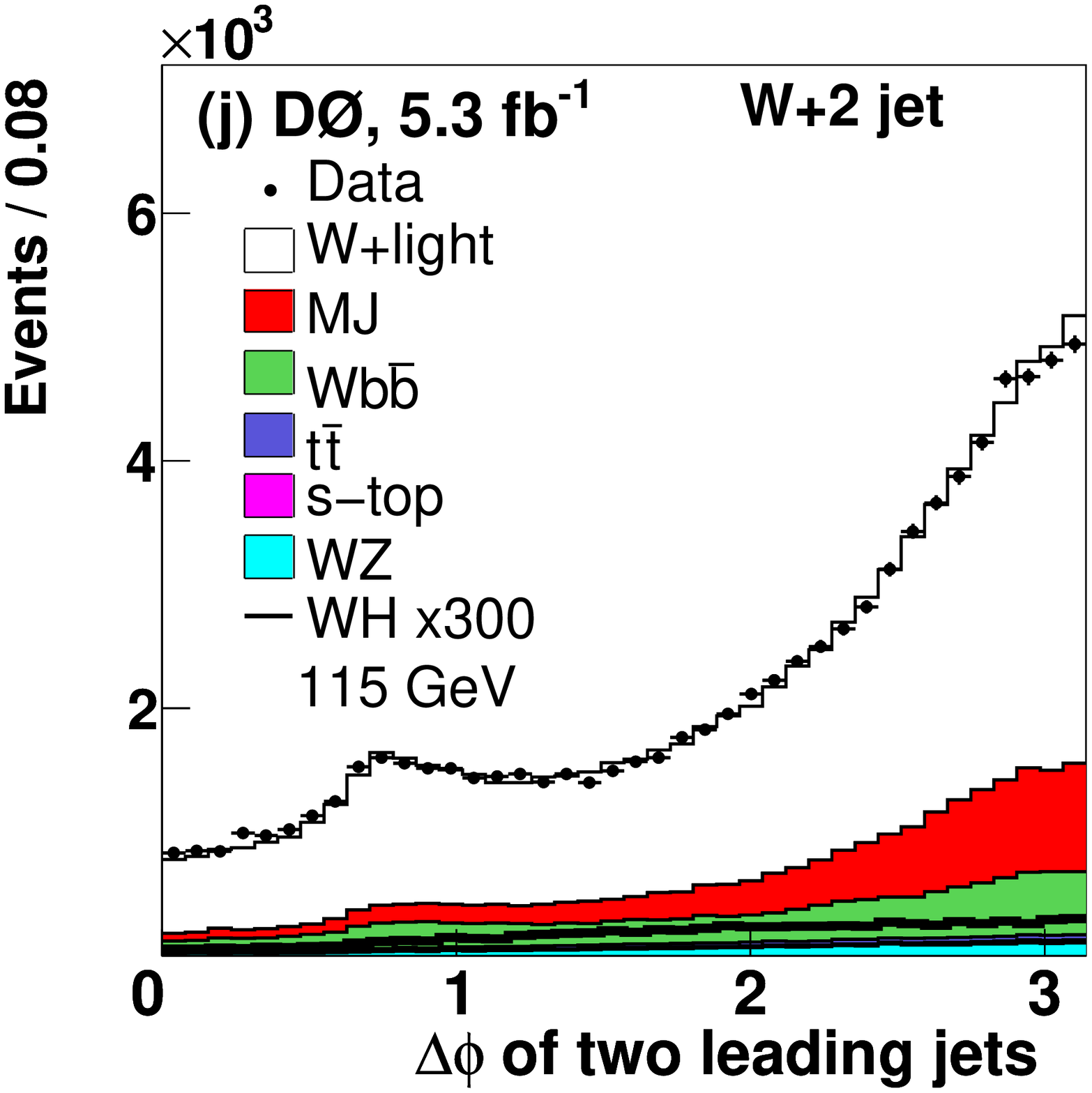}

\caption{[color online] Comparison of simulated events, including data-determined MJ background, to the $W$+2 jet selected data (black points)\
 for (a) isolated lepton $p_{T}$, (b) missing event transverse energy $E_{T}$, (c) transverse mass of the ($l,\MET$) system, (d) $p_{T}$ of $W\
$ boson candidates, (e) leading jet $p_{T}$, (f) $p_{T}$ of the second leading jet, (g) scalar sum of the $p_{T}$
of jets in the event ($H_{T}$), (h) transverse momentum of the dijet system, (i) separation $\Delta R$, and (j) azimuthal separation
$\Delta \phi$ for the two jets. The expectation for a $WH$ signal at $M_{H} = 115 ~\rm GeV$  has been scaled up by a factor of 300.  The
electron and muon selected samples are
combined in the figures.}
\label{fig:fig3}
\vskip -0.2cm
\end{figure*}

\begin{figure*}[t]
\clearpage
\hskip -0.2cm \includegraphics[width=2.3in]{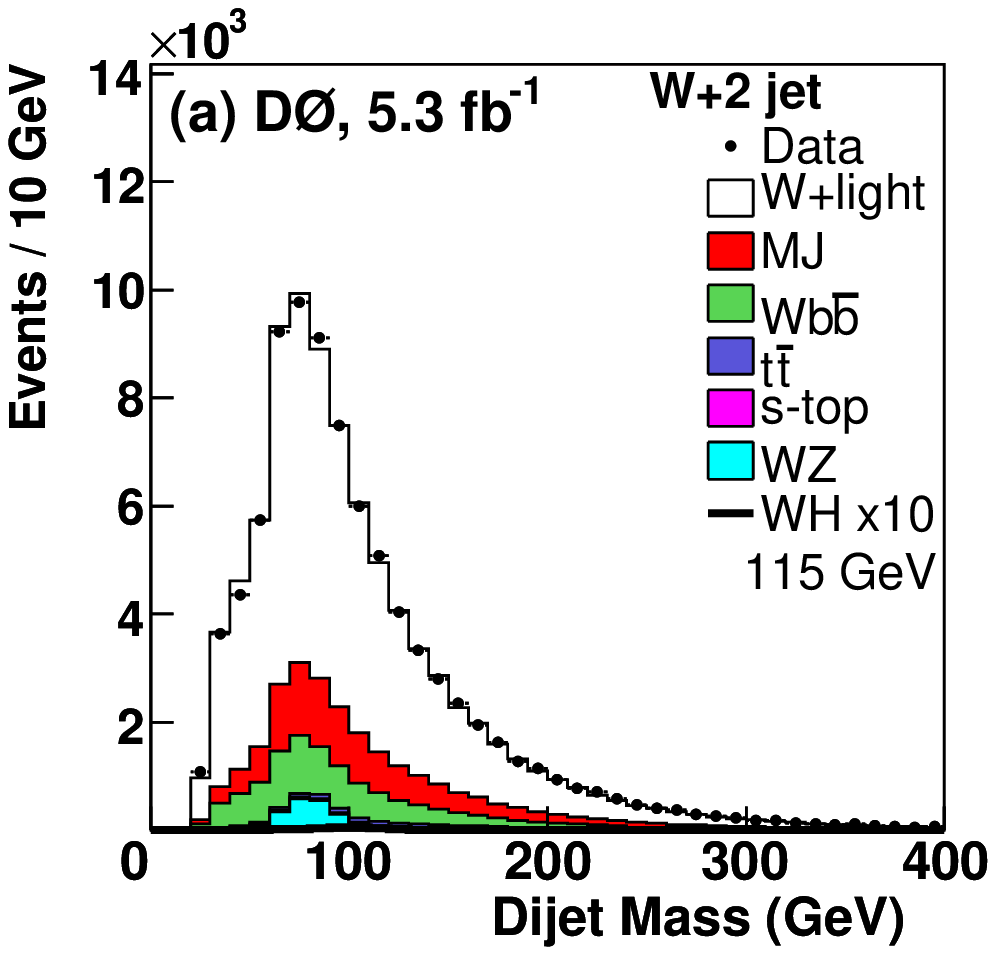}
\hskip 0.2cm  \includegraphics[width=2.3in]{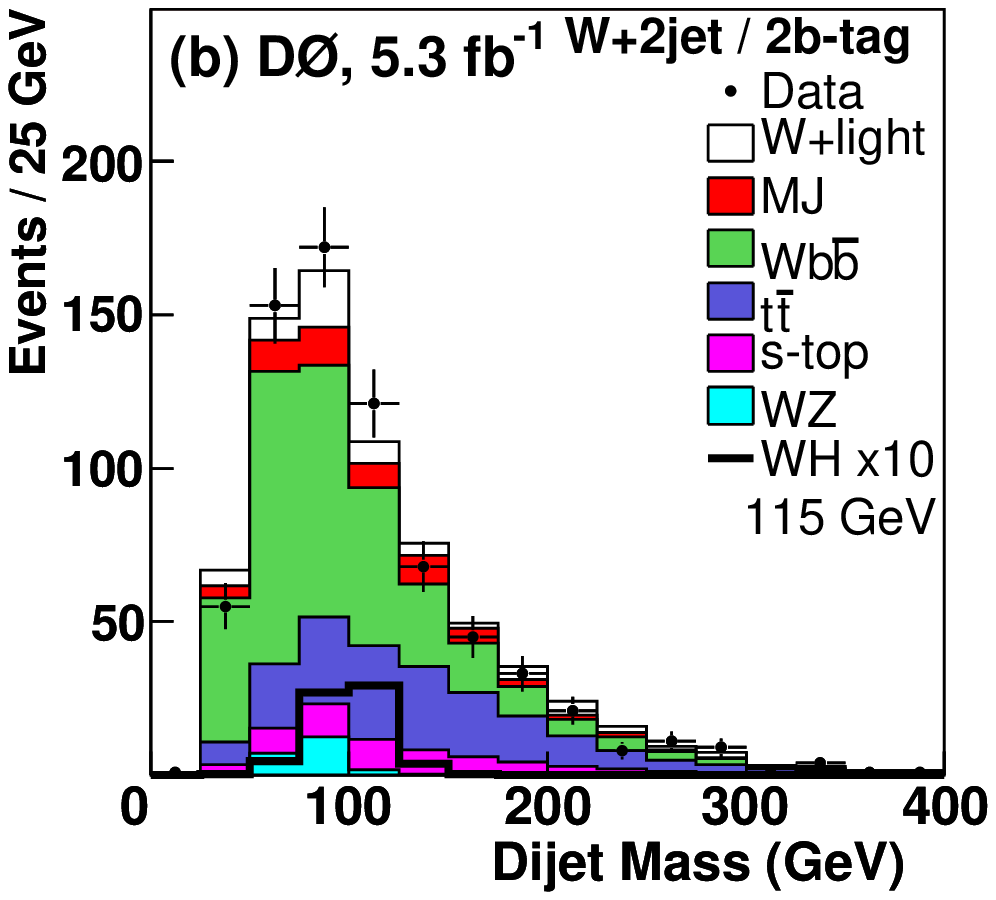}
\hskip 0.2cm  \includegraphics[width=2.3in]{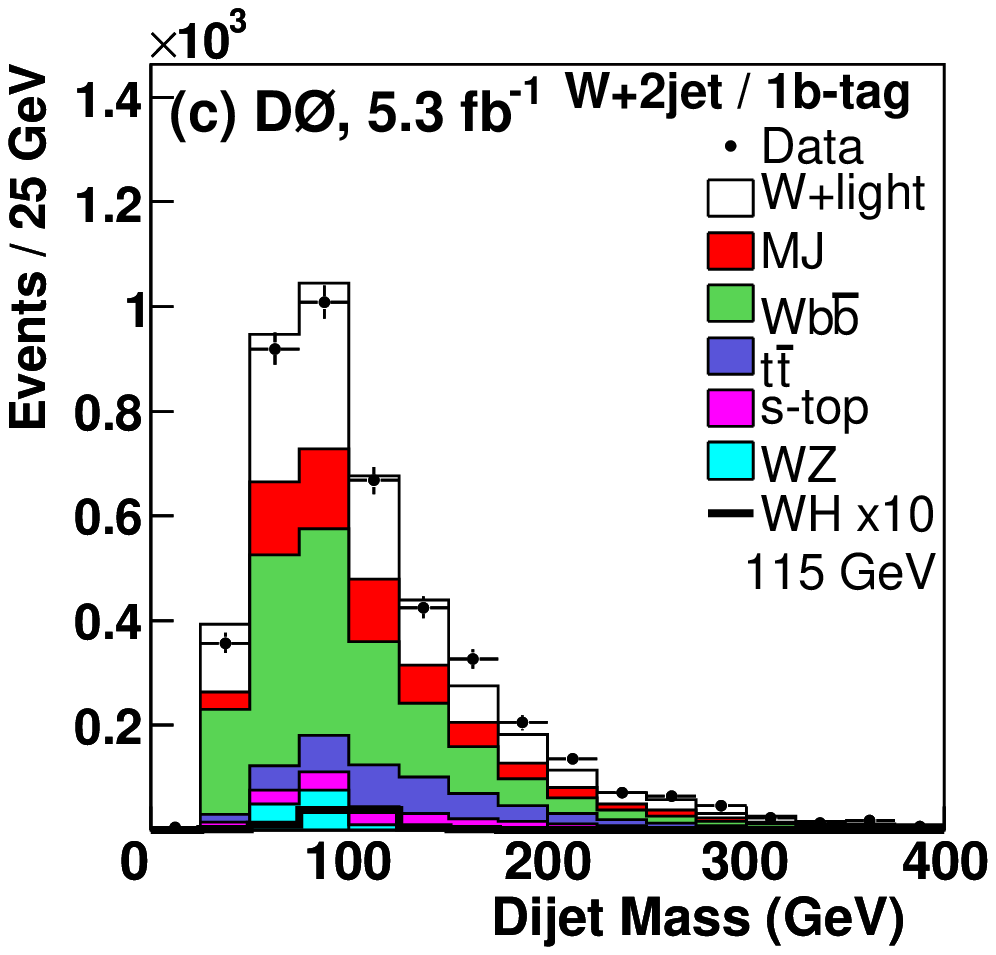}
\clearpage
\hskip -0.3cm \includegraphics[width=2.44in]{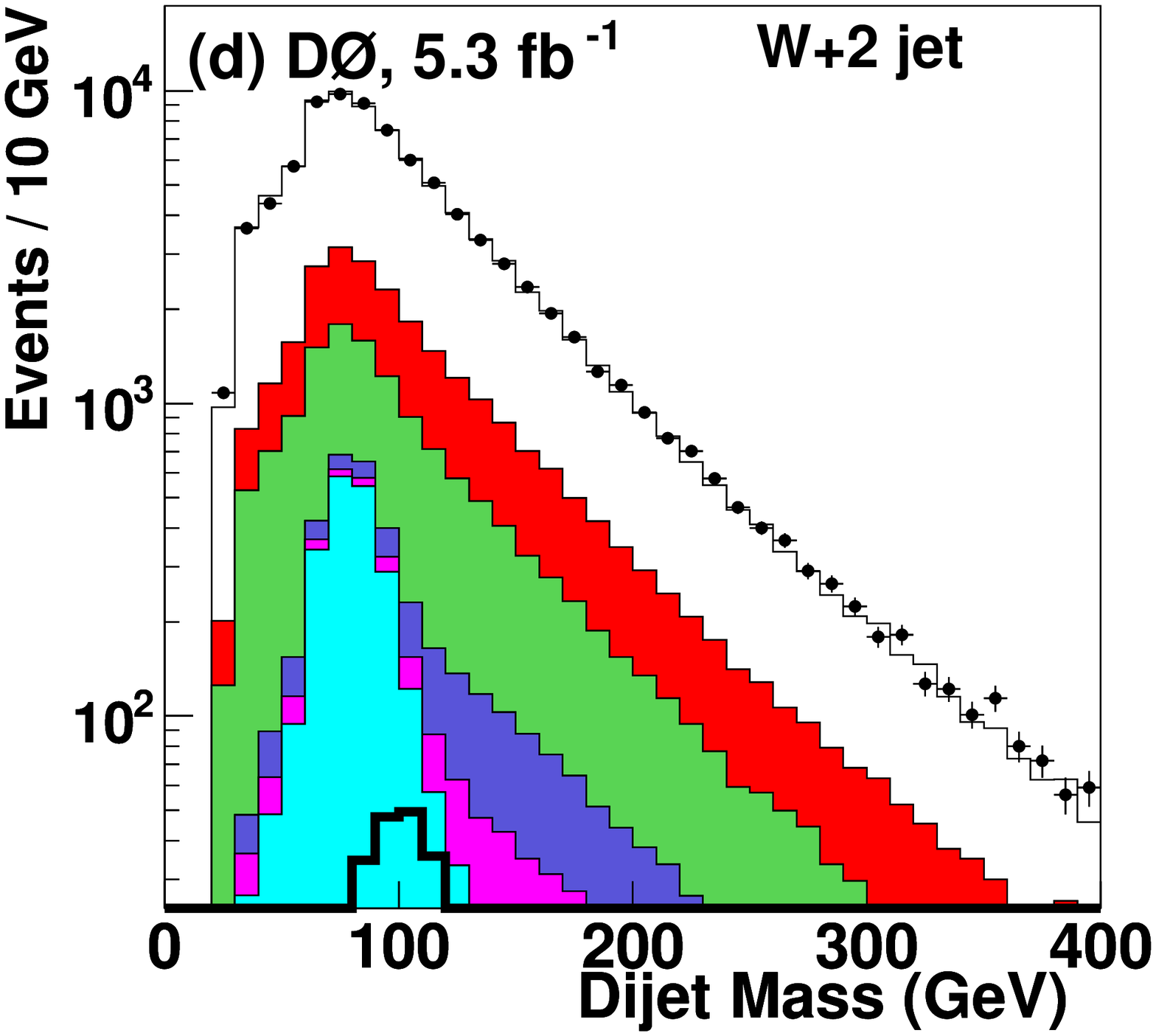}
\hskip -0.1cm \includegraphics[width=2.3in]{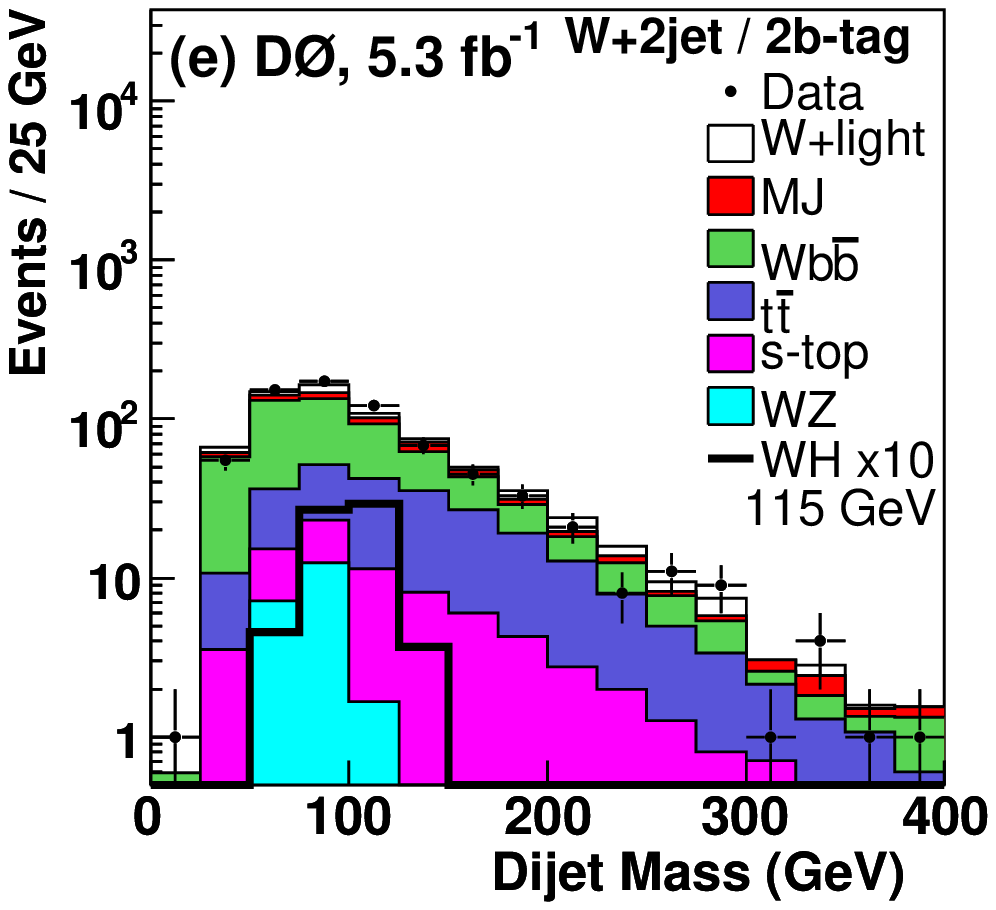}
\hskip 0.2cm  \includegraphics[width=2.3in]{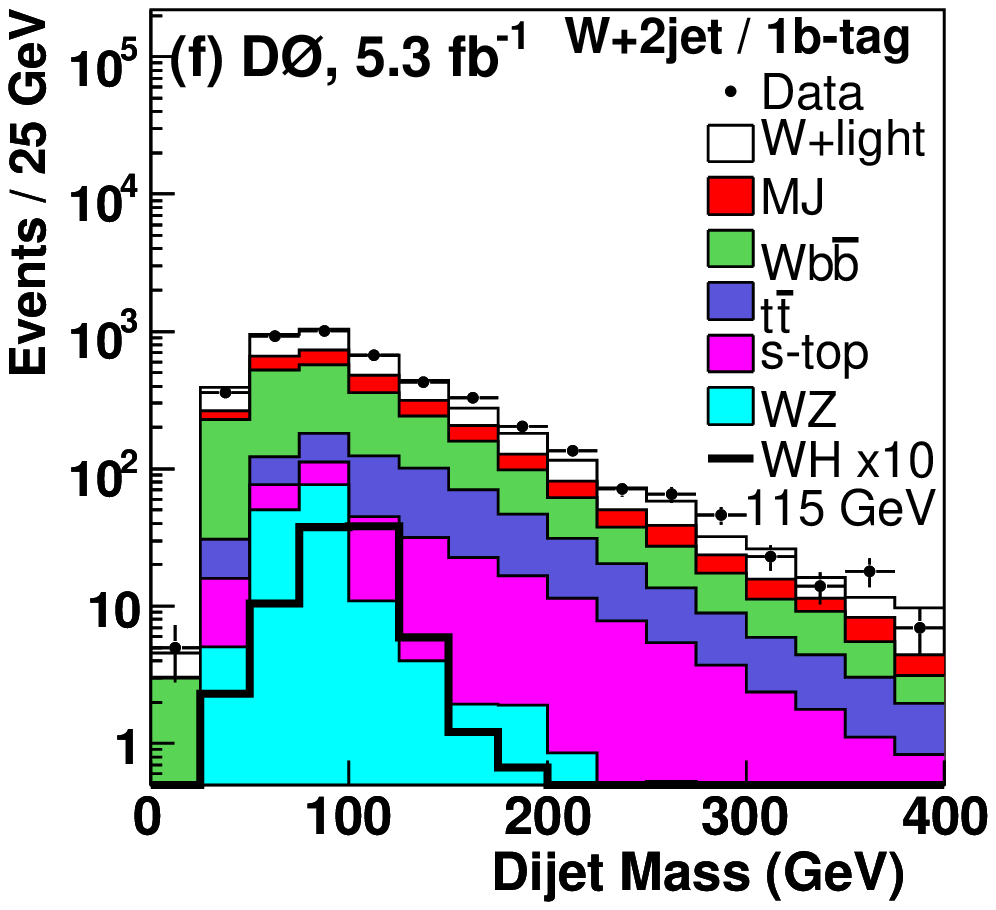}

\caption{[color online] Dijet invariant mass distribution for the $W$+2 jet selected
samples on linear and logarithmic scales for (a), (d) no $b$-tagging, (b), (e)
events that contain two $b$-tagged jets, and (c), (f) events that fail the
two-tagged requirement but contain a single NN $b$-tagged jet. The expectation for a $WH$ signal at $M_{H} = 115 ~\rm GeV$  has been scaled
up by a factor of 10.
}

\label{fig:two_taggedmass}
\end{figure*}

The measured probability for MJ events to enter the final electron plus two-jet 
selection sample is shown as a function of electron $p_{T}$ in Fig.\ \ref{fig:efakerate}. 
The MJ contribution in the electron channel arises from jets with a high 
enough fraction of energy 
deposited within the EM section of the calorimeter that they satisfy the 
electron identification criteria.
Additional MJ backgrounds in the electron channel originate from the semileptonic 
decays of hadrons and from photons that are misidentified as electrons.
The probability is measured separately in two CC regions ($|\eta|<0.7$ and $0.7<|\eta|<1.1$)
and separately in the EC region ($1.5<|\eta|<2.5$). In each range of $|\eta|$, the misidentified jet probability 
is parameterized as a function of electron $p_{T}$ in four intervals of the 
azimuthal separation $\Delta \phi (\MET, e)$ of the electron and the $\MET$ vector (the four regions are shown
combined for each $|\eta|$ interval in  Fig.\ \ref{fig:efakerate}). In the CC region, the  
probabilities are parameterized as sums of exponentials and 
first-order polynomials in electron $p_{T}$, whereas only a first-order polynomial 
in electron  $p_{T}$ is used in the smaller statistics EC region. For the smaller statistics
electron $W$+3 jet sample, the probabilities are determined once for each $|\eta|$ 
region, and are applied to each $\Delta \phi (\MET, e)$ interval separately after 
scaling to the average contribution obtained in that interval.

The measured probability for MJ background events to enter the final $\mu+2$jet 
sample is shown as a function of muon $|\eta|$ in Fig.\ \ref{fig:mufakerate}. 
The primary source of MJ background in the muon channel is from semileptonic decays of 
heavy quarks in which the decay muon satisfies the muon isolation criteria.
The contribution of MJ events entering the loose sample is smaller in the 
muon channel than in the electron channel. Consequently, the misidentified jet probability is parameterized in only two regions 
[$|\Delta\phi(\MET,\mu)| < \pi/2$ and 
$\pi/2 < |\Delta\phi(\MET,\mu)| < \pi$] of azimuthal 
separation  $\Delta \phi (\MET, \mu)$ between the muon $p_{T}$ and the $\MET$ vectors. In both pre-2006 and post-2006 upgrade data, the 
misidentification probability 
is parameterized using a third-order polynomial in muon $|\eta|^{2}$. 
The same functions are applied to both the muon $W$+2 jet and $W$+3 jet selected samples.

\section{Event Selection}

This section describes the selection of data samples containing events with a single reconstructed lepton, $\MET$, and either two or three jets 
of transverse momentum $p_{T} > 20 ~\rm GeV$, at least one of which is required to be consistent with having evolved from a $b$ quark. 
The samples are from data collected between 2002 and June 2009 at $\sqrt{s}=1.96~\rm TeV$. Candidate $W$ 
bosons are selected by requiring an electron or a muon with transverse momenta 
$p_{T}>15~\rm GeV$ and $\MET>20~\rm GeV$. 
Electrons are required to be
in the pseudorapidity region  $|\eta|<1.1$ or $1.5 \le |\eta| \le 2.5$ and muons in the region $|\eta| < 1.6$.  The selected $W\rightarrow e \nu$ 
and $W\rightarrow \mu \nu$  candidate events are divided into samples containing exactly two or exactly three reconstructed jets. Jets are required
to be in the region $|\eta| < 2.5$. A selection on the $H_{T}$ of the jets, $H_{T} > 60$ and $> 80 ~\rm GeV$, is also applied to 
the $W$+2 jet and $W$+3 jet samples, respectively, and the event PV is required to be reconstructed within $z_{PV}=\pm 40~\rm cm$ of the center 
of the detector. At least three charged tracks are required 
to be associated with that vertex. 

\begin{figure*}[t]
\clearpage
\hskip -0.2cm \includegraphics[width=2.3in]{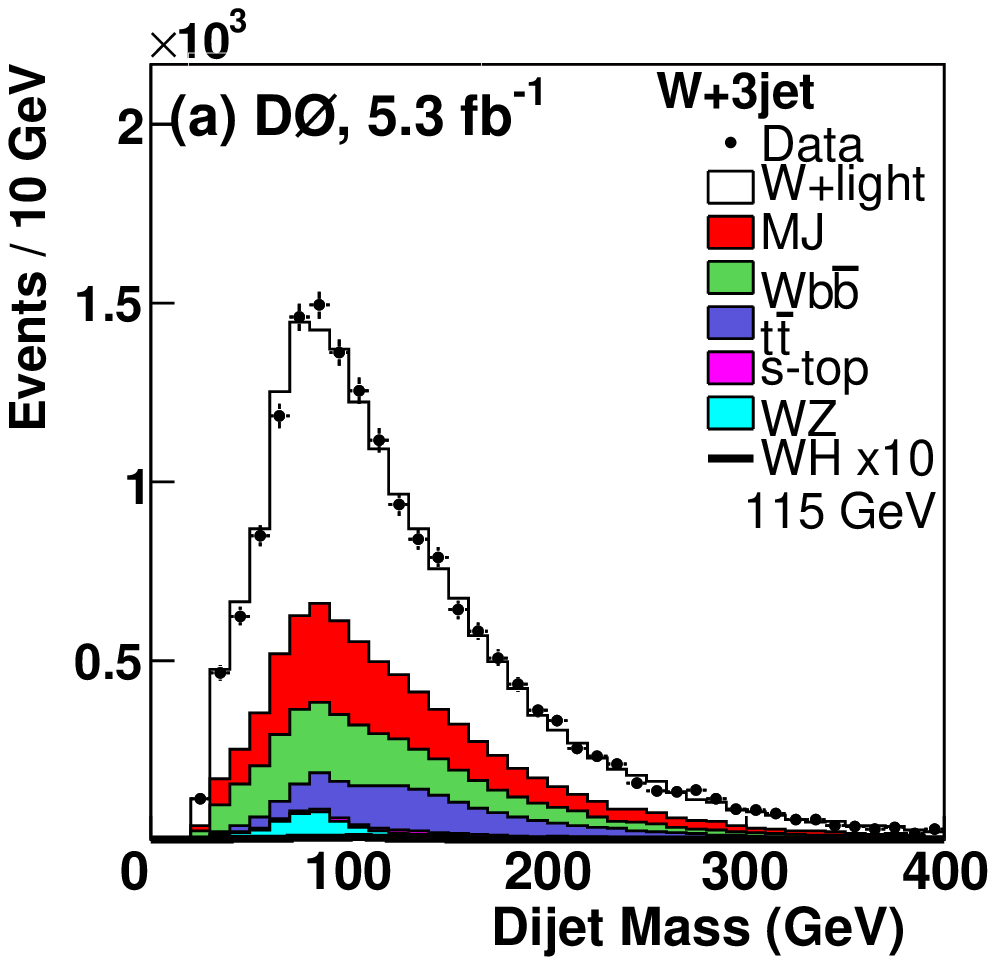}
\hskip 0.2cm   \includegraphics[width=2.3in]{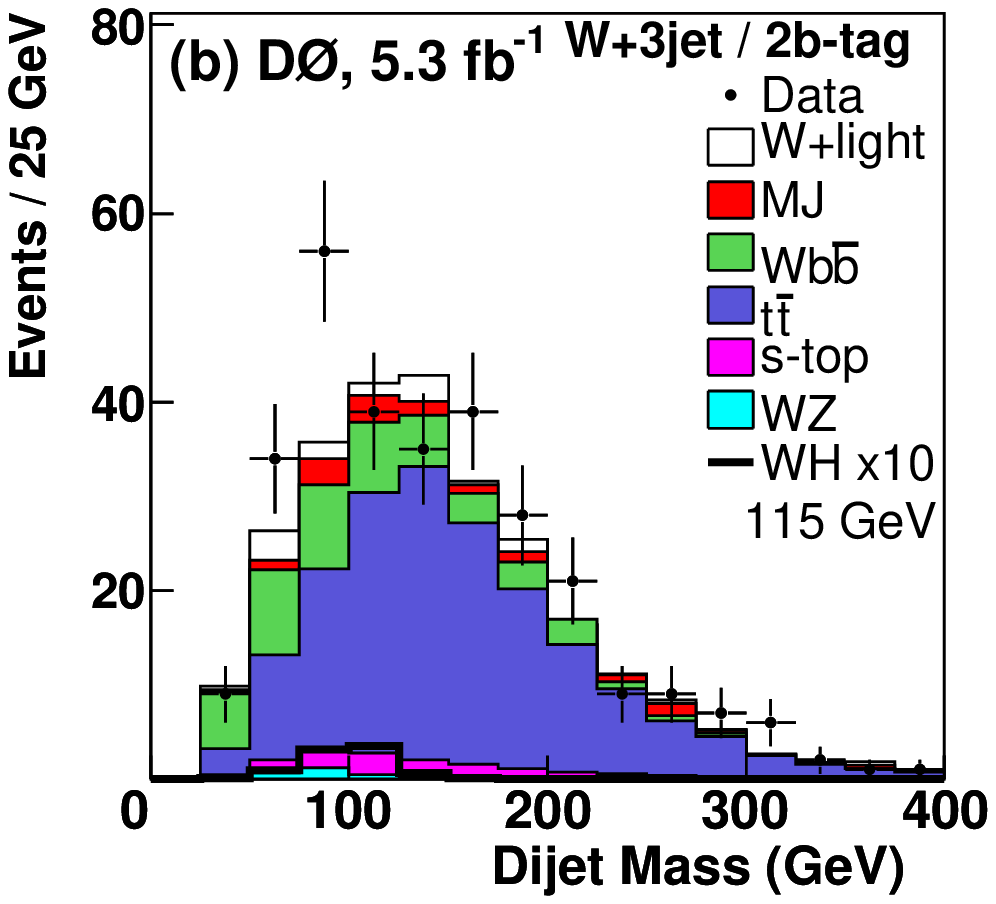}
\hskip 0.2cm \includegraphics[width=2.3in]{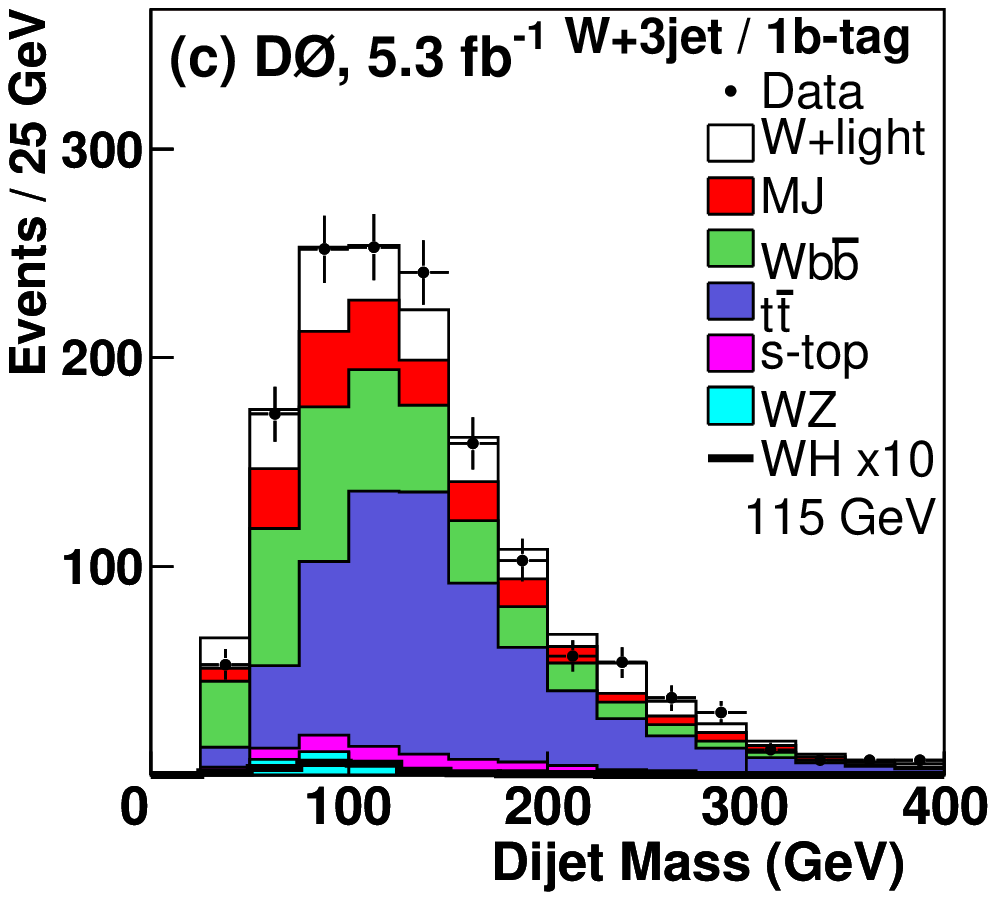}
\clearpage
\hskip -0.3cm \includegraphics[width=2.44in]{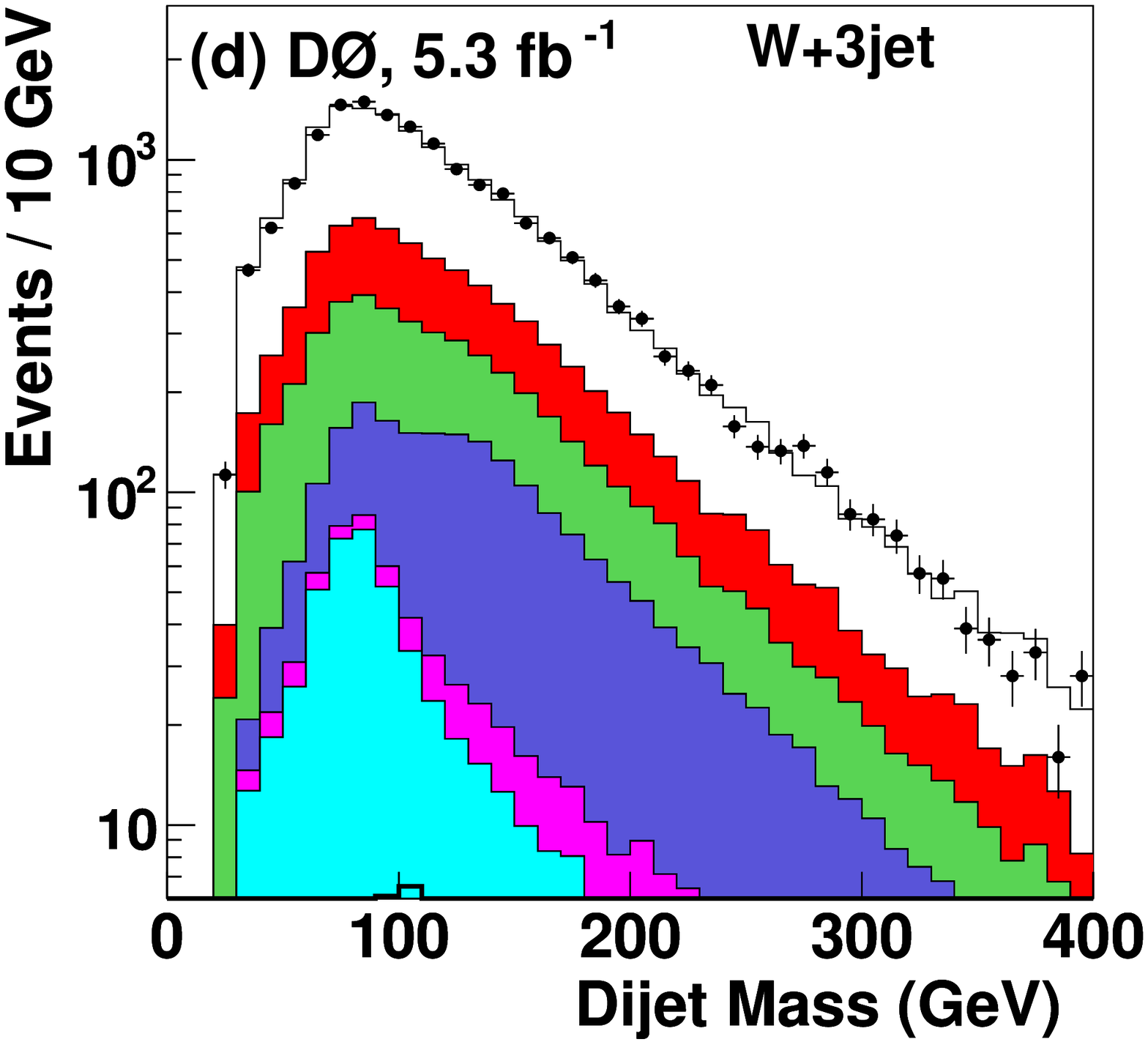}
\hskip -0.1cm \includegraphics[width=2.3in]{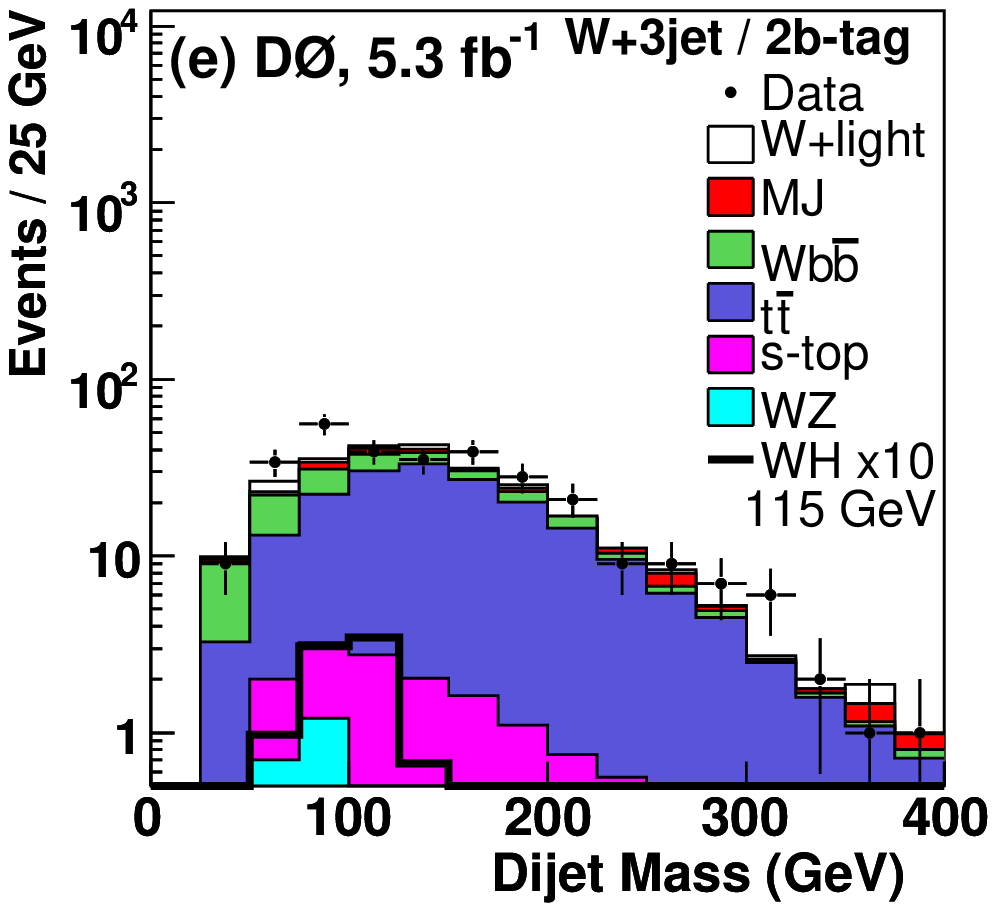}
\hskip 0.2cm \includegraphics[width=2.3in]{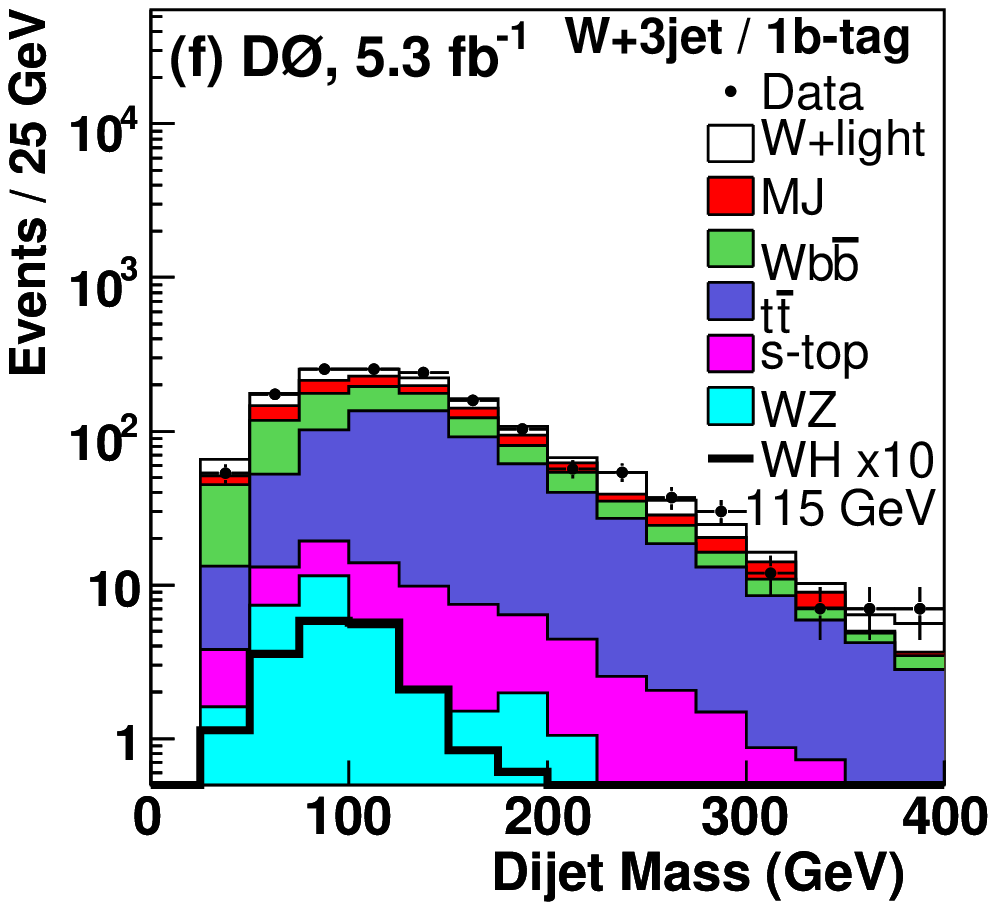}
\clearpage
\caption{[color online] Dijet invariant mass distribution for the $W$+3 jet selected
samples on linear and logarithmic scales for (a), (d) no $b$-tagging, (b), (e)
events that contain two $b$-tagged jets, and (c), (f) events that fail the
two-tagged requirement but contain a single NN $b$-tagged jet (the two
leading jets in the $W$+3 jet samples are used to form the dijet invariant mass). The expectation for a $WH$ signal at 
$M_{H} = 115 ~\rm GeV$ has been scaled up by a factor of 10. }
\label{fig:three_taggedmass}
\end{figure*}

Distributions of lepton $p_{T}$ and $\MET$ are compared to the sum of the expected SM background contributions and data-determined MJ background 
for the $W$+2 jet selected sample, which has the
largest statistics of all selected samples, in 
Figs.\ \ref{fig:fig3}(a) and (b).
The electron and muon decay channel samples are combined in the figures, and all
corrections to the background simulations have been applied. 
Details of the background estimates are given in Secs.\ VI and VII.

To suppress $Z/\gamma^{*} \rightarrow \ell^{+}\ell^{-}$ and $t\bar{t}$ background events and to avoid double counting 
events in Higgs searches based on dilepton final states, events with additional electrons or 
muons isolated from jets that pass 
$p_{T}^{~e} > 20~\rm GeV$ and $p_{T}^{\mu} > 15 ~\rm GeV$ are rejected. Events 
containing isolated high-$p_{T}$ $\tau$ leptons that decay hadronically are also rejected by 
requiring $p_{T}^{\tau} < 10 ~\rm  GeV$ or $p_{T}^{\tau} < 15 ~\rm GeV$, depending on the $\tau$ decay channel \cite{tauid}.

The transverse mass of the $W$ boson candidates ($M_{T}^{W}$) is reconstructed 
from the ($\ell, \MET$) system using the lepton transverse energy ($E_{T}^{\ell}$), $\MET$, and the azimuthal separation $\Delta \phi(\ell,\MET)$ 
between the isolated lepton and the $\MET$ vector:
\begin{equation}
M_{T}^{W} = [2 E_{T}^{\ell} \MET [1-\cos\Delta \phi(\ell,\MET)]]^{\frac{1}{2}}.
\end{equation}
The distribution of $M_{T}^{W}$ for selected $W$ boson candidates is shown in 
Fig.\ \ref{fig:fig3}(c). In addition to the dominant
contribution from events with real $W$ boson decays, there is a
significant component from MJ events that
contributes mainly at small values of $M_{T}$. Consequently the lower signal-to-background region at low $\MET$ is rejected by requiring
\begin{equation}
M_{T}^{W} > 40 {\rm (GeV)} -0.5 \MET. 
\end{equation}

The $p_{T}^{W}$ distribution for the $W$ boson candidates is 
compared to the sum of the expected SM and MJ background 
contributions, prior to the requirement in Eq.\ 7, in Fig.\ \ref{fig:fig3}(d). 

Kinematic jet properties for the selected $W$+2 jet sample are also compared 
to the sum of the expected SM background contributions, including MJ background, in 
Fig.\ \ref{fig:fig3}. The corrected electron
and muon channel samples are combined in the figure. 
The background prediction provides an adequate description of the data 
for all the distributions. 

\begin{figure*}[t]
\clearpage
\hskip -0.2cm \includegraphics[width=2.2in]{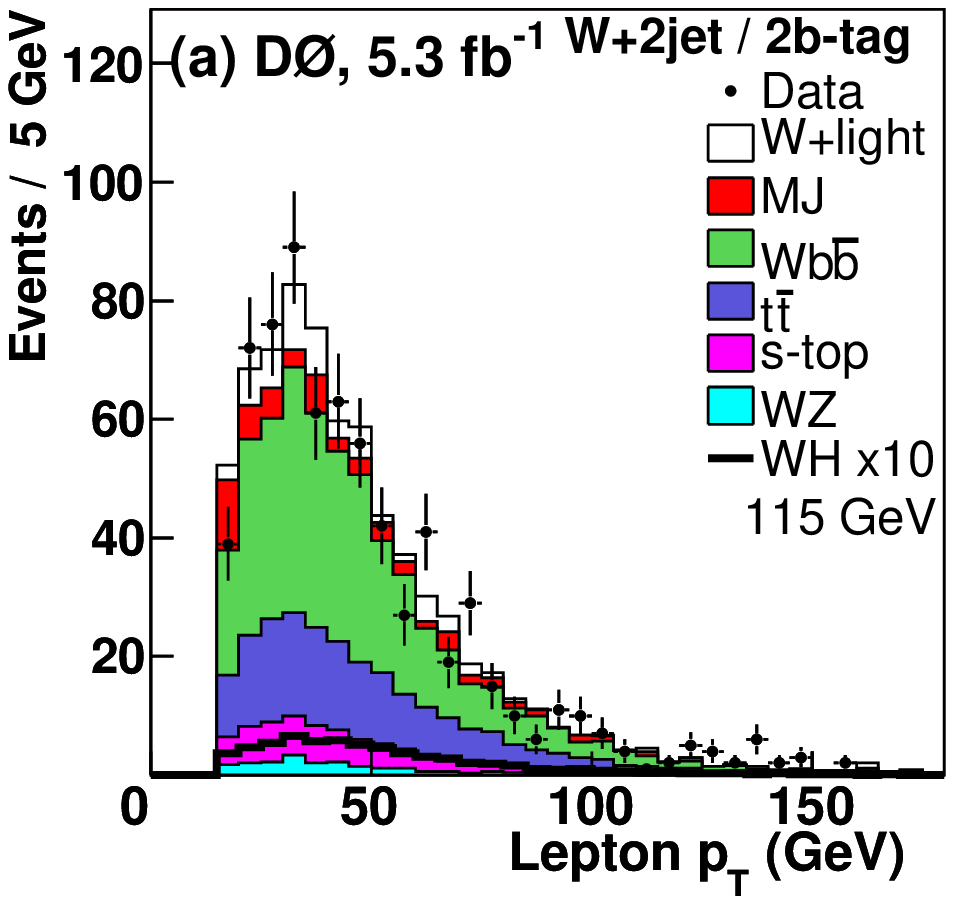}
\hskip 0.2cm \includegraphics[width=2.2in]{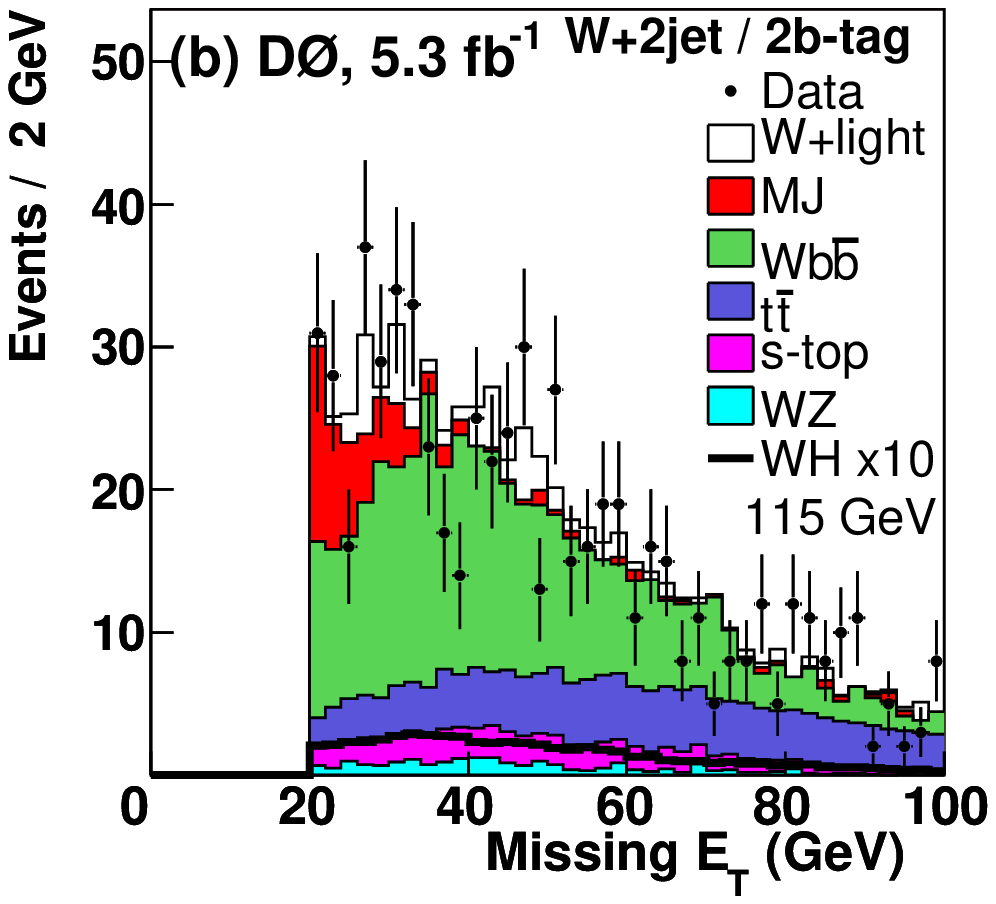}
\hskip 0.2cm \includegraphics[width=2.2in]{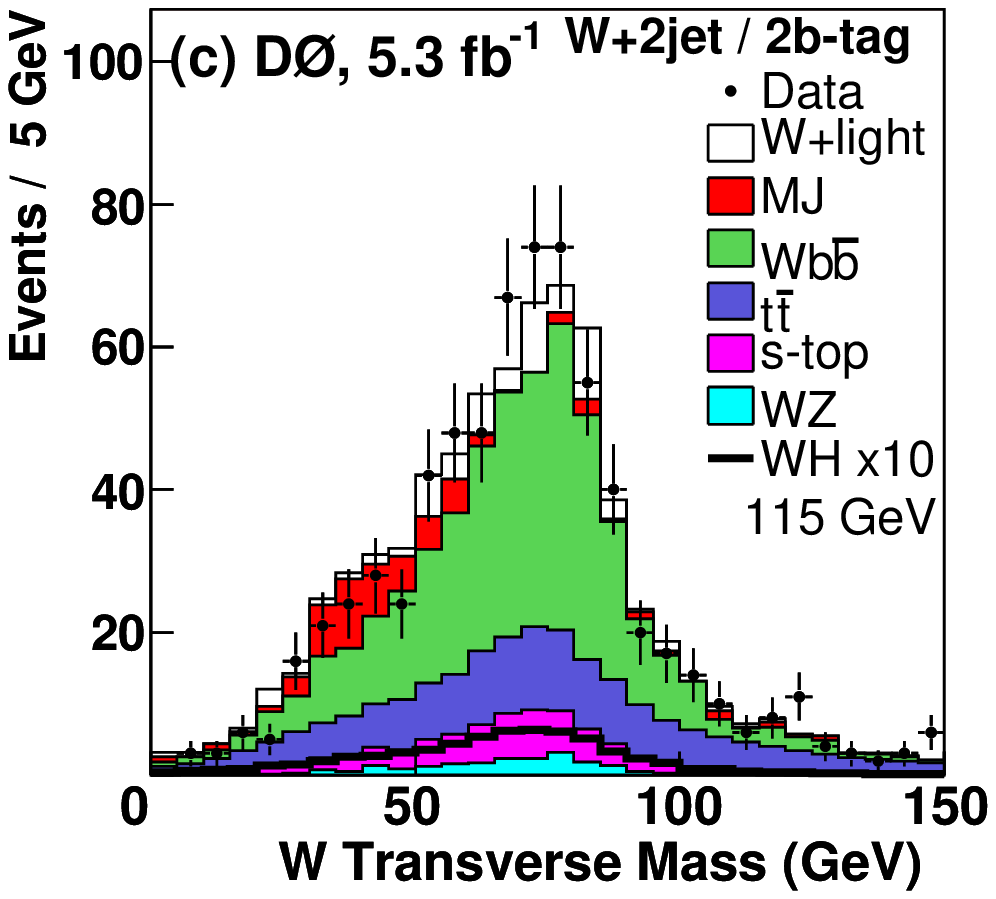}
\clearpage
\hskip -0.2cm \includegraphics[width=2.2in]{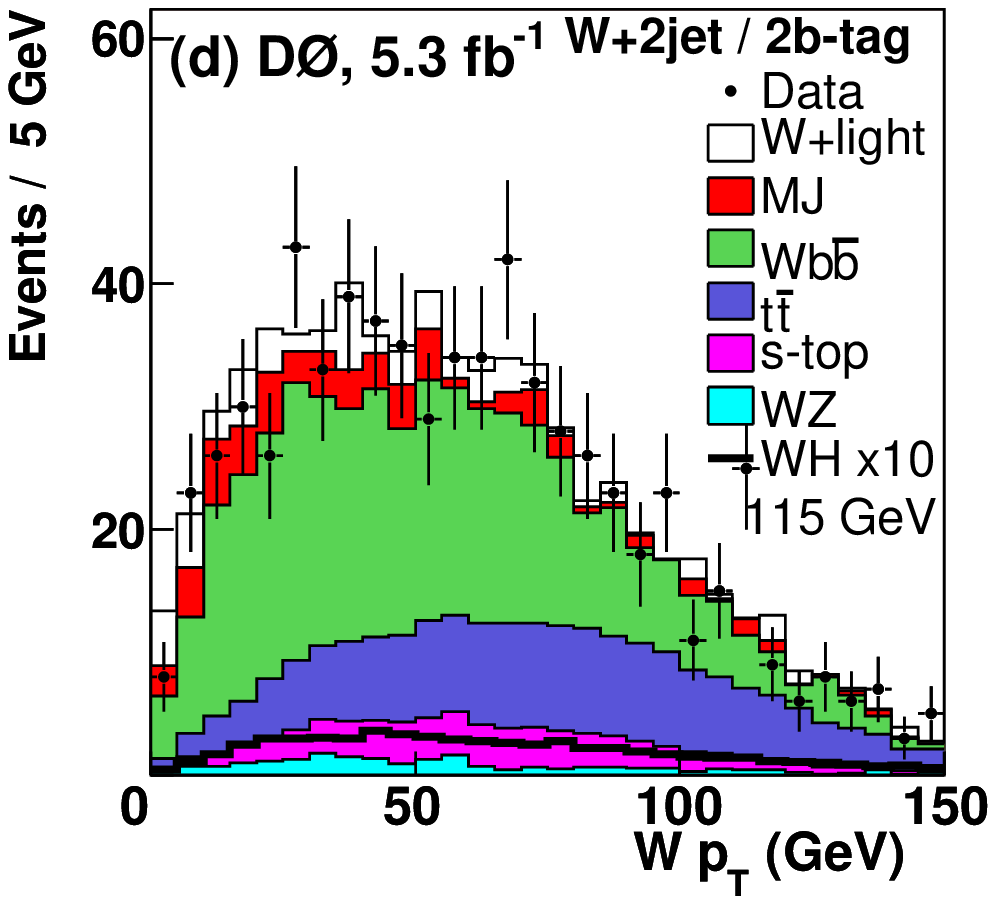}
\hskip 0.2cm \includegraphics[width=2.2in]{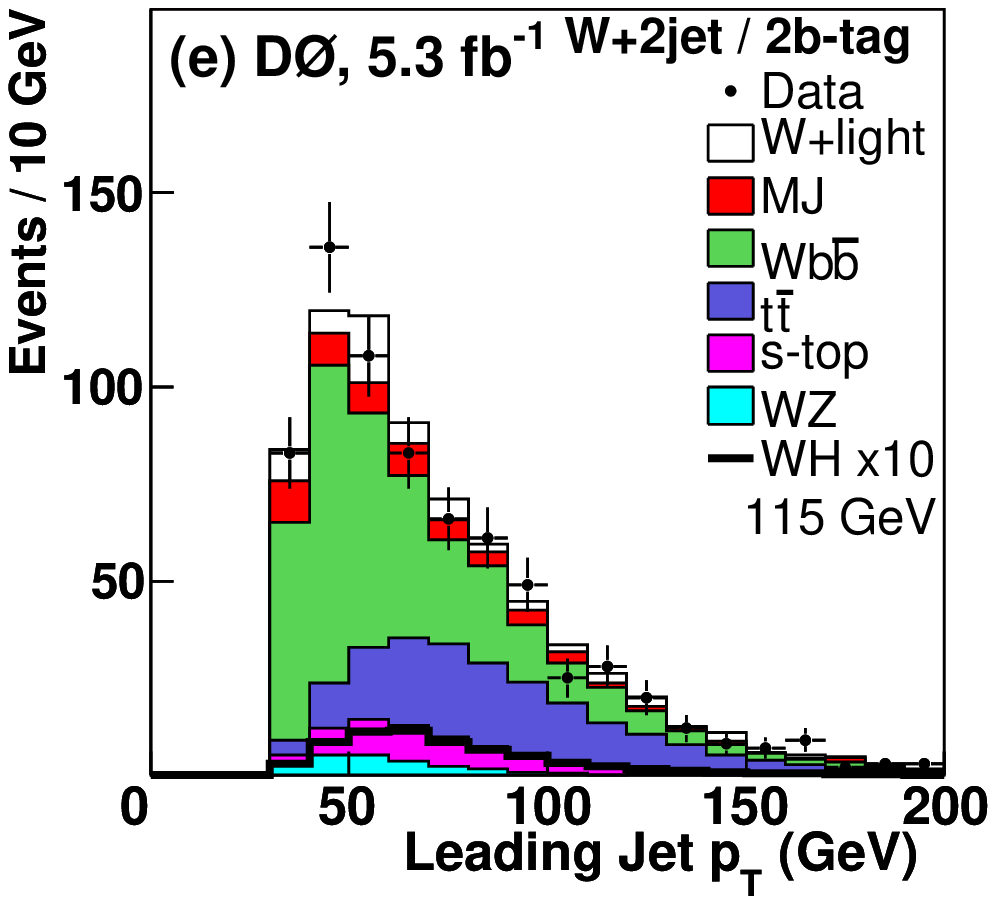}
\hskip 0.2cm \includegraphics[width=2.2in]{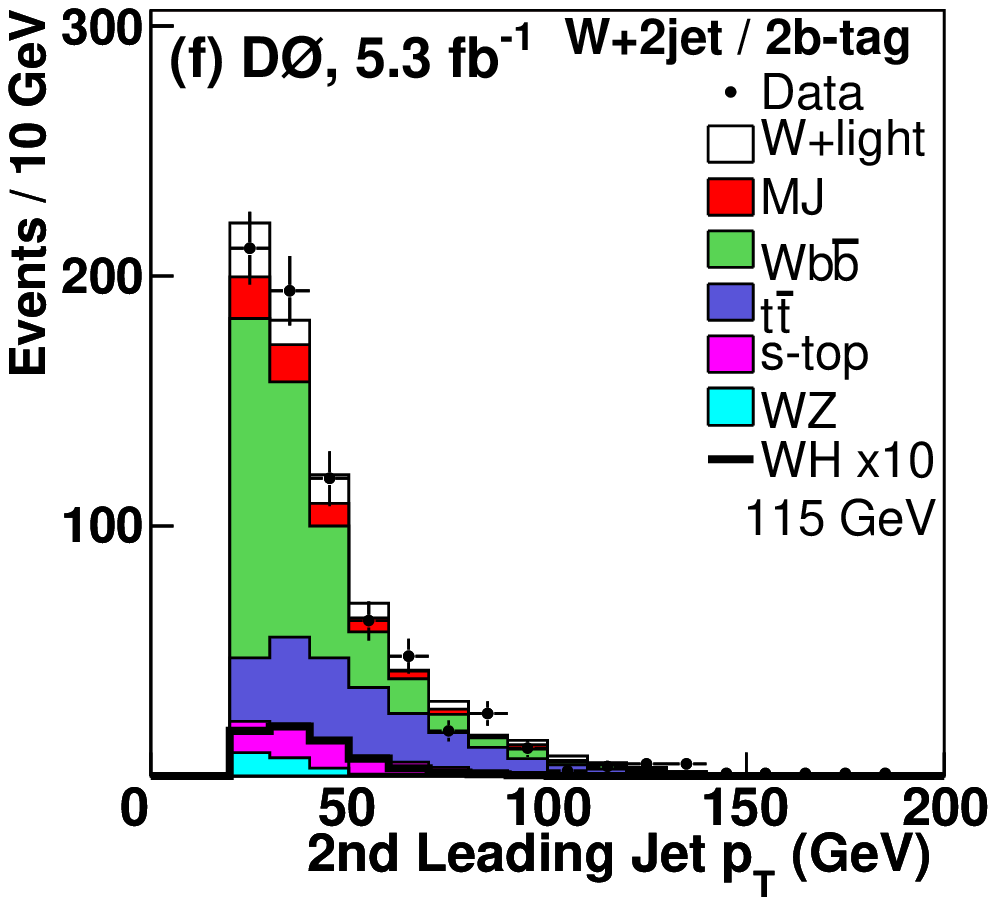}
\clearpage
\hskip -0.2cm \includegraphics[width=2.2in]{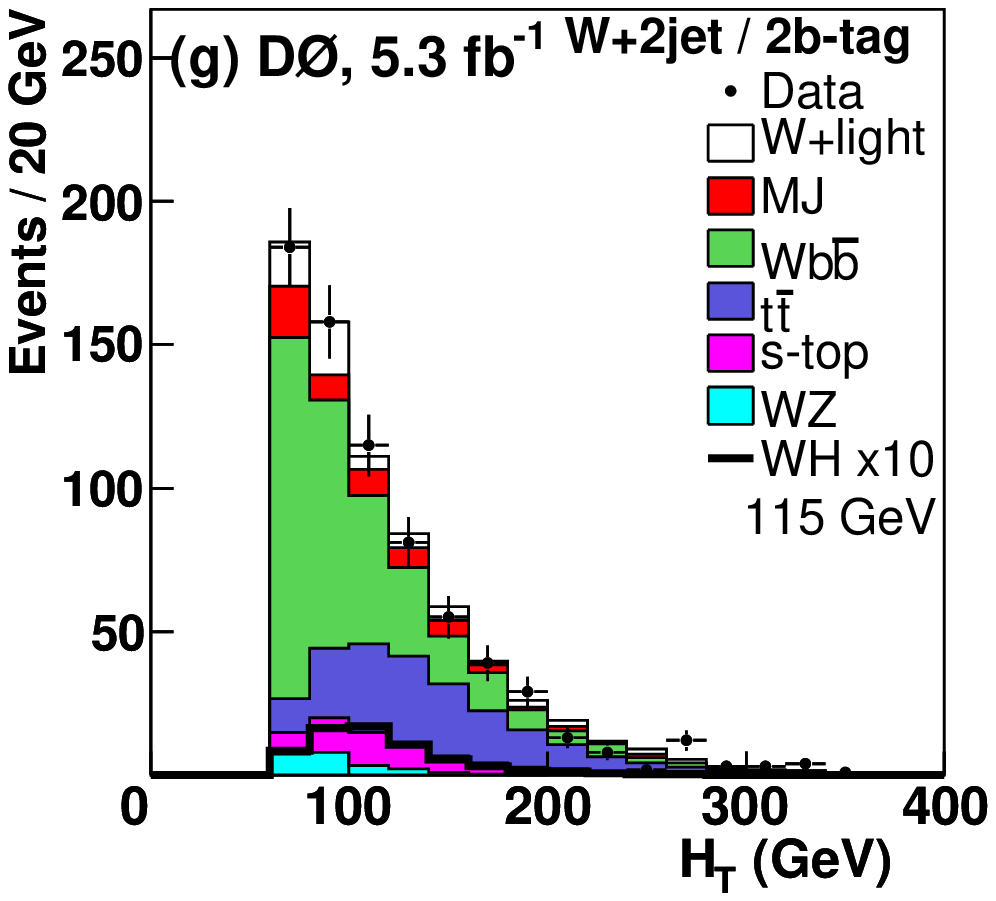}
\hskip 0.2cm \includegraphics[width=2.2in]{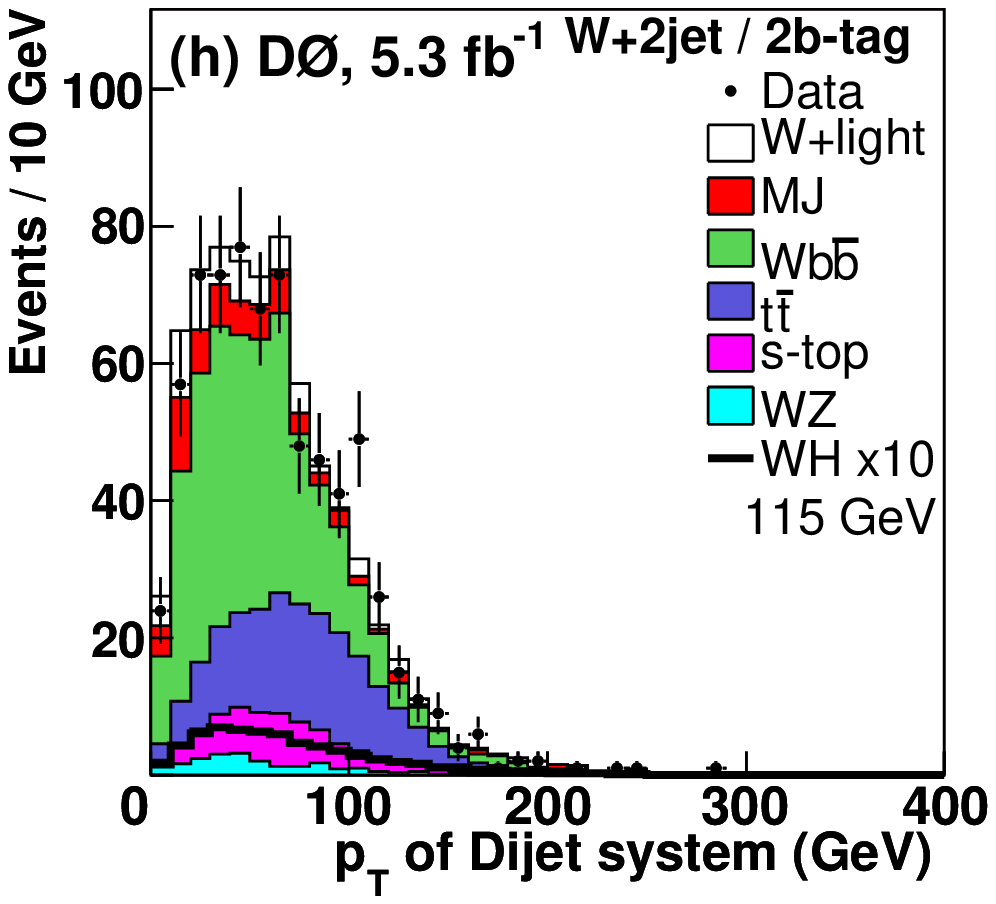}
\clearpage
\hskip -0.2cm  \includegraphics[width=2.2in]{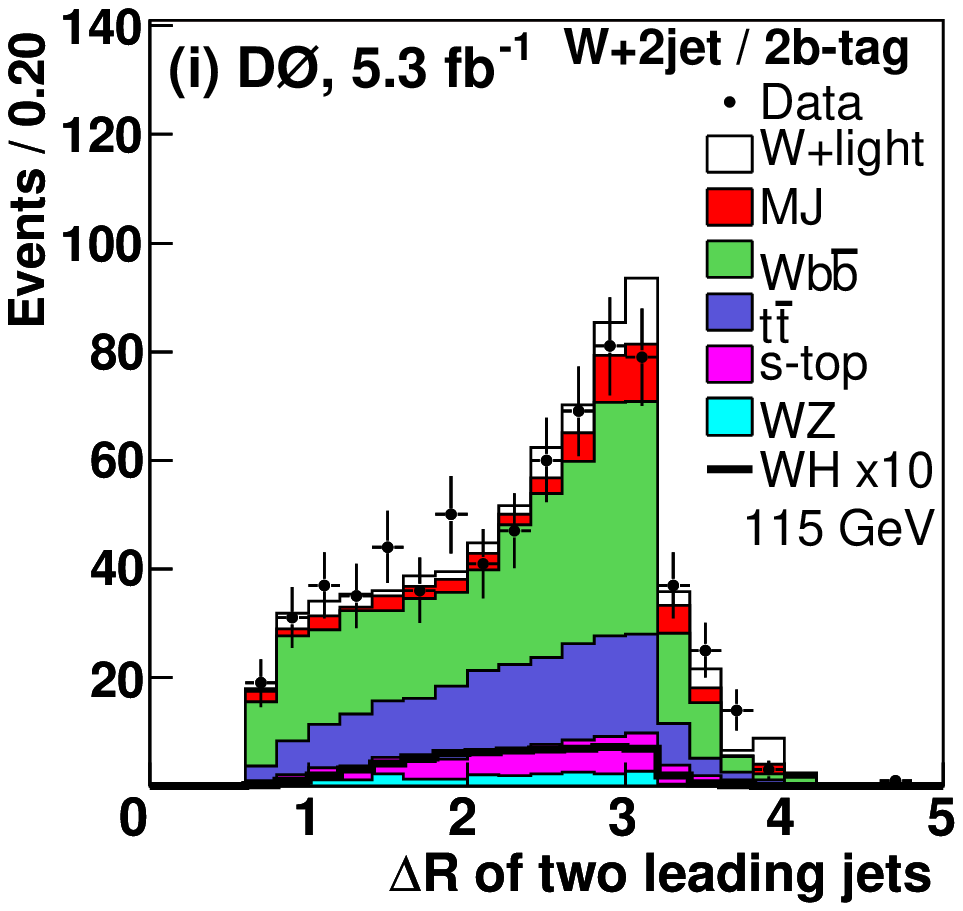}
\hskip 0.2cm \includegraphics[width=2.2in]{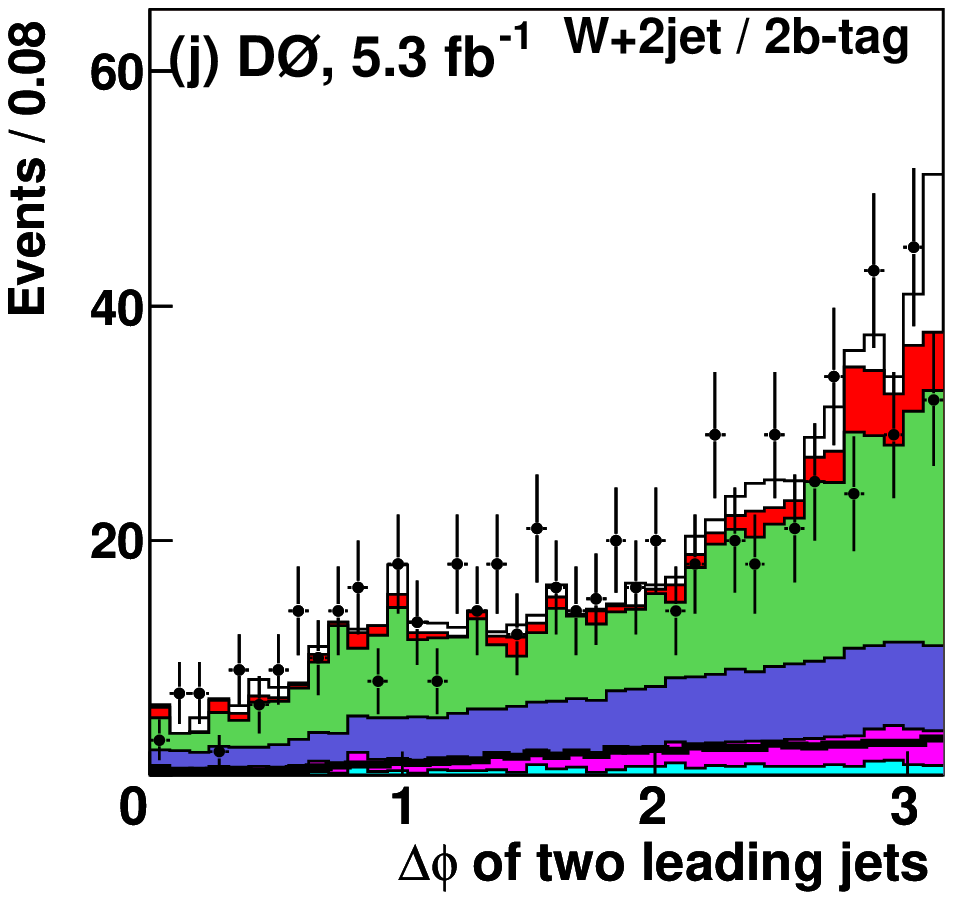}

\caption{[color online] Comparison of the expected backgrounds to the two-$b$-tagged jet data sample in $W$+2 jet selected events. The expectation for a $WH$ signal at $M_{H} = 115 ~\rm GeV$  has been scaled up by a factor of 10.}
\label{fig:btagtwo}
\end{figure*}

To increase the final sensitivity, both the $W$+2 jet and $W$+3 jet samples are subdivided into statistically 
independent samples 
based on whether one or two of the leading jets in the event are consistent with having been initiated by a $b$ quark, 
as discussed in Sec.\ V.
The first sample requires two jets, both with NN output values larger than 
a ``loose'' requirement (``loose-tag''). The second sample, selected from events 
that fail the two-tag requirement, requires a single jet with a NN output above
a larger ``tight'' value requirement (``tight-tag''). 
In two-$b$-tagged jet events, the typical efficiency for identifying a $p_T = 50~\rm GeV$ jet that contains a $b$ hadron is $(59\pm 1)$\% with a 
misidentification probability of 1.5\%\ for light parton ($u,d,s,g$) initiated jets.  
In the single-$b$-tagged jet event sample, the typical efficiency for identifying a $p_T = 50~\rm GeV$ jet 
that contains a $b$ hadron is $(48\pm 1)$\%, with a lower misidentification probability of 0.5\%\ for light parton ($u,d,s,g$) initiated jets. 
The $b$-tagging efficiency is treated separately from the jet taggability
  efficiency.
Events that do not satisfy either of these tagging requirements are not 
considered further in the analysis.

The tagging efficiencies for jets that have passed the taggability requirements are studied in data and
the efficiencies are applied to the simulation via event weights. These weights depend on the $p_{T}$, $\eta$, 
and partonic flavor of each tagged jet. In two $b$-tagged events, the event weights are given by the product
of the weights of the two $b$-tagged jets. In single-$b$-tagged jet events, the final event weight also accounts for 
the simulated contribution of two-$b$-tagged jet events that ``migrate'' to 
the simulated single-$b$-tagged jet samples.

Distributions of dijet invariant mass, prior to $b$-tagging, after requiring 
two $b$-tags, and for single-$b$-tagged events, are shown for the 
$W$+2 jet and $W$+3 jet selections in Figs.\ \ref{fig:two_taggedmass}
and \ref{fig:three_taggedmass}, respectively. The sums of the expected backgrounds are compared to the data, and the electron and 
muon channel samples are again shown combined in each figure. Comparisons of 
kinematic properties in $W$+2 jet events are shown in Figs.\  \ref{fig:btagtwo}
and \ref{fig:btagone} 
for the two- and single-$b$-tagged samples, respectively. 
The expected signal contribution at $M_{H}=115~\rm GeV$ is shown scaled by a factor of 10 in each figure.
 
The total event yields for each of the $b$-tagged samples, in data 
and in simulation, are summarized in Table \ref{yield}. In two-$b$-tagged 
jet events, the dominant backgrounds are from $Wb\bar{b}$ and $t\bar{t}$ 
processes. In single-$b$-tagged jet events, the dominant 
backgrounds are $W$ boson production in association with light or $c$-quark 
jets as well as $t\bar{t}$ production and MJ events.  The expected number of signal 
events in each sample is listed for an assumed Higgs mass $M_{H}=115 \rm~GeV$. 
The uncertainties quoted are the combined  statistical and systematic 
uncertainties, and the systematic uncertainties are those prior to the 
application of the fitting procedure applied when determining cross section 
upper limits described in Sec.\ XI.

{
\begin{table}
\vskip -0.1cm
\caption{ \label{yield} Event yields for the $W$+2 jet and $W$+3 jet samples after requiring two $b$-tagged jets or a single $b$-tagged jet in\
 the events. The expected contributions to the total background from the simulated $W$+light,
data-derived MJ, and simulated $Wb\bar{b}$, $t\bar{t}$, single top quark, and $WZ$ diboson samples are also listed.
The uncertainties quoted are the combined statistical and systematic
uncertainties (prior to the application of the fitting procedure applied when determining
cross section upper limits). The expected signal contribution is shown for an assumed Higgs mass $M_{H}=115 \rm~GeV$.\\}
\begin{tabular}{lrclrclrclrcl}
\hline
\hline
     & \multicolumn{3}{c} { $W$+2 jet }& \multicolumn{3}{c} { $W$+2 jet }
               & \multicolumn{3}{c} { $W$+3 jet } & \multicolumn{3}{c} { $W$+3 jet }\\
     & \multicolumn{3}{c} { 2 $b$-tag }& \multicolumn{3}{c} {  1 $b$-tag  }
               & \multicolumn{3}{c} { 2 $b$-tag } & \multicolumn{3}{c} { 1 $b$-tag }\\
\hline

$W$+light       & 57.5  &$\pm$& 9.2             & 1290  &$\pm$& 201             &       12.1    &$\pm$& 1.8     &       210     &$\pm$& 35    \
  \\
MJ              & 56.5  &$\pm$& 4.2     &               663     &$\pm$& 43      &12.7   &$\pm$& 1.0     &               186     &$\pm$& 13    \
  \\
$Wb\bar{b}$     &       346     &$\pm$& 93      &       1601    &$\pm$& 383     & 47.8  &$\pm$& 12.9            &               358     &$\pm$\
& 90\\
$t\bar{t}$      & 177   &$\pm$& 35      &               417     &$\pm$& 54      &  176  &$\pm$& 35      &               633     &$\pm$& 96    \
  \\
s-top   & 58.3  &$\pm$& 11.4    &               203     &$\pm$& 33      & 13.0  &$\pm$& 2.7     &               53.6    &$\pm$& 9.1     \\
$WZ$            & 22.5  &$\pm$& 3.3     &               152.6   &$\pm$& 17.6    & 2.6   &$\pm$& 1.1     &               33.9    &$\pm$& 4.8   \
  \\
\hline
Total   &  718  &$\pm$& 120     &               4326    &$\pm$& 501     &       264     &$\pm$& 44      &       1474    &$\pm$& 160     \\

\multicolumn{1}{l} {Data} & \multicolumn{3}{c}{ 709 }& \multicolumn{3}{c} {  4316 }
               & \multicolumn{3}{c} {301 } & \multicolumn{3}{c} {1463 }\\
\hline
$WH$            &  6.5  &$\pm$& 1.0             &       9.7     &$\pm$& 0.9     &       0.8     &$\pm$& 0.2     &       2.1     &$\pm$& 0.3   \
  \\

\hline
\hline
\end{tabular}
\\
\end{table}
}

\section{Multivariate Discriminant}

To separate the remaining background from the signal, a multivariate random forest (RF)
discriminant technique \cite{RF1,RF2} is applied independently to each of the 16 subsamples, defined by categorizing events by lepton flavor
(electron or muon), jet multiplicity (2 jets or 3 jets), $b$-tag
multiplicity (single- or two-$b$-tagged), and pre- and post-upgrade data.
The RF technique employs a set of decision trees, each of which applies a series of consecutive
binary decisions trained on simulated events of known origin until a predefined stopping configuration
is reached. Half of the simulated events are used for training and validation, and the remaining half are used to estimate the relative 
contributions of signal and background in the data.

Each individual decision tree examines an initial input event training sample and 
applies selection criteria on a list of potentially discriminating
variables to subdivide the training sample into smaller signal or
background regions referred to as nodes.  At each step, the selection
criterion is chosen to maximize the positive cross entropy ``figure of merit''
value

\begin{equation}
Q=-p\ln p-q\ln q,
\end{equation}
where $p$ ($q$) is the fraction of correctly (incorrectly) classified events at each stage.
The process is continued until a pure signal or pure background node is obtained, and the remaining node regions can no longer be further maximized and 
split without leaving fewer than a prespecified minimum number of events in 
the other daughter samples. The resulting output 
nodes are referred to as leaves. 

\begin{figure*}[t]

\clearpage
\hskip -0.2cm \includegraphics[width=2.2in]{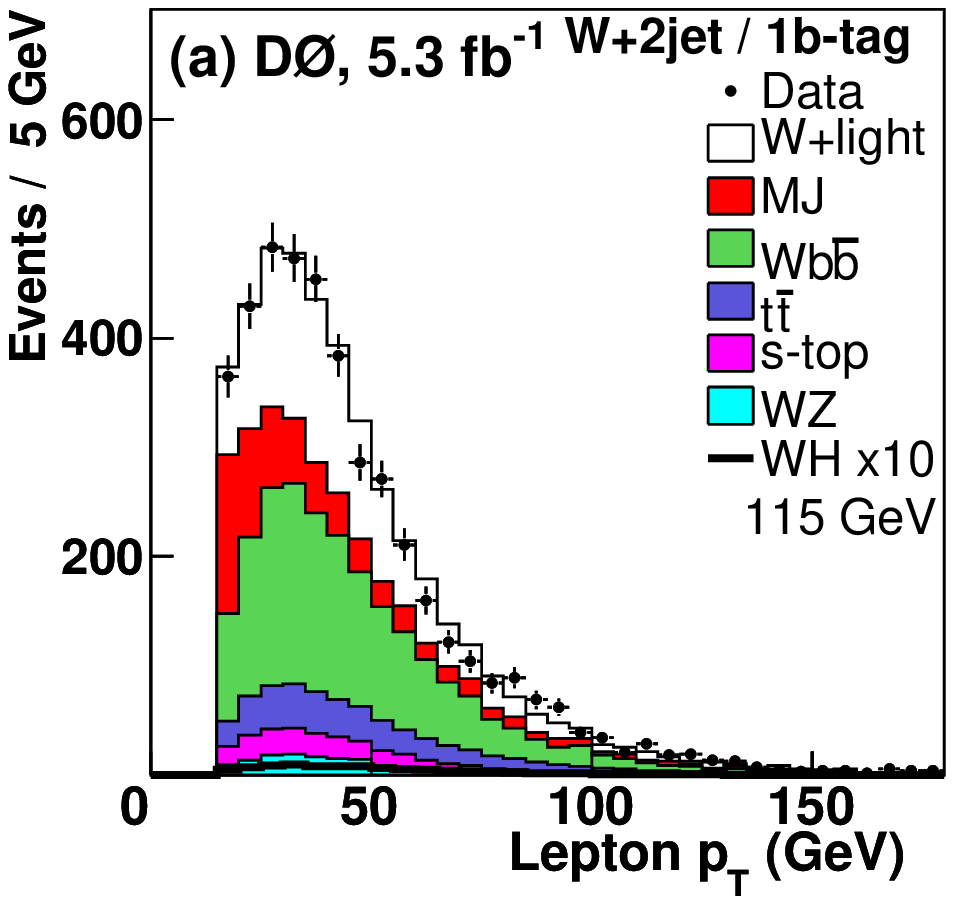}
\hskip 0.2cm \includegraphics[width=2.2in]{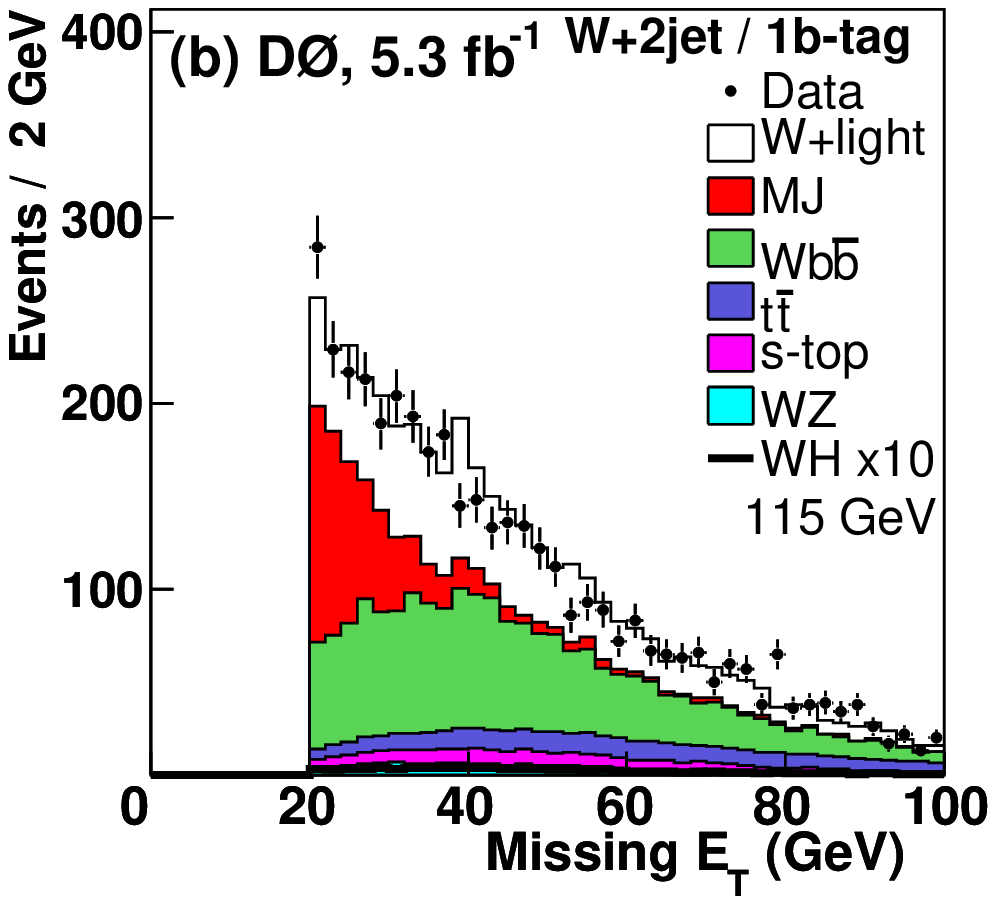}
\hskip 0.2cm \includegraphics[width=2.2in]{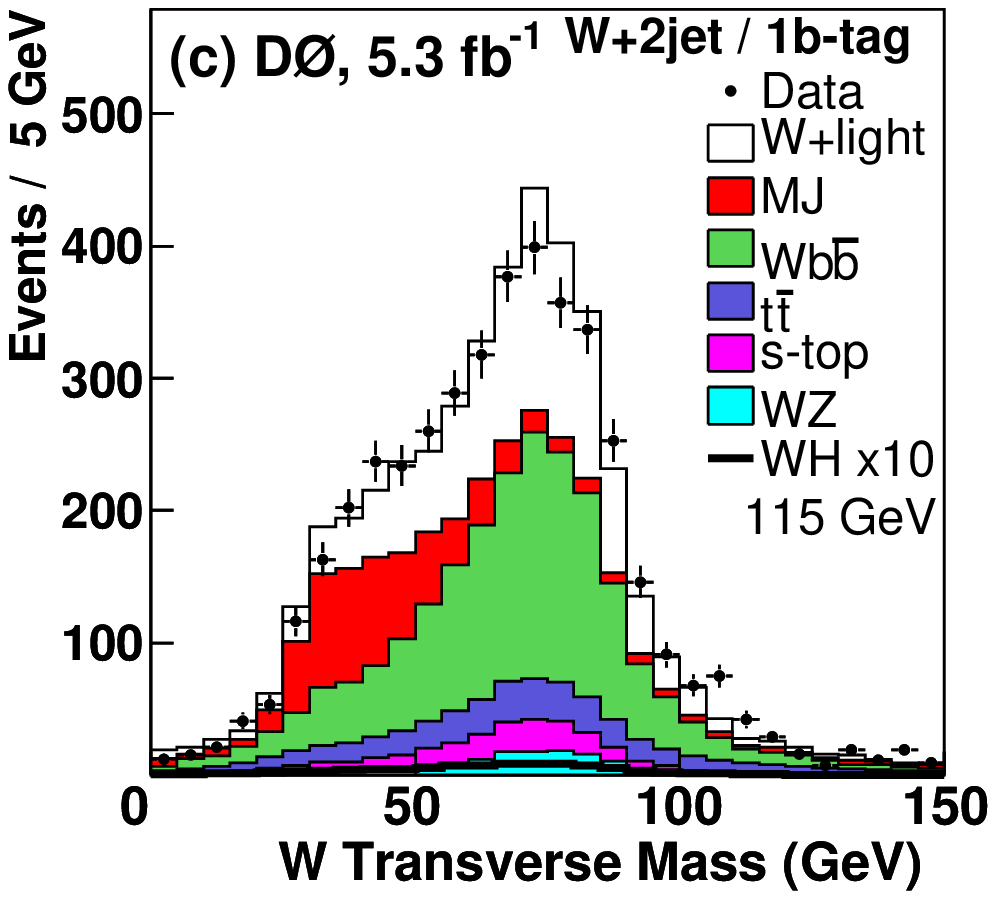}
\clearpage
\hskip -0.2cm \includegraphics[width=2.2in]{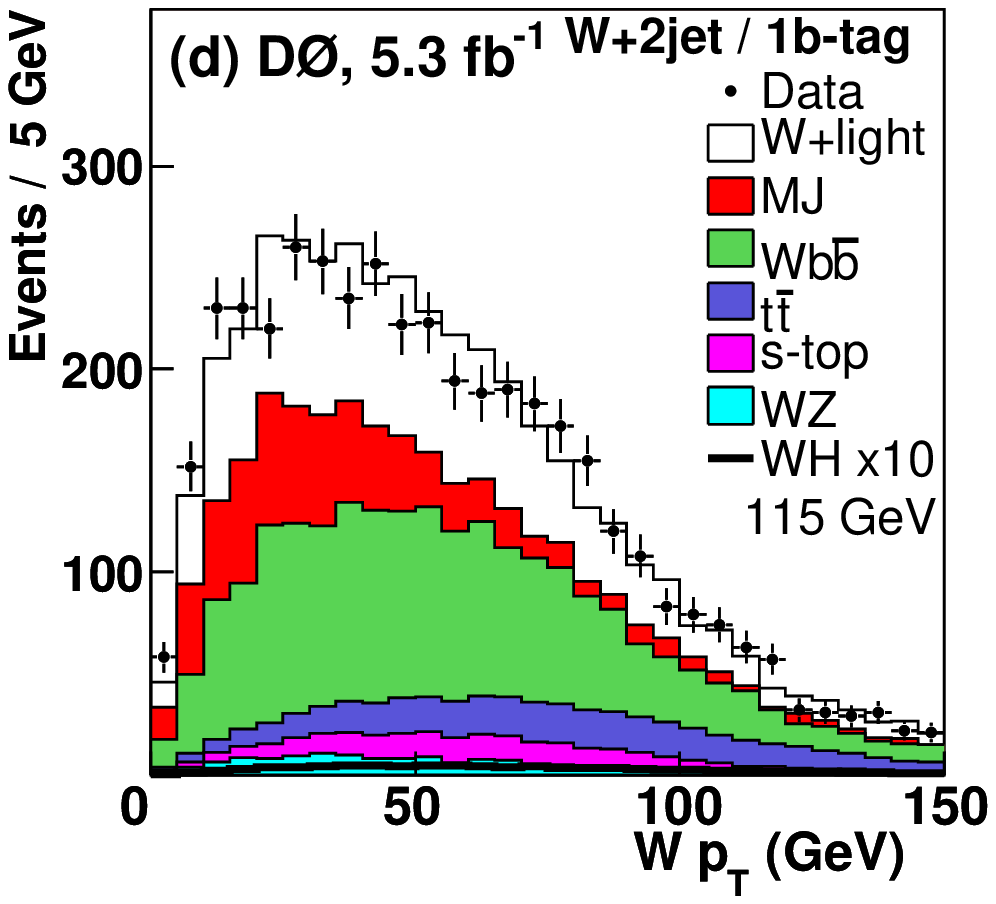}
\hskip 0.2cm \includegraphics[width=2.2in]{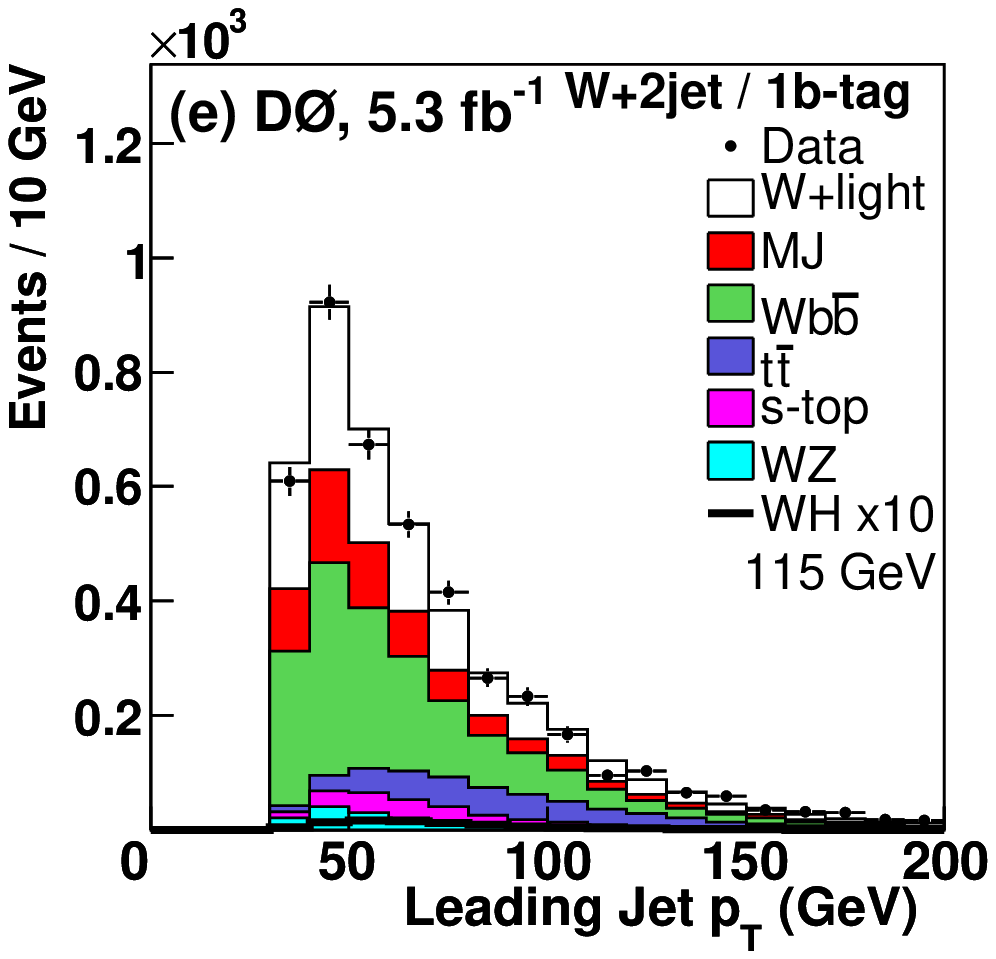}
\hskip 0.2cm \includegraphics[width=2.2in]{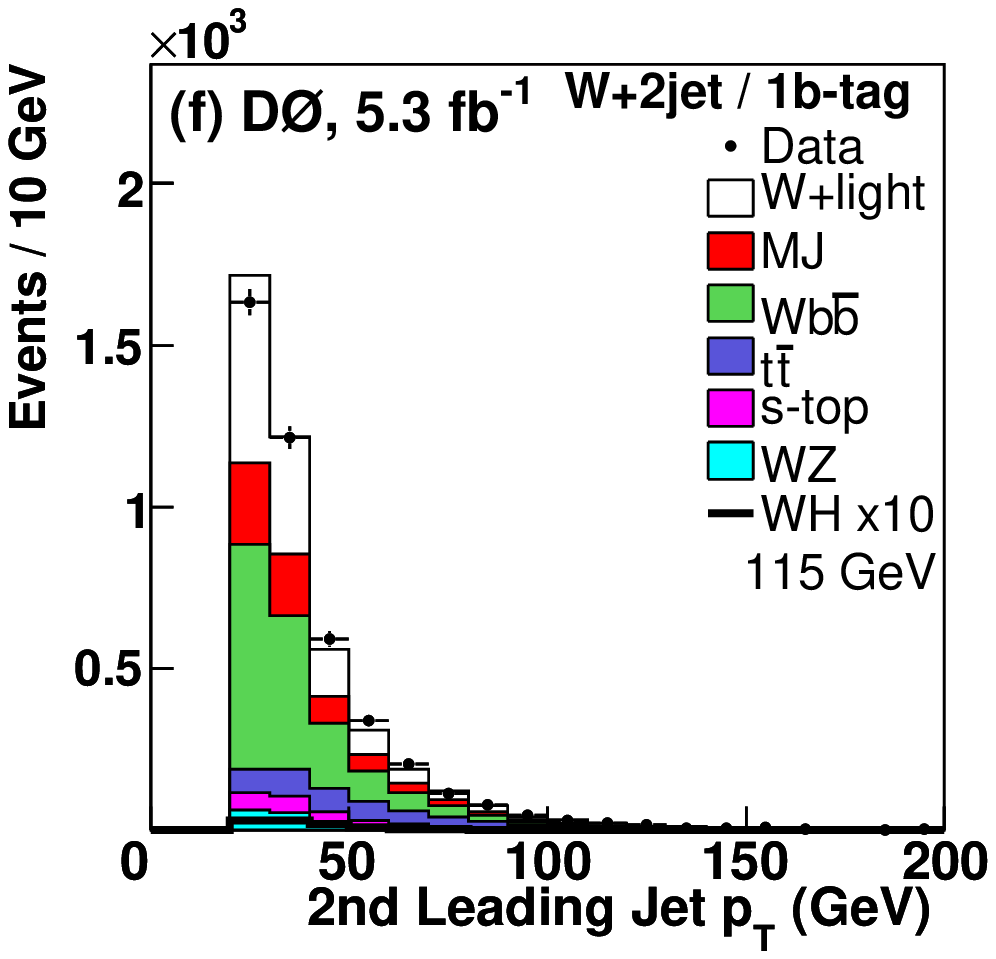}
\clearpage

\hskip -0.2cm \includegraphics[width=2.2in]{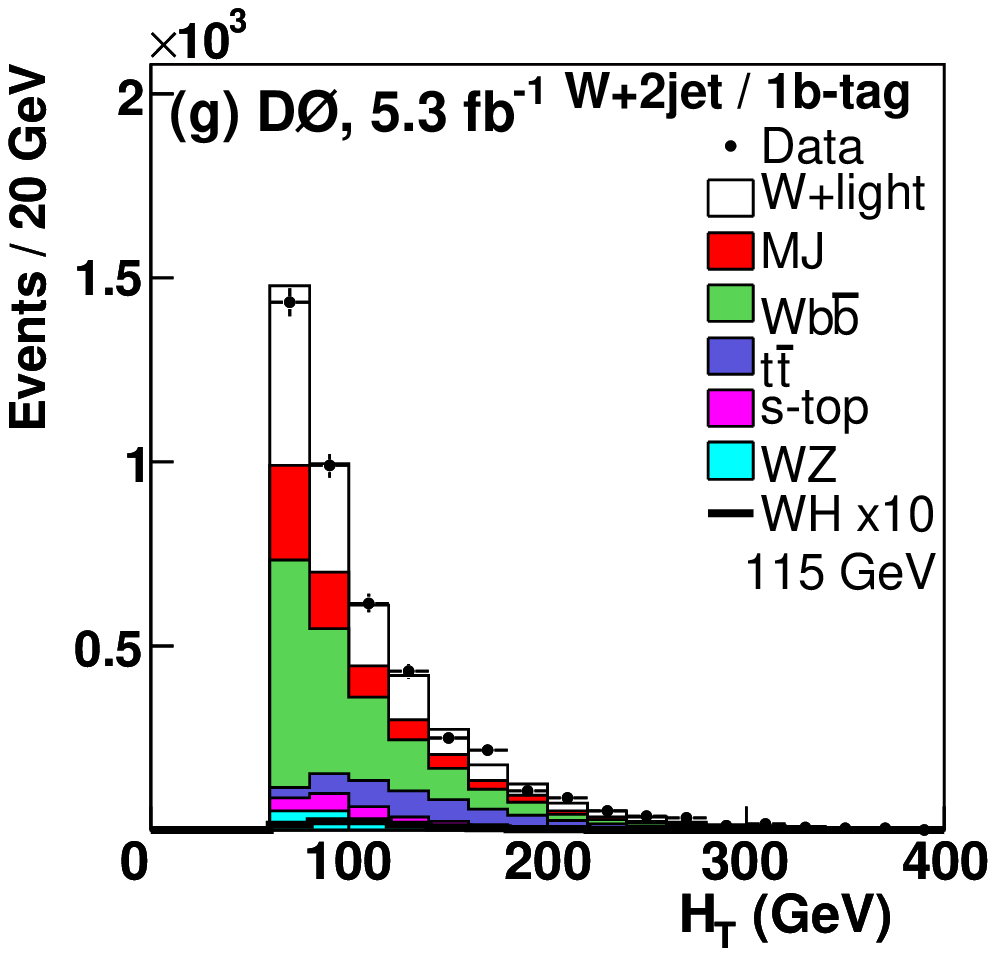}
\hskip 0.2cm \includegraphics[width=2.2in]{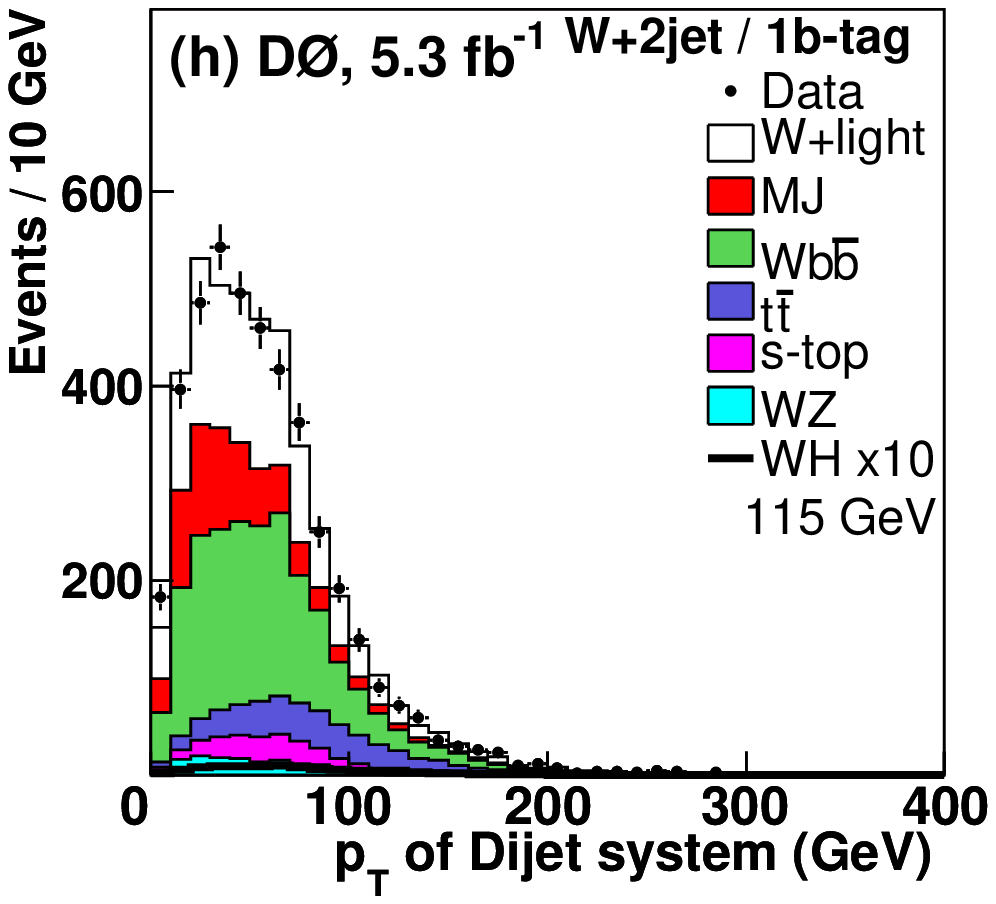}
\clearpage
\hskip -0.2cm \includegraphics[width=2.2in]{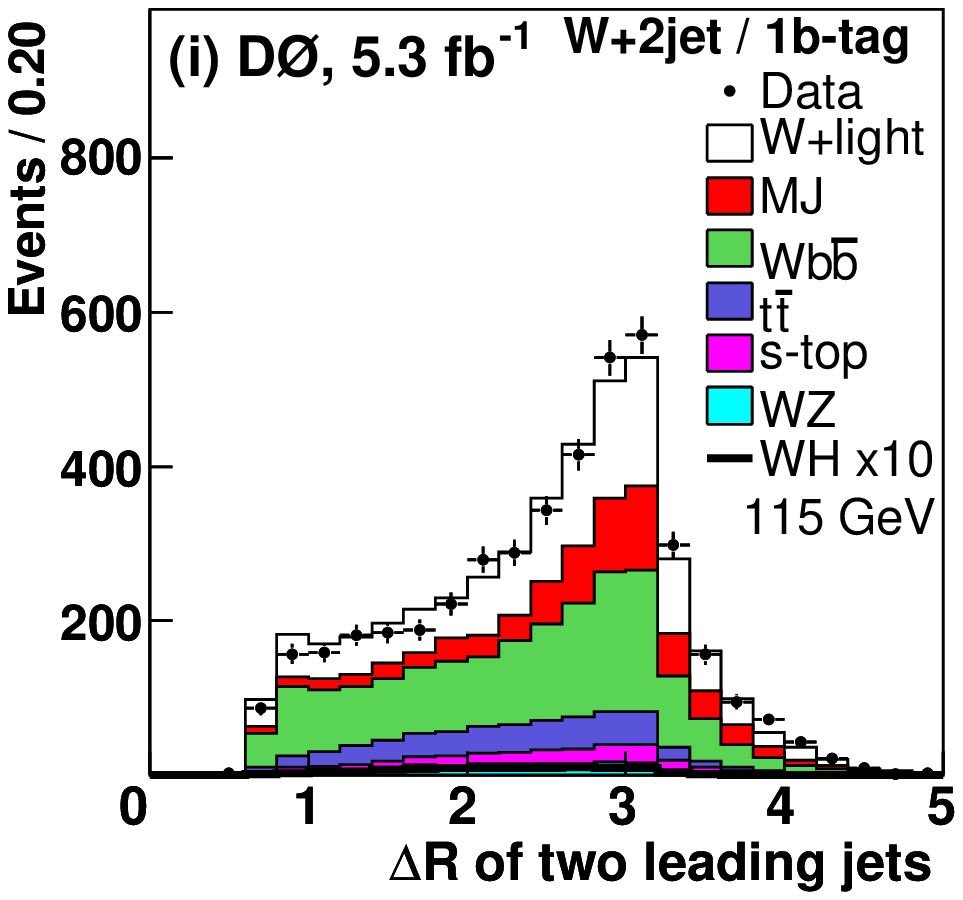}
\hskip 0.2cm \includegraphics[width=2.2in]{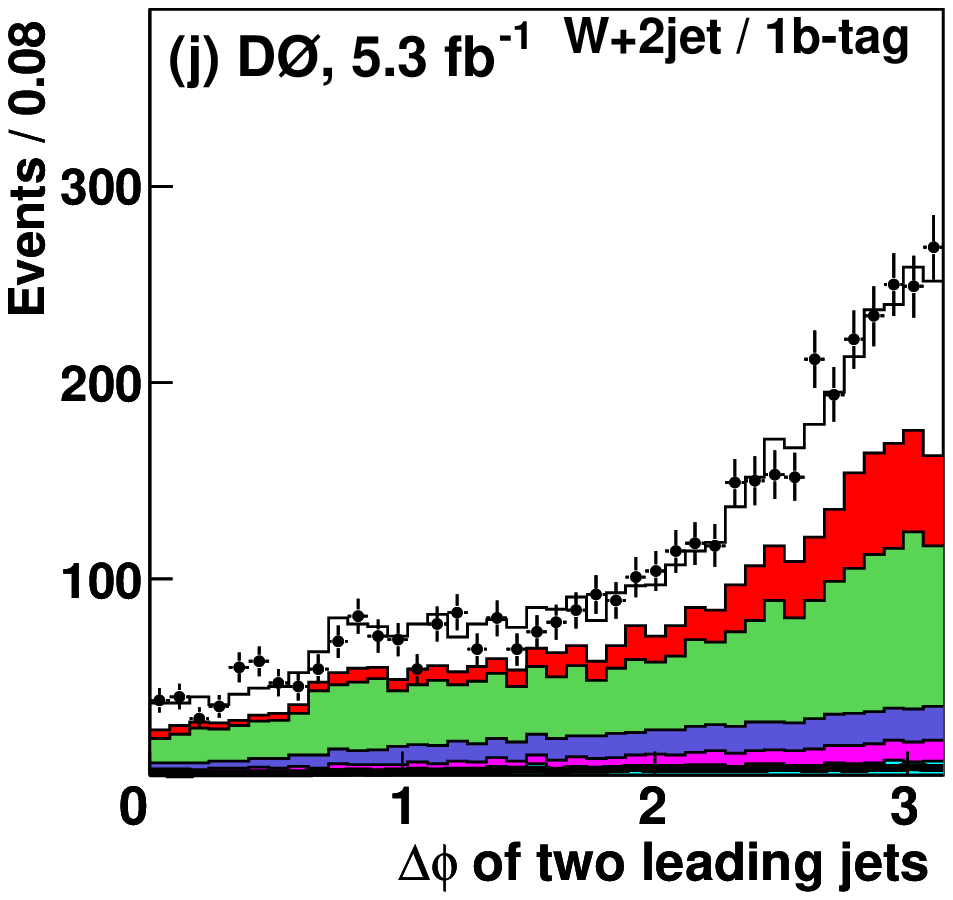}

\caption{[color online] Comparison of the expected backgrounds to the single-$b$-tagged jet data sample in $W$+2 jet selected events. The 
expectation for a $WH$ signal at $M_{H} = 115 ~\rm GeV$  has been scaled up by a factor of 10.}
\label{fig:btagone}
\end{figure*}

\begin{table}
\caption{\label{rf_variables} Description of the 20 kinematic input quantities provided to each random forest discriminant.\\}
\begin{tabular}{l l}
\hline \hline
RF input variable & Description\\
\hline
\MET{} 				 & Missing transverse energy \\
$M_W^T$ 			 & Lepton-\MET{} transverse mass  \\
$p_{T}$($\ell$-\MET{} system)    & $p_{T}$ of $W$ boson candidate\\ 
$p_{T}$($j_1$)	 	         & Leading jet $p_{T}$ \\
$p_{T}$($j_2$) 		         & Subleading jet $p_{T}$ \\
$m_{jj}$ 			 & Dijet invariant mass \\
$p_{T}$(dijet system) 	 	 & $p_{T}$ of dijet system \\
$\Delta R$($j_1$,$j_2$) & $\Delta R$ between the two leading jets \\
$\Delta \phi$($j_1$,$j_2$) 	 & $\Delta \phi$ between the two leading jets \\
$H_{T}$		            	 & Scalar sum of the transverse \\
				 & momenta of all jets in the event\\
$H_{Z}$				 & Scalar sum of the longitudinal \\
 				 & momenta  of all jets in the event \\
$\Delta \phi$($j_1$, $\ell$) 	 & $\Delta \phi$ between the leading jet \\ 
                                 & and the lepton \\

$E(j_{2})$  		         & Second leading jet energy \\

$\sqrt{\hat{s}}_{1} = \Sigma E (\nu_{1} + \ell \rm  +jets)$ 		 & Center-of-mass energy of the \\ 
                                 & $\nu$+$\ell$+dijet system with larger \\  
                                 & solution for the longitudinal   \\
                                 & momentum of the $\nu$ candidate\\ 
$\sqrt{\hat{s}}_{2} = \Sigma E (\nu_{2} + \ell \rm +jets)$ 		 & Center-of-mass energy of the \\ 
                                 & $\nu$+$\ell$+dijet system with smaller \\
                                 & solution for the longitudinal \\
                                 & momentum of the $\nu$ candidate \\
$\Delta R$(dijet,$\ell+\nu_{1}$) & $\Delta R$ between the dijet system \\
 				 & and the $\ell+\nu$ system with larger \\ 
                                 & solution for the longitudinal \\ 
                                 & momentum of the $\nu$ candidate \\
$\Delta R$(dijet,$\ell+\nu_{2}$) & $\Delta R$ between the dijet system \\
 				 & and the $\ell+\nu$ system with smaller \\
                                 & solution for the longitudinal \\
                                 & momentum of the $\nu$ candidate \\
Aplanarity                       & $\frac{3}{2} \lambda_{3}$, where $\lambda_{3}$ is the 
smallest \\
 & eigenvalue of the normalized \\
& momentum tensor: \\
& $M_{ij} = \frac{(\Sigma_{\mu} p^{\mu}_{i} p^{\mu}_{j})}{\Sigma_{\mu} |\bar{p}^{\mu}_{i}|^{2}}$ \\
& where $\mu$ runs over jets and \\ 
& the charged lepton and $p^{\mu}_{i}$ is \\
& the $i$th 3-momentum component \\
& of the $\mu$th physics object.\\

$\cos (\theta^*)$         	 & Cosine of angle between the $W$ \\ 
				 & candidate and nominal proton \\
                                 & beam direction in the zero \\
                                 & momentum frame (see Ref. \cite{spincorr}) \\
$\cos (\chi)$ 			 & Cosine of angle between lepton \\
                                 & and rotated 3-momentum vector  \\
                                 & of the dijet system in the production\\
                                 & plane of the $W$ boson rest frame \cite{spincorr}\\
\hline \hline
\end{tabular}
\end{table}

For each of the subsamples, the decision tree algorithm is run multiple times to create
the forest and variants of the default training sample are used for each decision tree within each RF.
The outputs of the decision trees within each RF are combined to yield final RF output distributions.
The decision tree samples are obtained using bootstrap aggregation (``bagging''), and a random
subset of 13 of the 20 input discriminating variable distributions are assigned within each decision tree to create the forest. Varying
the number of input variables used by $\pm 1$ is found to have a negligible
effect on the RF output.

\begin{figure*}[t]
\clearpage
\hskip -0.2cm \includegraphics[width=2.2in]{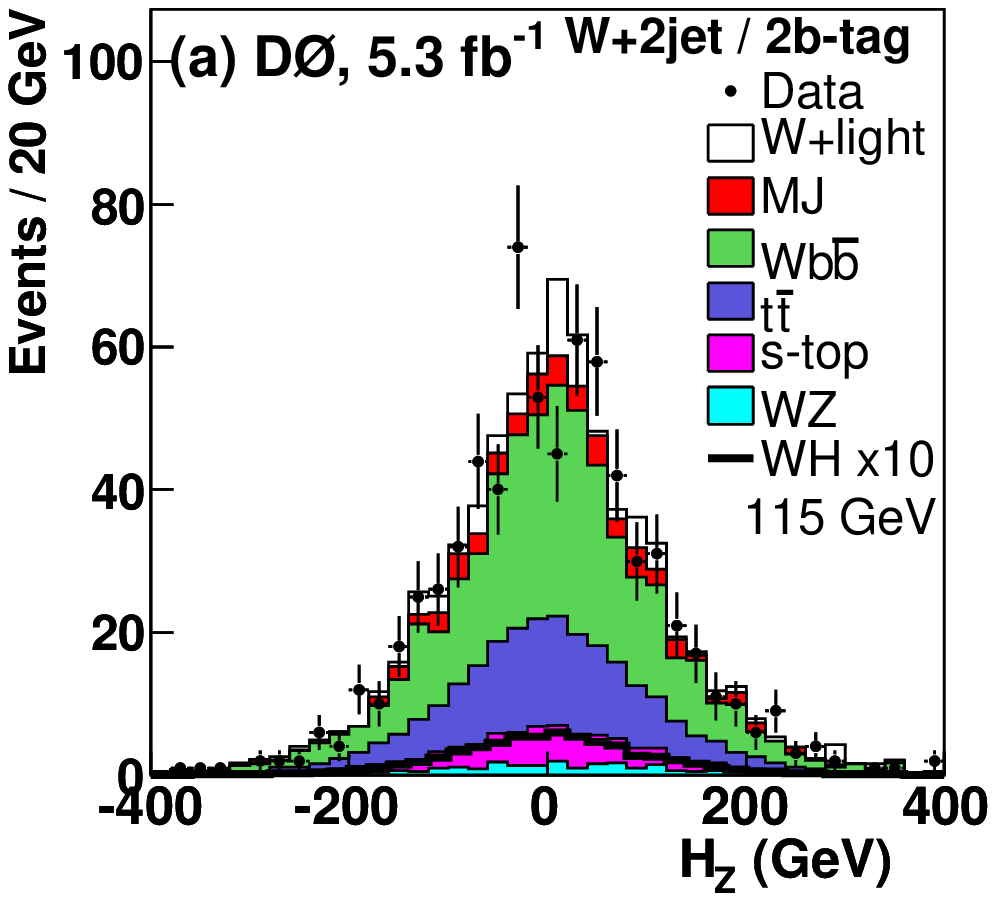}
\hskip 0.2cm \includegraphics[width=2.31in]{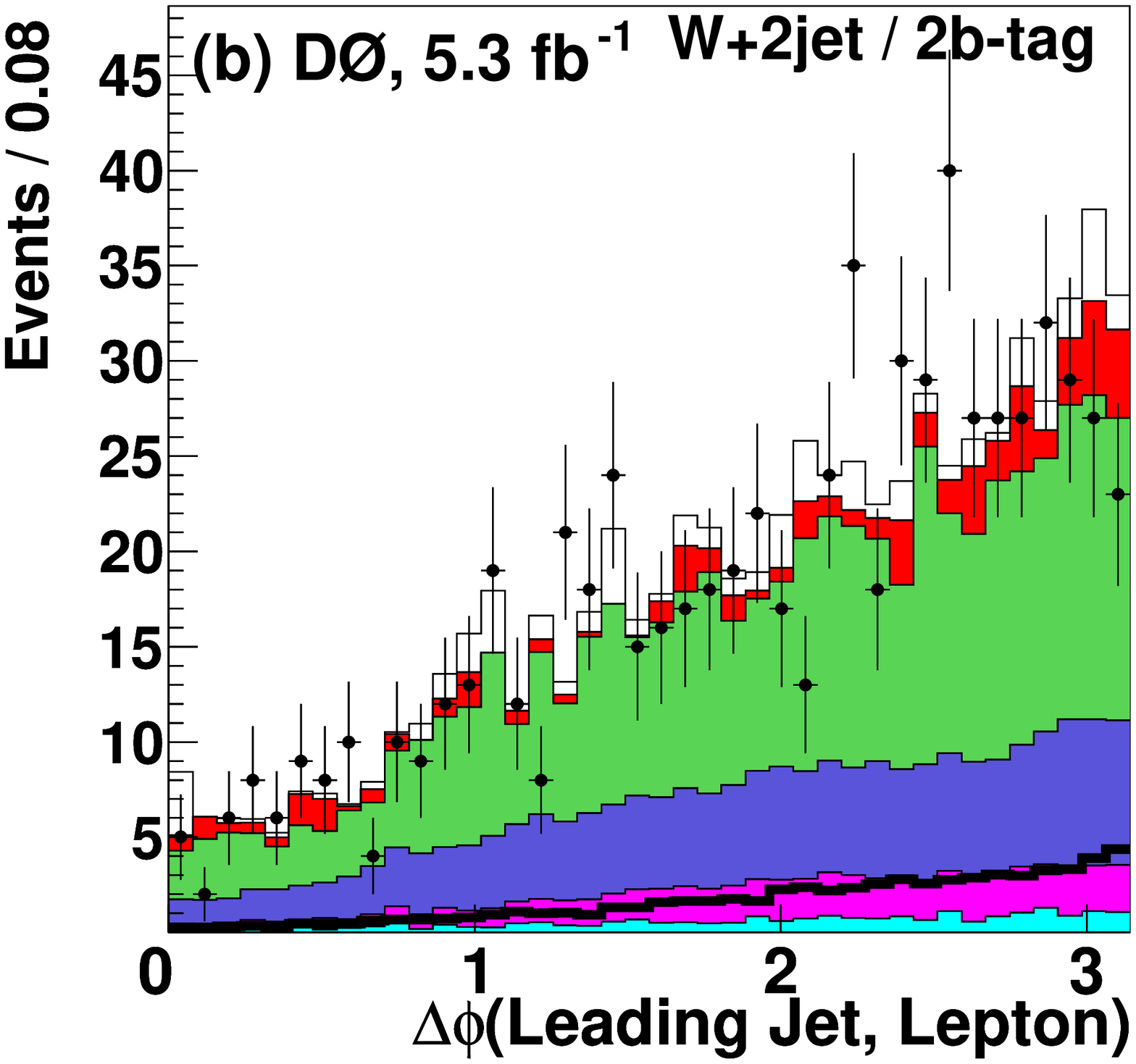}
\includegraphics[width=2.2in]{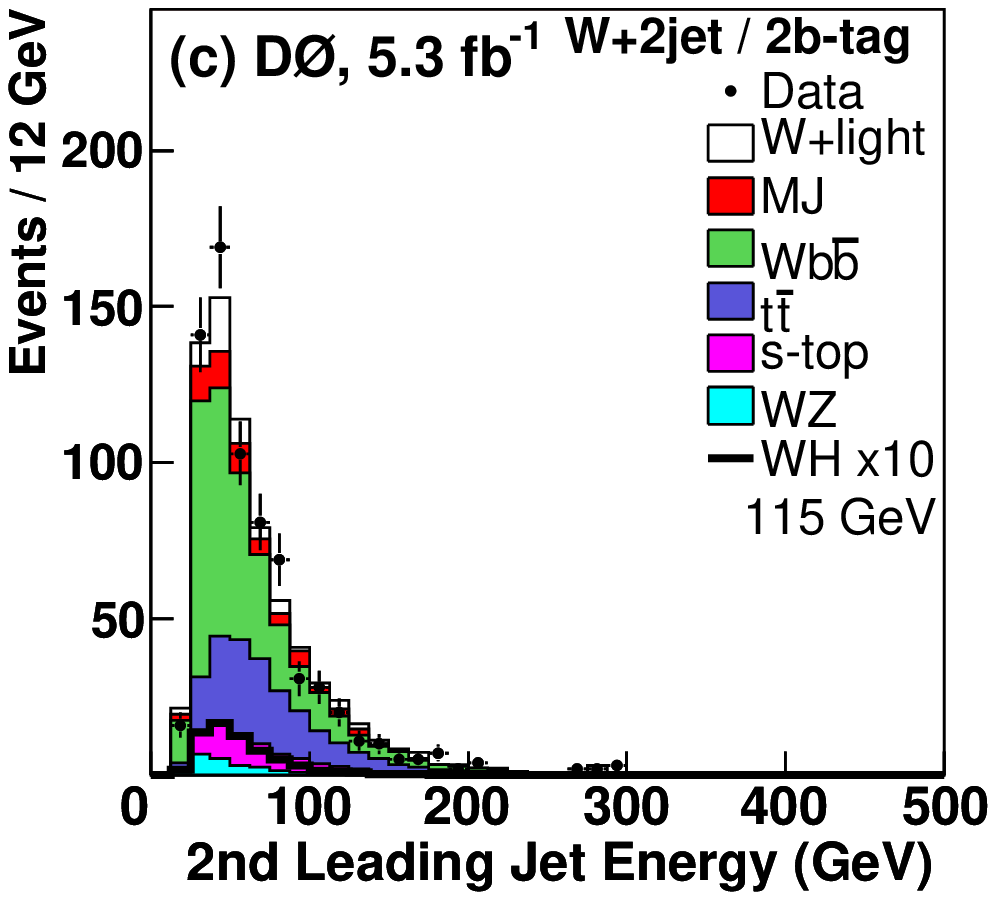}

\hskip -0.2cm \includegraphics[width=2.2in]{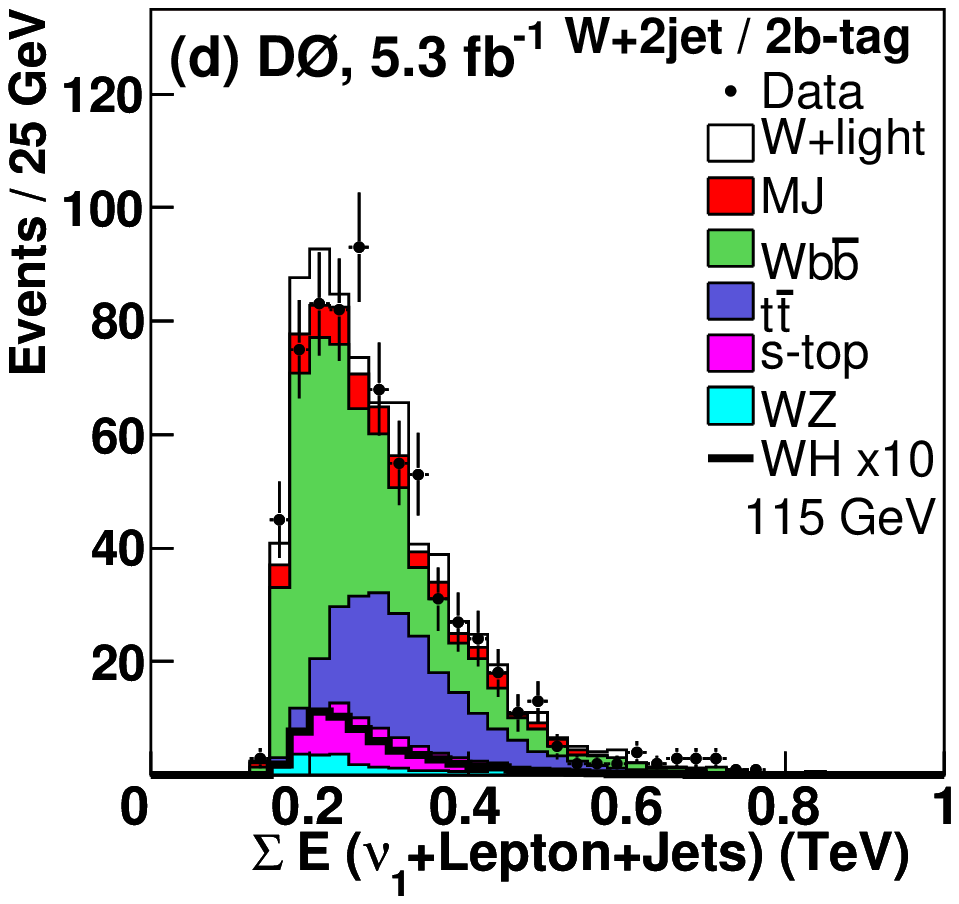}
\hskip 0.2cm \includegraphics[width=2.2in]{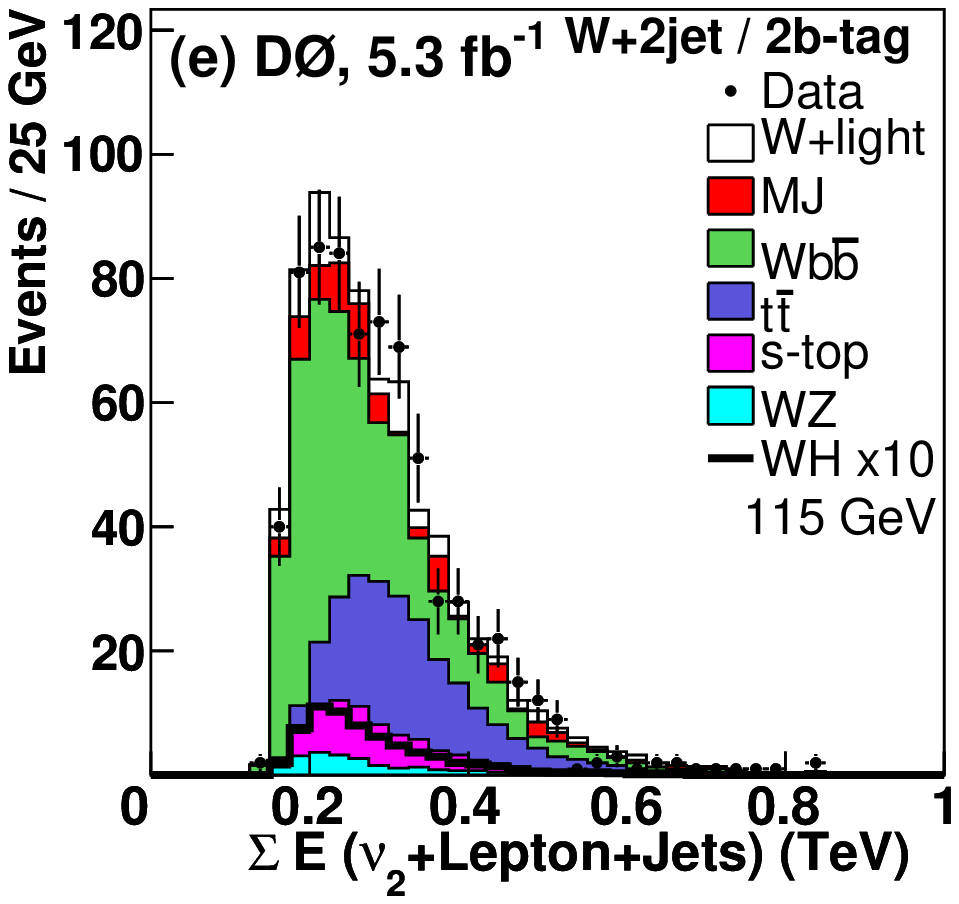}
\hskip 0.2cm \includegraphics[width=2.2in]{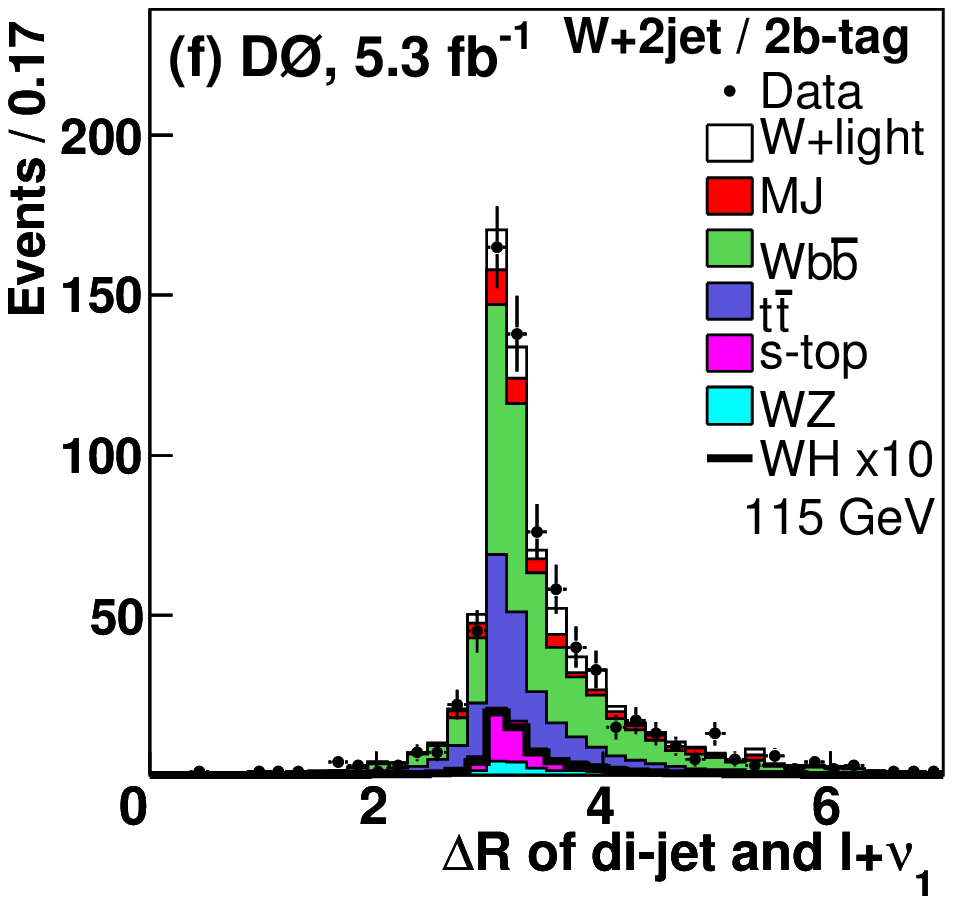}

\hskip -0.2cm \includegraphics[width=2.2in]{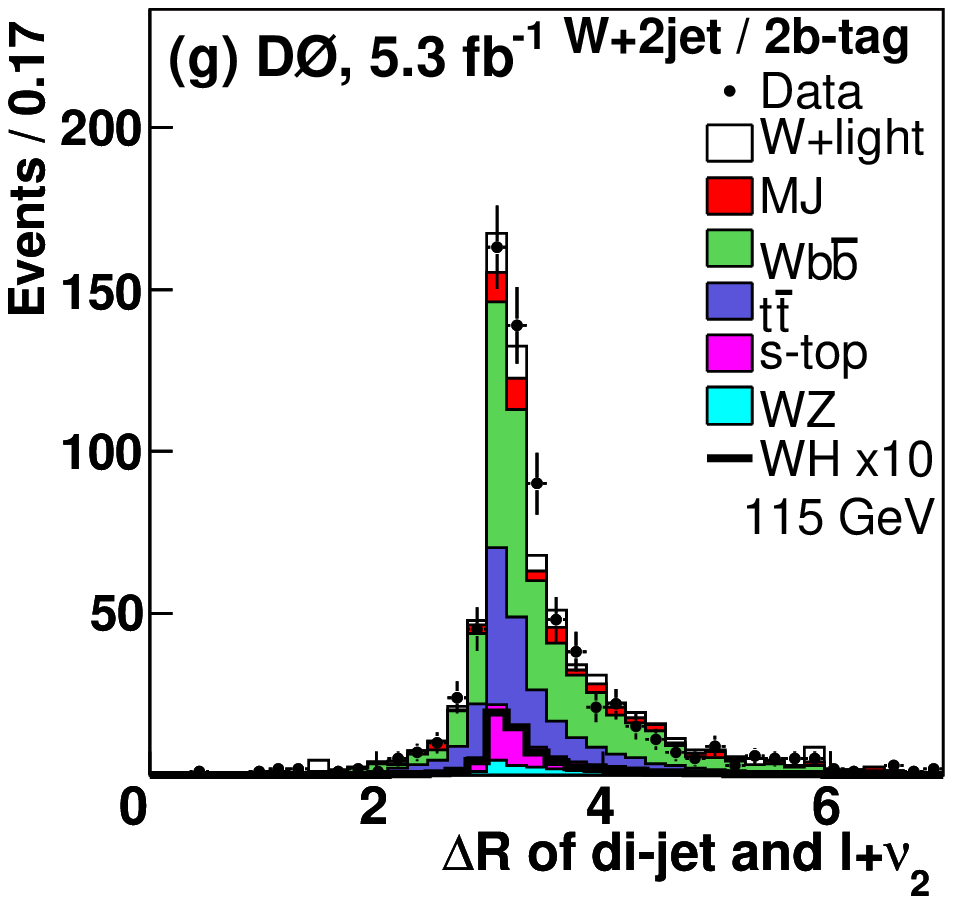}
\hskip 0.2cm \includegraphics[width=2.2in]{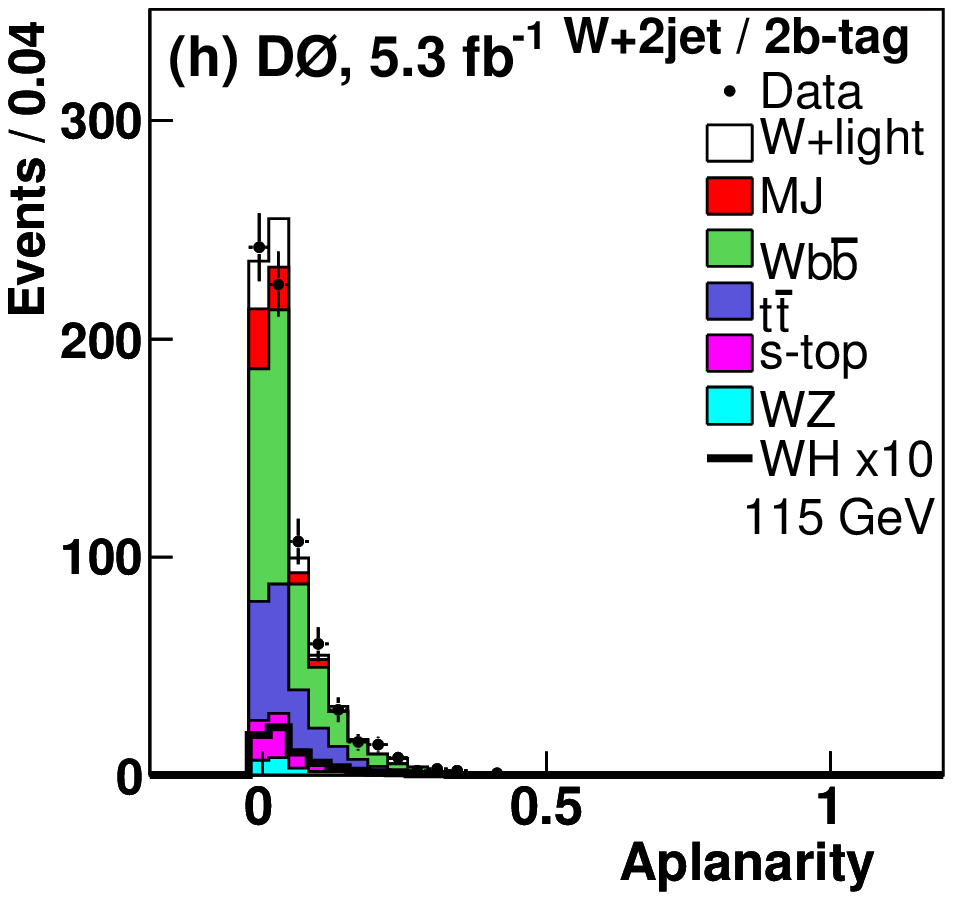}

\hskip -0.2cm \includegraphics[width=2.2in]{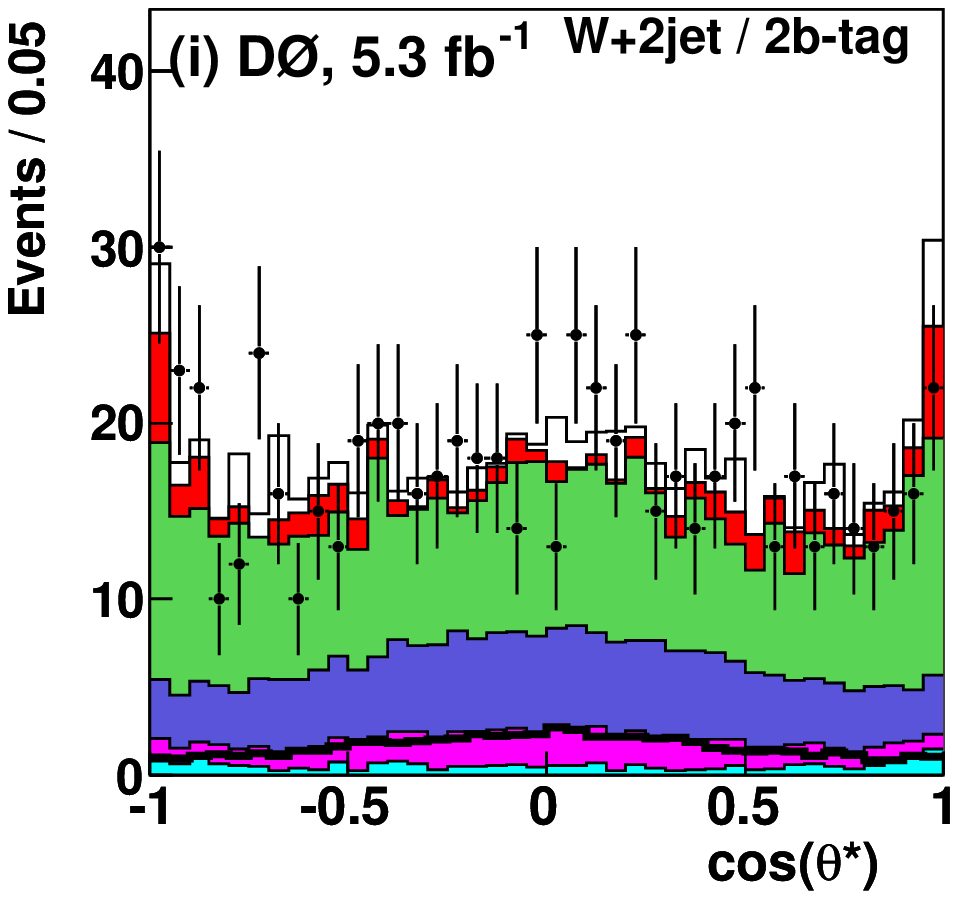}
\hskip 0.2cm \includegraphics[width=2.2in]{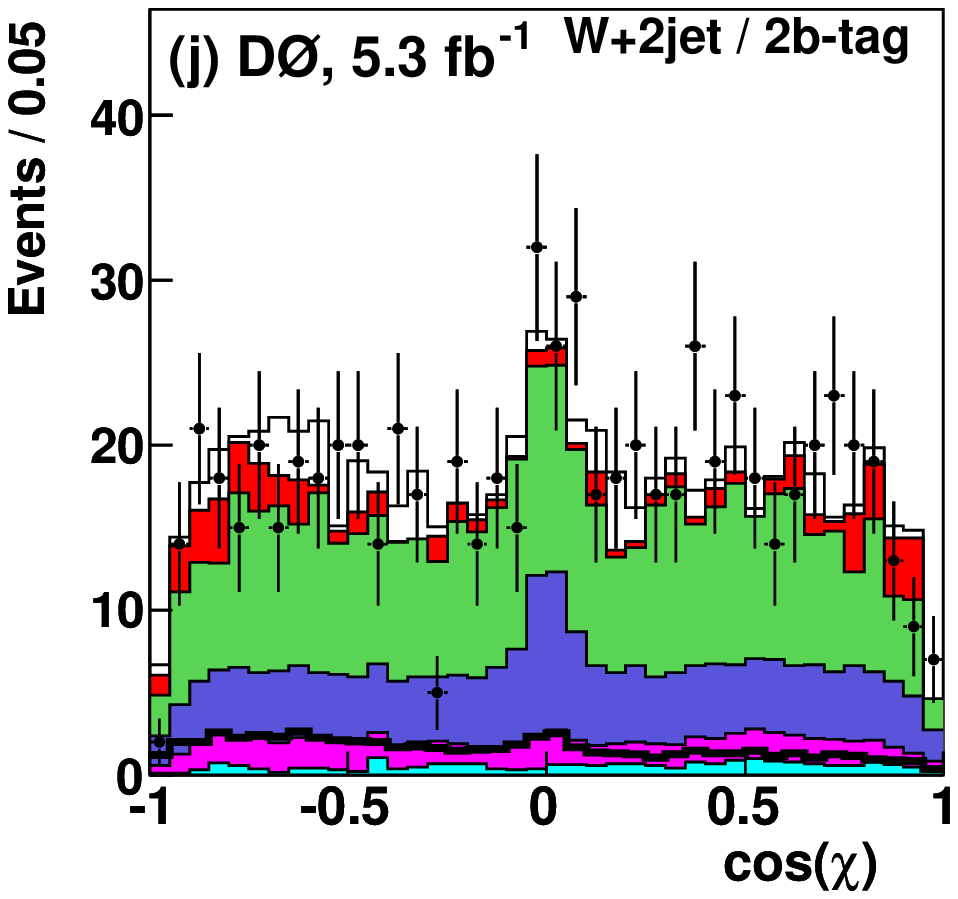}

\caption{Comparison of the total backgrounds to data
for the additional variables provided as inputs to the random forests. The distributions are
compared after requiring two-$b$-tagged jets in $W$+2 jet events. Each variable
is defined in Table \ref{rf_variables}. The expectation for a $WH$ signal at $M_{H} = 115 ~\rm GeV$  has been scaled up by a factor of 10.}
\label{fig:rfinputstwo}
\end{figure*}

\begin{figure*}[t]
\hskip -0.2cm \includegraphics[width=2.2in]{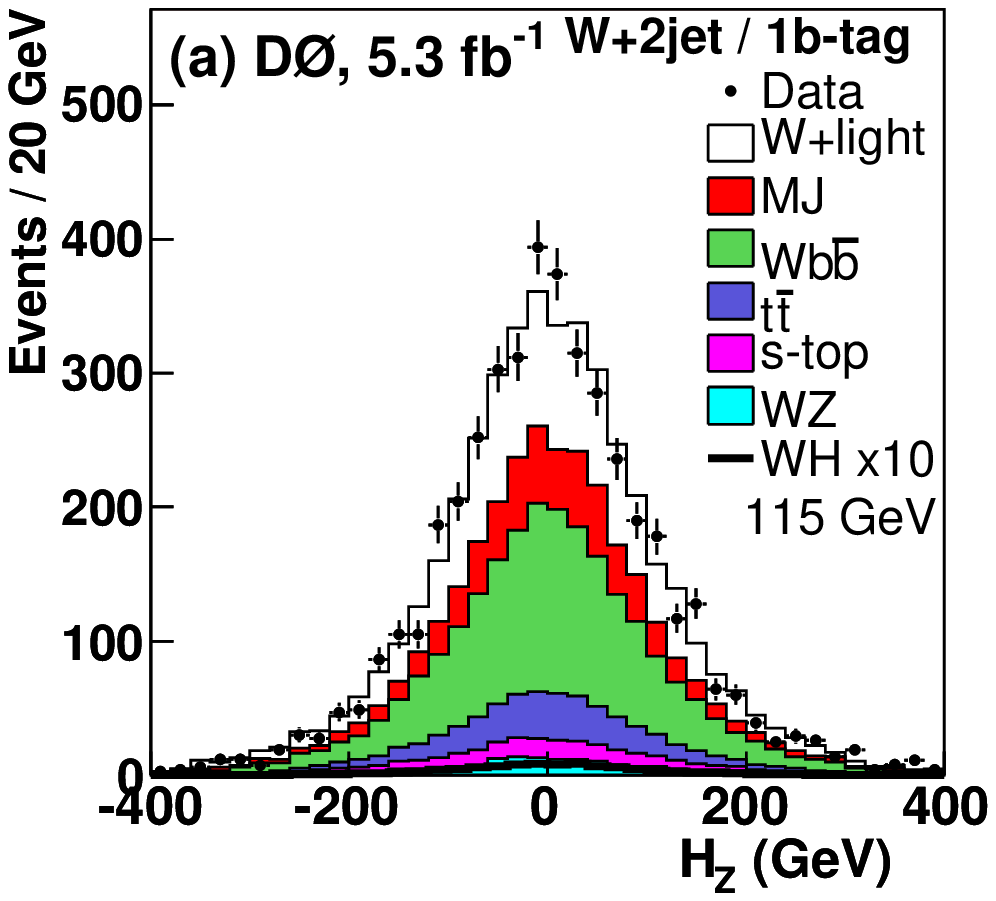}
\hskip 0.2cm \includegraphics[width=2.31in]{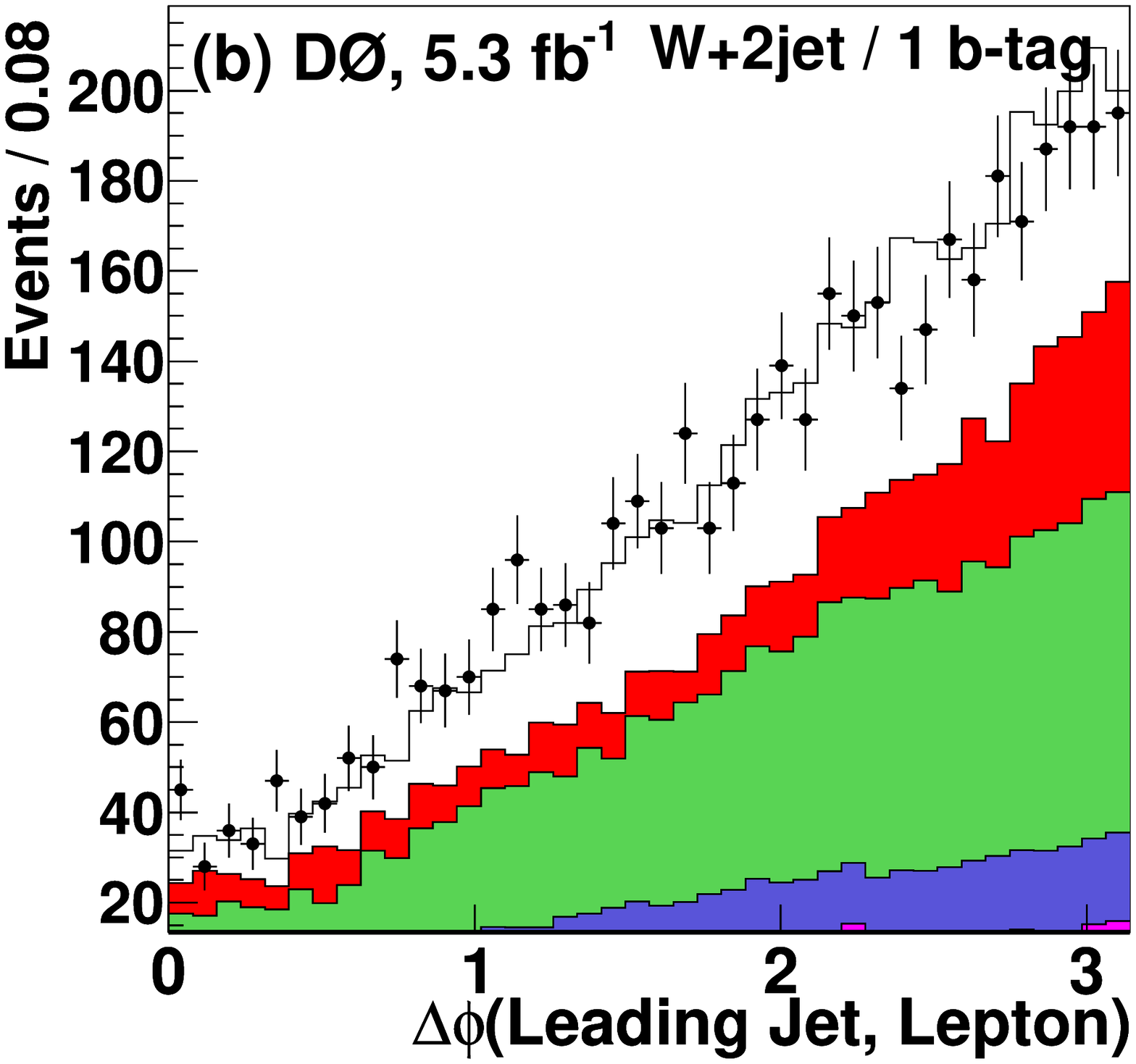}
 \includegraphics[width=2.2in]{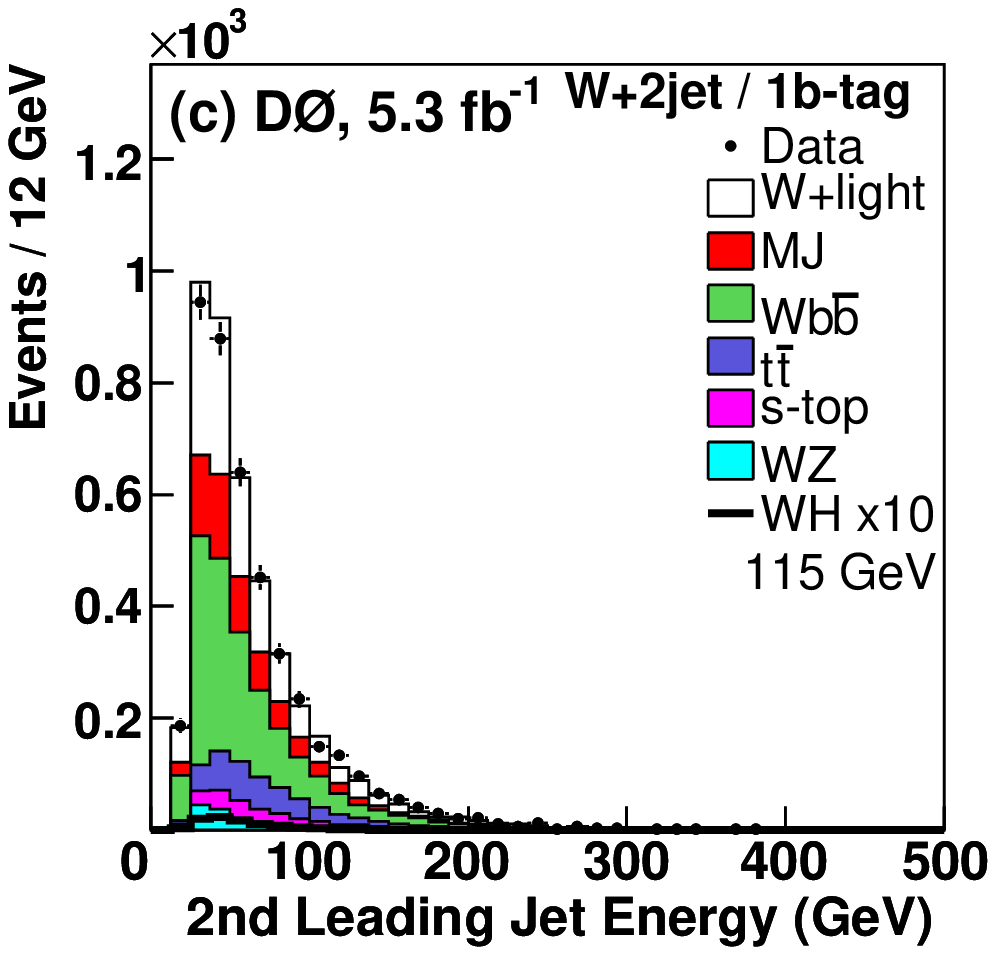}

\hskip -0.2cm \includegraphics[width=2.2in]{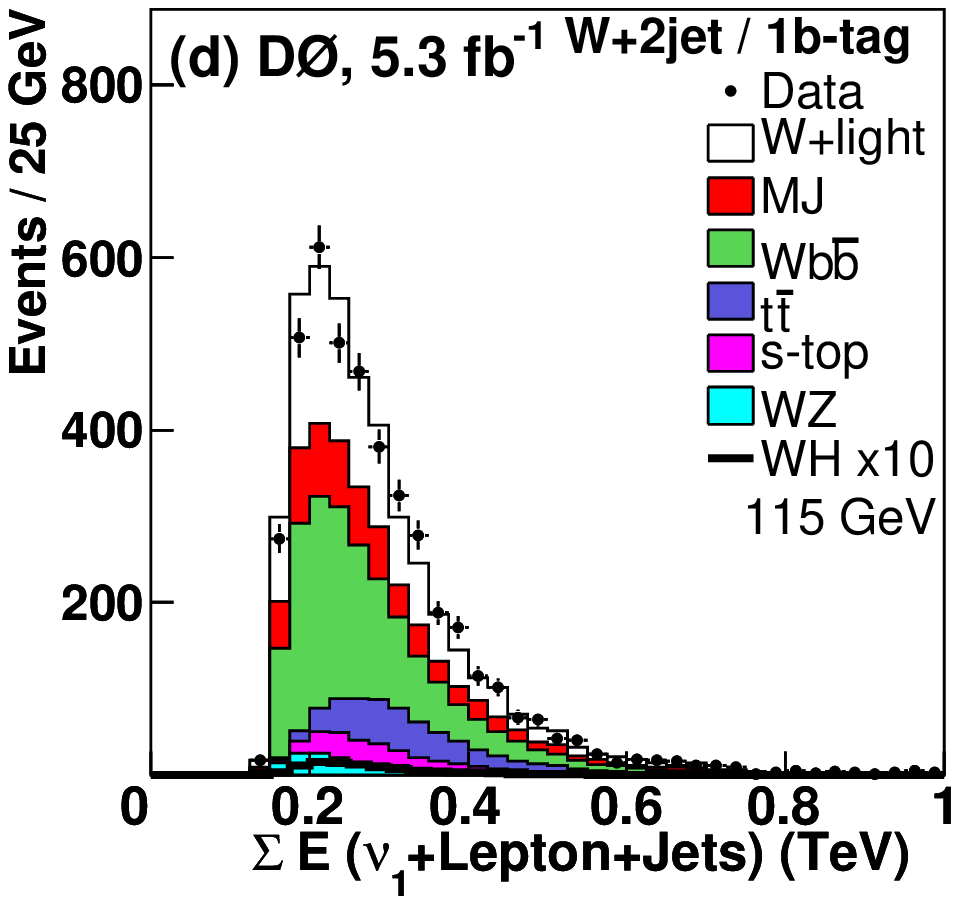}
\hskip 0.2cm \includegraphics[width=2.2in]{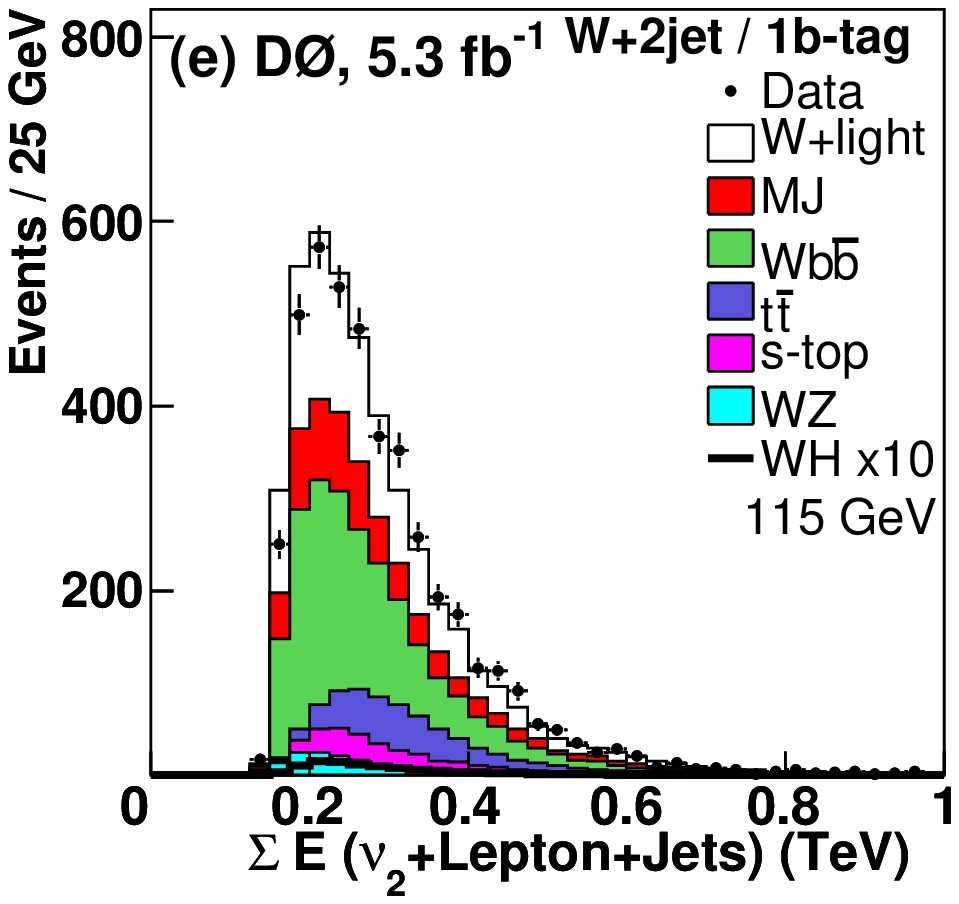}
\hskip 0.2cm \includegraphics[width=2.2in]{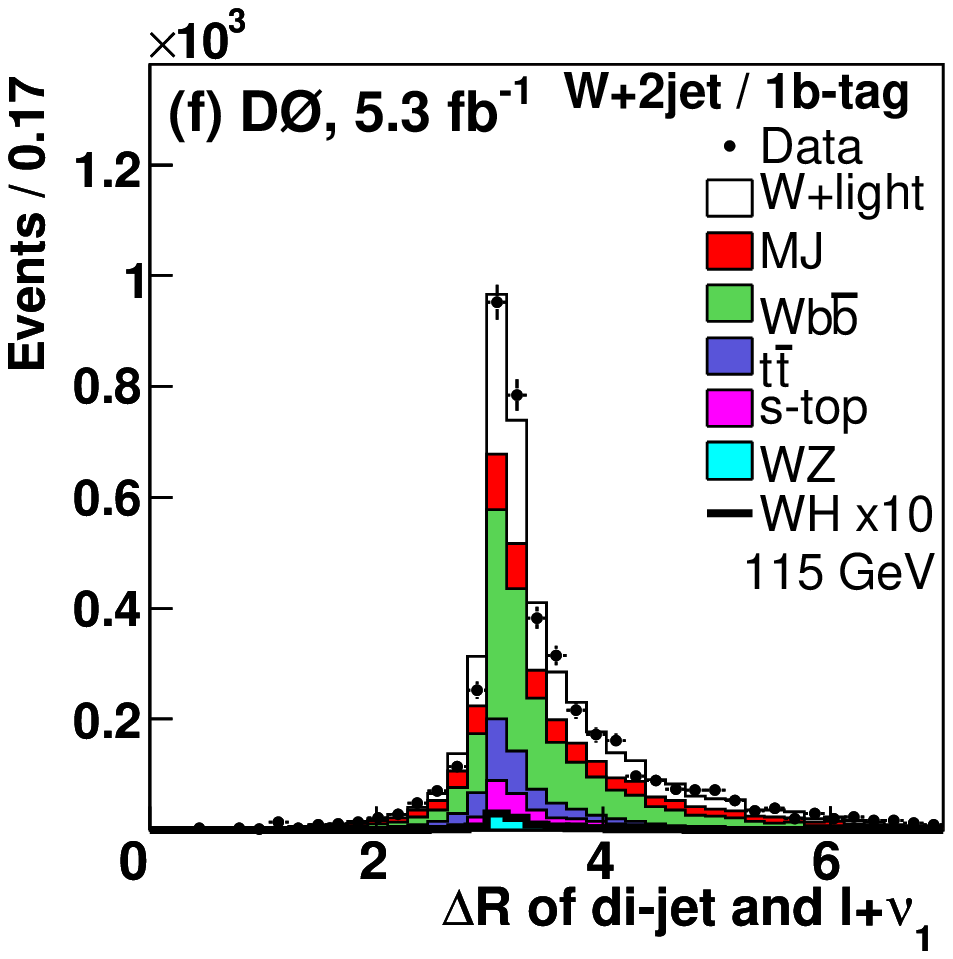}

\hskip -0.2cm \includegraphics[width=2.2in]{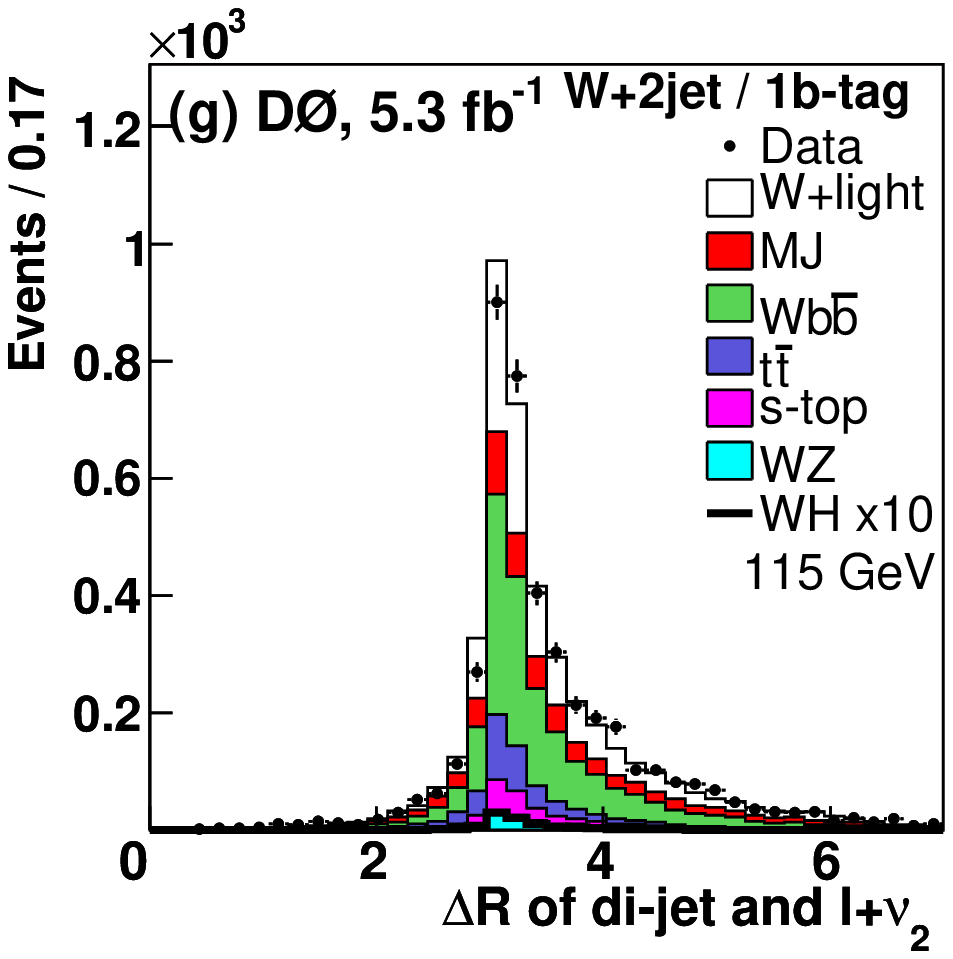} 
\hskip 0.2cm\includegraphics[width=2.2in]{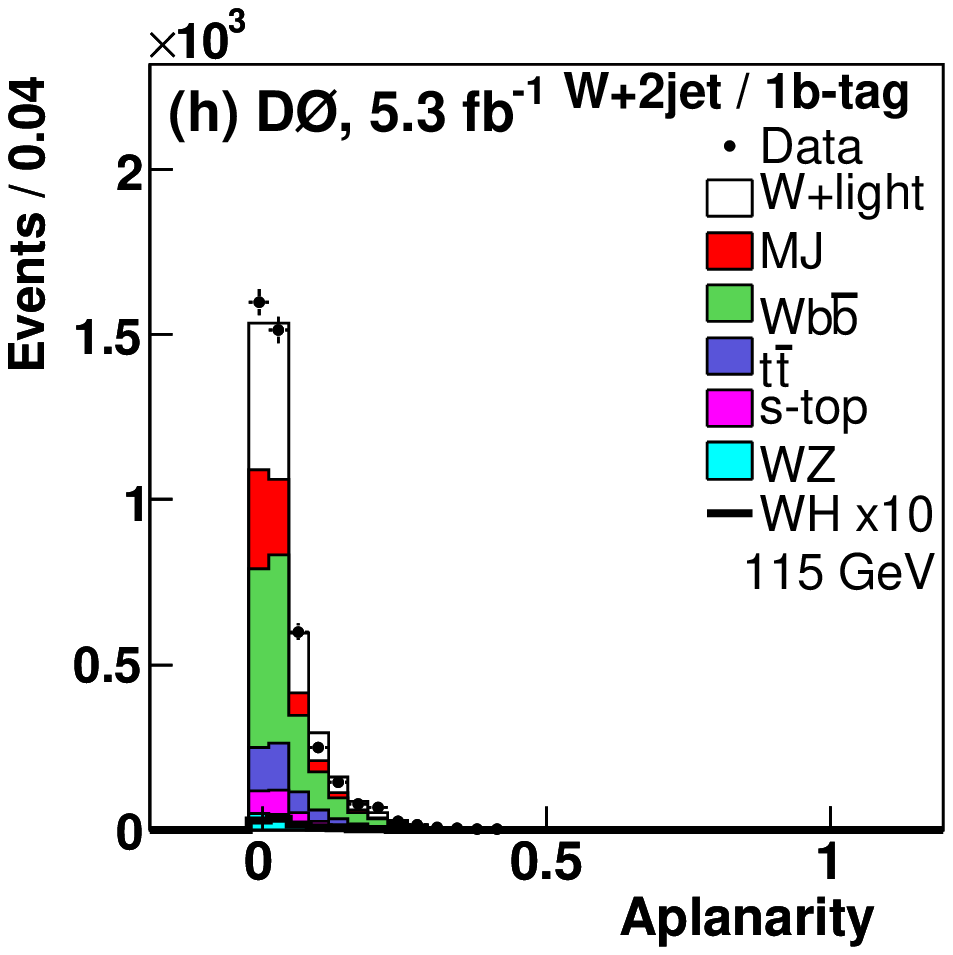}

\hskip -0.2cm \includegraphics[width=2.2in]{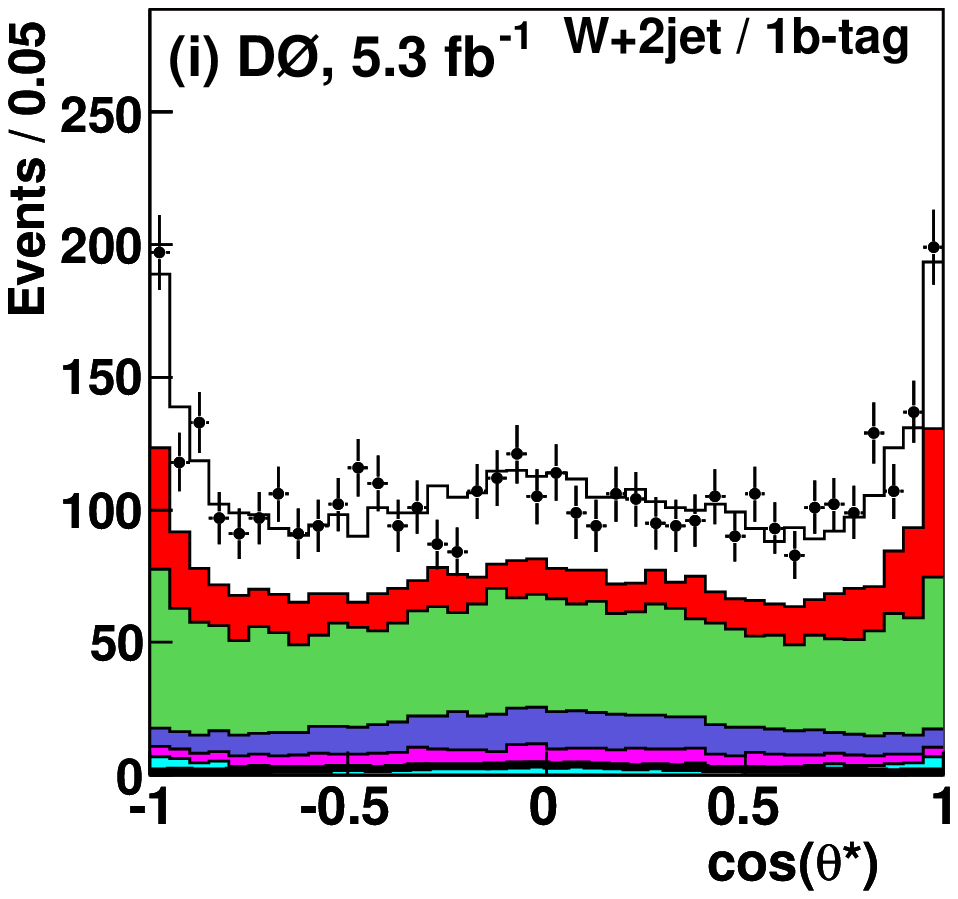}
\hskip 0.2cm \includegraphics[width=2.2in]{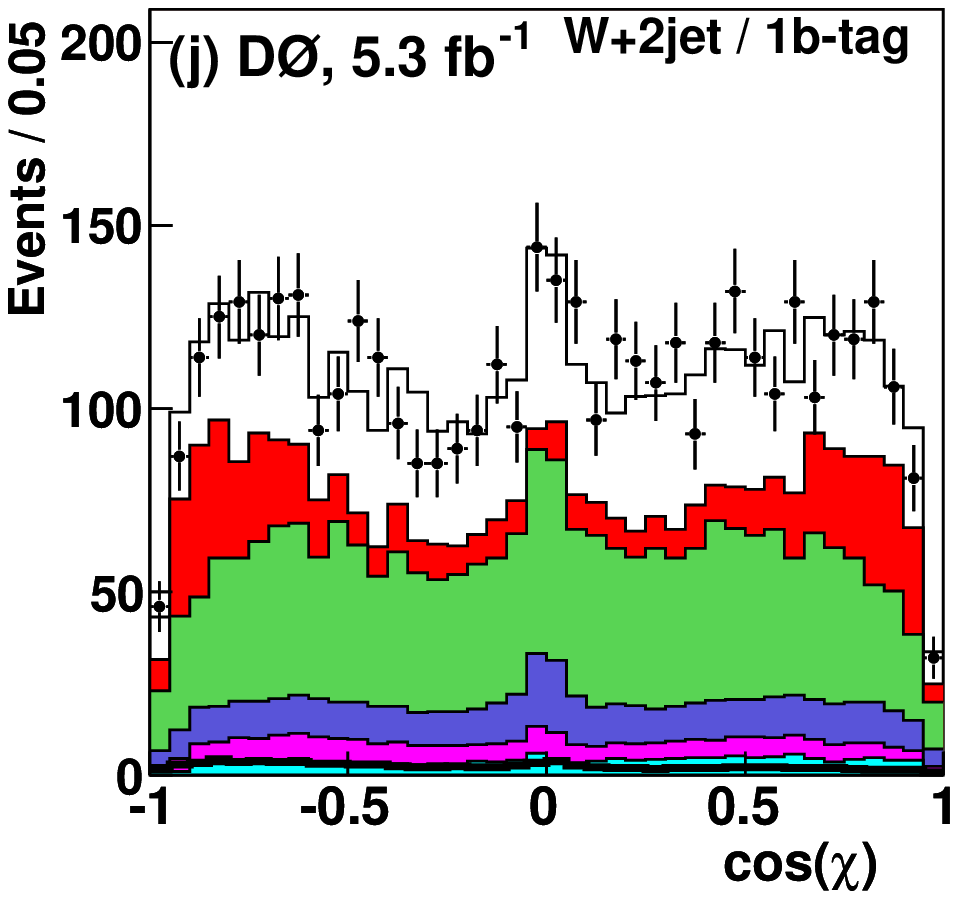}

\caption{Comparison of the total backgrounds to data
for the additional variables provided as inputs to the random forests. The distributions are
compared after requiring a single-$b$-tagged jet in $W$+2 jet events. Each variable 
is defined in Table \ref{rf_variables}. The expectation for a $WH$ signal at $M_{H} = 115 ~\rm GeV$  has been scaled up by a factor of 10.}
\label{fig:rfinputsone}
\end{figure*}

The 20 input variables used to build the RF decision are optimized in dedicated studies of their discriminating power and are listed,
together with their definitions, in Table \ref{rf_variables}.
Agreement between the data and the total MC and data-determined
background estimates are obtained for each input variable distribution
for both the two-$b$-tagged and single-$b$-tagged samples as well as for the full sample prior to the application of $b$-tagging. The same set
of input variables is used for the $W$+2 jet and $W$+3 jet samples.
In addition to the ten variables already discussed in
Sec. VIII, and displayed in Figs.\  \ref{fig:btagtwo}
and \ref{fig:btagone}, a further ten discriminating variables
are provided to each RF and these are shown for the $W$+2 jet sample, after the application of
two and one $b$-tag requirements to the events, in Figs.\ \ref{fig:rfinputstwo} and \ref{fig:rfinputsone}, respectively.

Two input distributions are provided for $\sqrt{\hat{s}}$ and $\Delta R$(dijet,$\ell+\nu$) corresponding to each of the two solutions for the
longitudinal momentum component of the missing energy vector
(assuming the lepton and \MET\ are decay products of an on-shell $W$ boson).
The angles $\theta^{*}$ and $\chi$ are included to exploit kinematic differences arising from
the expected spin-0 nature of the Higgs and non-spin-0 nature of the
$Wb\bar{b}$ background. The angle  $\theta^{*}$ is the angle between the $W$ boson candidate
and the nominal proton beam direction in the zero momentum frame, and  $\chi$
is the angle between the charged decay lepton and rotated (production plane)
three-momentum vector of the dijet system after boosting to the $W$ boson rest frame \cite{spincorr}.

Each RF is trained simultaneously using all simulated backgrounds sources 
(the MJ contribution is excluded) for each simulated Higgs 
mass point, and the process is repeated for each of the 16 subsamples. The minimum number of events in a leaf is tuned in separate studies and 
the number that maximizes the sensitivity is
selected. 
The number of decision trees used within
each forest is also studied and tuned using the procedure of Ref.\ \cite{diboson}.

\begin{figure*}[t]
\hskip -0.2cm \includegraphics[width=2.2in]{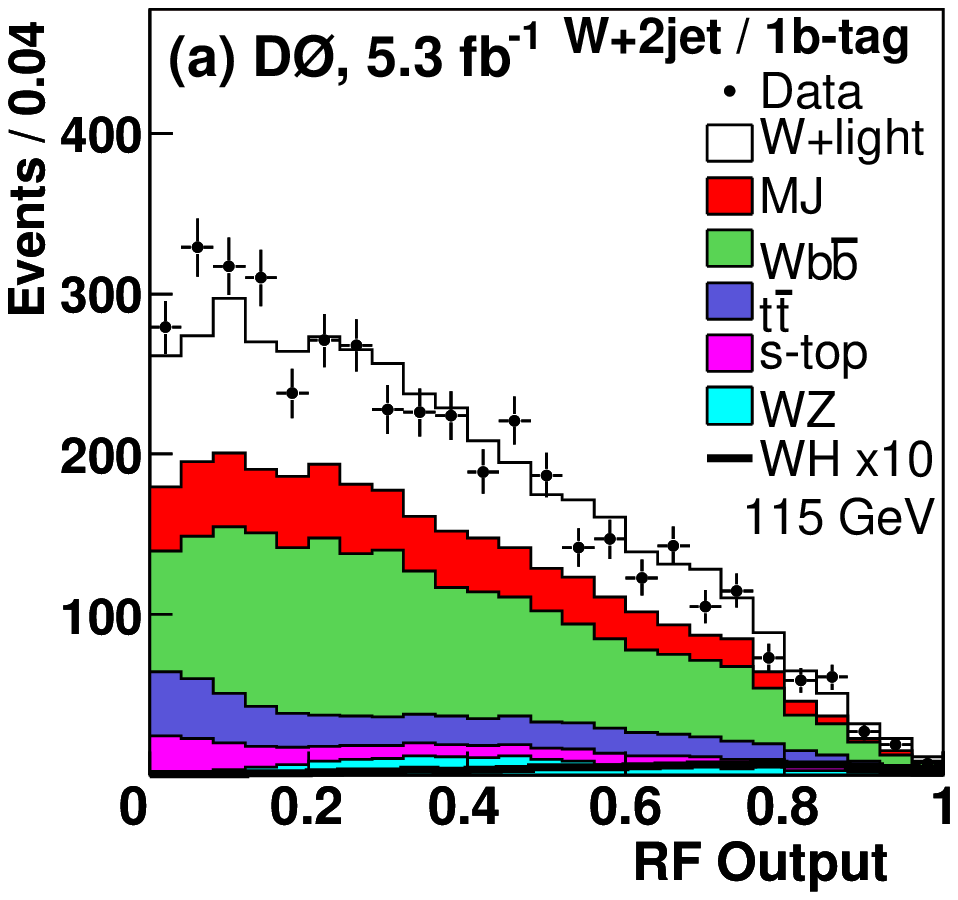}
\hskip 0.2cm \includegraphics[width=2.2in]{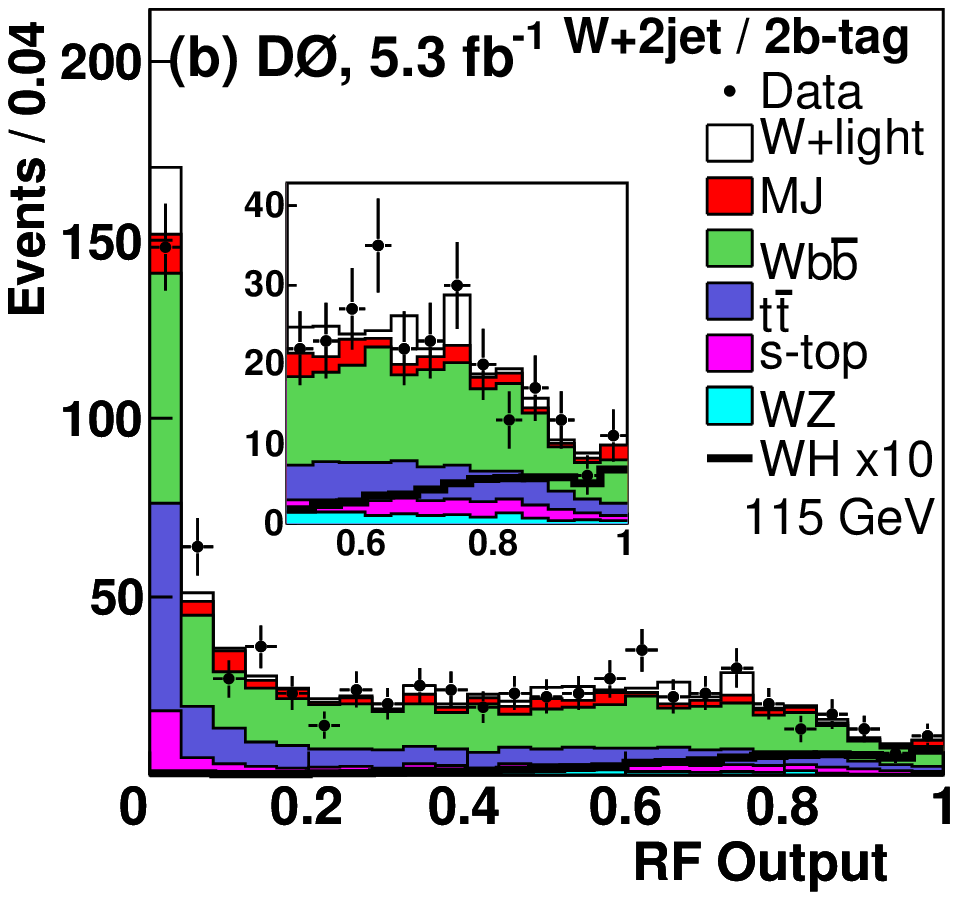}

\hskip -0.2cm \includegraphics[width=2.2in]{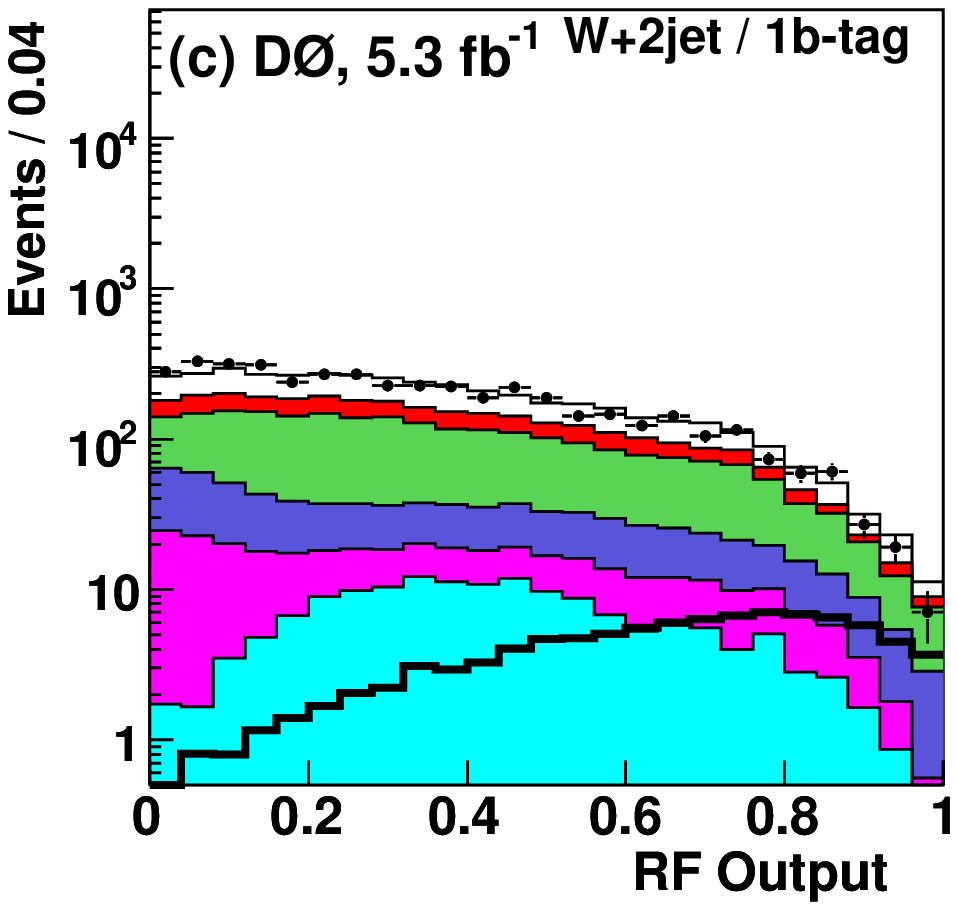}
\hskip 0.2cm \includegraphics[width=2.2in]{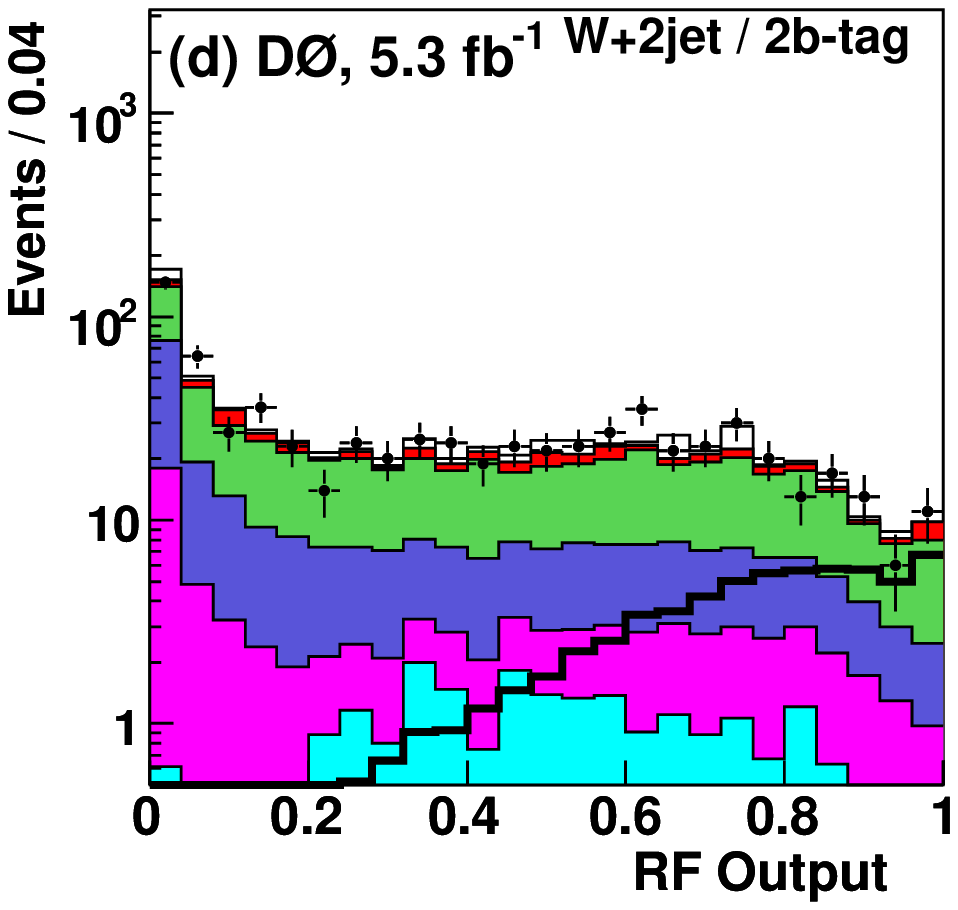}
\vskip 0.4cm
\caption{[color online] Output RF distributions on linear and logarithmic scales for (a), (c) single (one-b-tag) and (b), (d) two-$b$-tagged
$W$+2 jet events. The expectation for signal at $M_{H}=115 \rm ~GeV$ (solid black line) is scaled by a factor of 10.}
\label{fig:rf2jet}
 \end{figure*}

\begin{figure*}[t]
\hskip -0.2cm \includegraphics[width=2.2in]{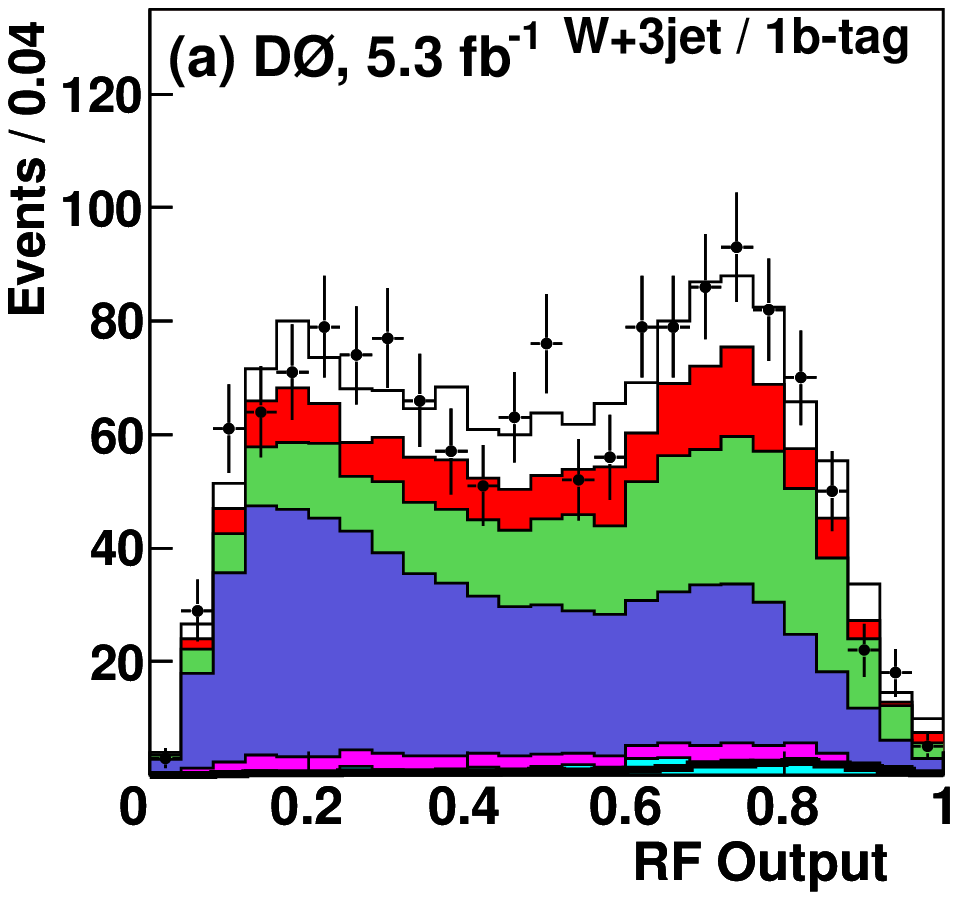}
\hskip 0.2cm \includegraphics[width=2.2in]{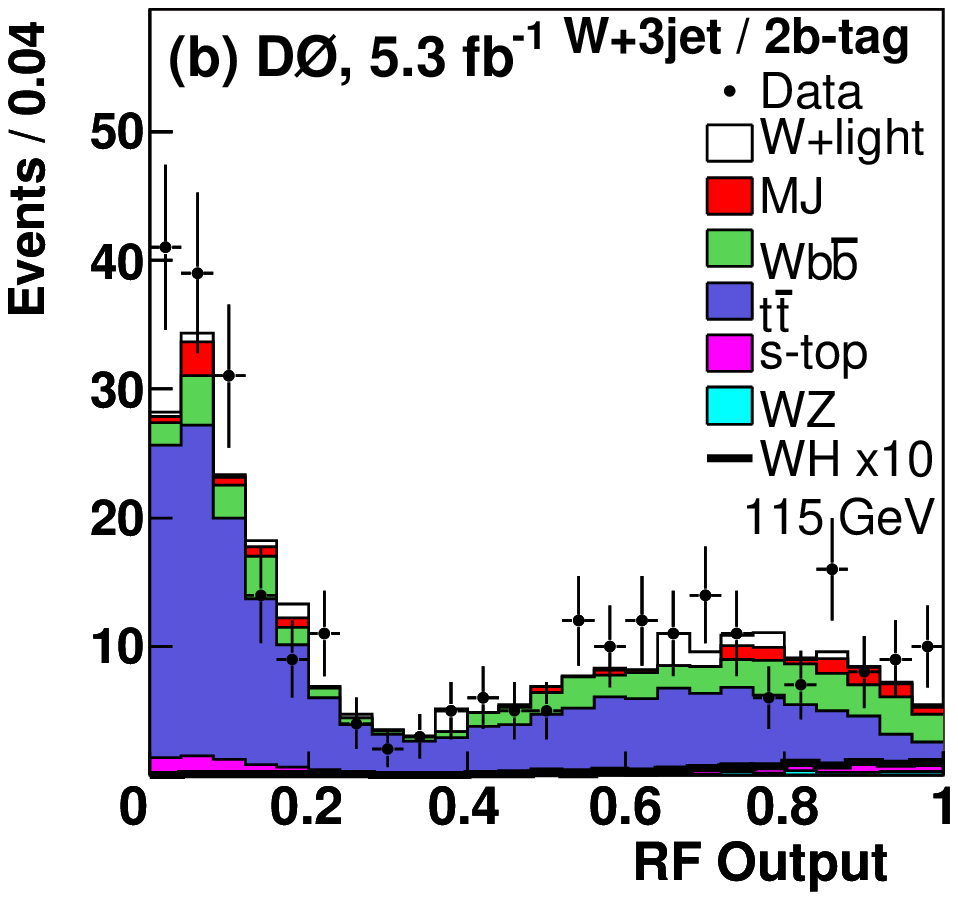}

\hskip -0.2cm \includegraphics[width=2.2in]{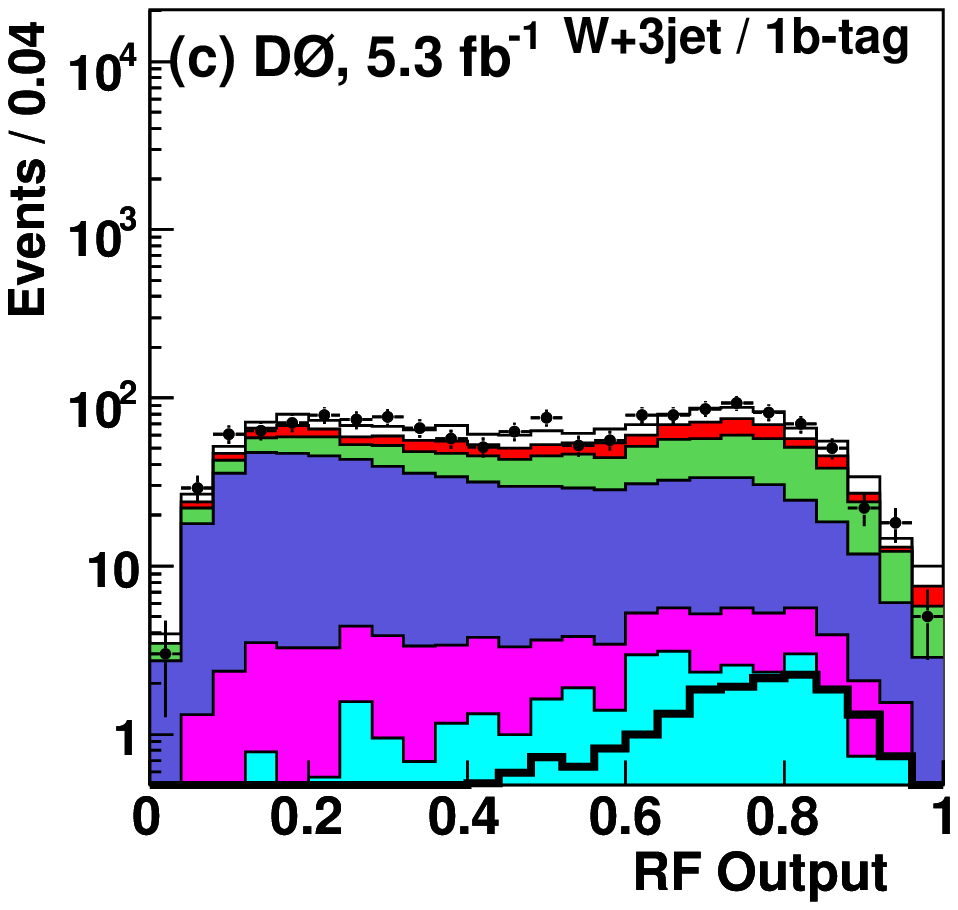}
\hskip 0.2cm \includegraphics[width=2.2in]{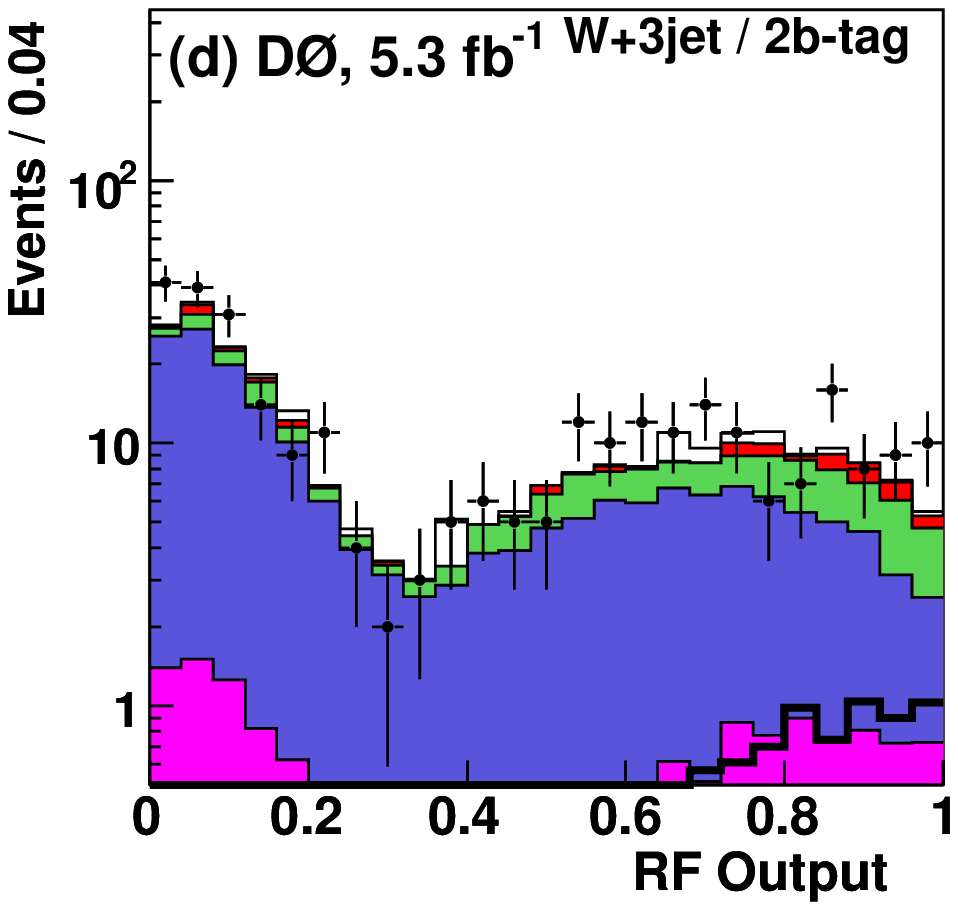}
\vskip 0.4cm
\caption{[color online] Output RF distributions on linear and logarithmic scales for  (a), (c) single  (one-b-tag) and (b), (d) two-$b$-tagged
$W$+3 jet events. The  expectation for signal at $M_{H}=115 \rm ~GeV$ (solid black line) is scaled by a factor of 10. }
\label{fig:rf3jet}
 \end{figure*}

The resulting RF output distributions are shown in Figs.\ \ref{fig:rf2jet} and \ref{fig:rf3jet} for the two- and single-$b$-tagged jet 
requirements in the final $W$+2 jet and $W$+3 jet samples, respectively. The electron and muon channel samples have been 
combined in the figures, and the preupgrade and postupgrade samples are also combined in the 
figures.
The figures show the results obtained using the $M_{H}=115 ~\rm GeV$ signal samples. An improved separation of simulated signal 
and background contributions, is obtained in all cases.

\section{Systematic Uncertainties}
The impact of each possible source of systematic uncertainty is assessed separately for the 
signal and for all backgrounds, for each of the 16 statistically independent samples, and categorized 
according to whether it affects the normalization and the shape (shape systematic) of the RF discriminant output
distributions or whether it only affects the normalization of signal and backgrounds. A full analysis 
is repeated after individually varying each 
source by $\pm 1$ standard deviation (s.d.) in the simulation, except where noted 
otherwise (the uncertainty in the MJ background modeling is determined 
separately from data). After each variation, the simulated and MJ background yields are normalized to the selected data samples prior to the 
application of $b$-tagging.

The systematic uncertainty assigned to the data-determined efficiency
of the triggers used in the electron channel is (3--5)\%. In the muon 
channel, where the full list of available triggers is used, a comparable 
uncertainty of (3--4)\% is assigned. In the muon channel, this 
uncertainty arises from
a normalization uncertainty of 2\%, obtained after comparing results 
using the single high-$p_{T}$ muon and the full list of triggers, and a 
shape systematic of (1--3)\% as a function of jet $p_{T}$, applied 
to the non-single-muon trigger efficiency. The shape systematic is 
obtained by comparison of the 
single high-$p_{T}$ muon and non-single-muon triggered components of the
dataset.

The uncertainty on the identification and reconstruction of isolated 
electrons, as well as their energies, affects the shapes of the electron 
channel RF distributions and is (5--6)\%. In the muon channel, the uncertainty 
comprises three contributing sources: an uncertainty of 0.8\% applied to the pre-upgrade 
muon identification efficiency (a 1.2\% uncertainty is applied to the 
post-upgrade samples, which is increased for muon $p_{T}<20 ~\rm GeV$ by 
adding 2\% in quadrature), an
uncertainty in the corresponding track reconstruction of 2.3\% (pre-upgrade) 
and 1.4\% (post-upgrade), and an uncertainty of 3.8\% (pre-upgrade) and 0.9\% 
(post-upgrade) on the scale factors used to correct the efficiencies for muons 
to pass isolation criteria in the MC to those measured in the data. 

Sources of systematic uncertainty on the selection and reconstruction 
of jets are the jet resolution and jet energy scale, as well as the 
jet identification efficiency and vertex confirmation requirement (applied to the post-upgrade part of the 
dataset). Shape uncertainties for jet resolution and jet energy
scale are determined by varying parameters in the jet resolution function 
and the energy scale correction and repeating the analysis using the kinematics
of the modified jets. The size of this effect on the RF distribution depends on the sample and 
process and is in the range 15\%-30\%. The jet identification and vertex 
confirmation uncertainties are each 
determined by randomly reducing the number of jets that remain in simulation
(the $+1$ s.d.~ result is then obtained by inverting the $-1$ s.d.~ result).
The resulting RF shape systematic uncertainty is about 5\%.  Because of low 
statistics after $b$-tagging for the $W$+light and $WZ$ samples, the jet
systematic uncertainties applied for these
backgrounds are determined prior to $b$-tagging.

The uncertainty on the jet taggability requirement is determined 
by varying the jet taggability correction factors. The taggability uncertainty affects 
the shapes of the RF output distributions and is about 3\%.
The RF shape uncertainty for the response of the $b$-tagging 
algorithm is applied 
separately for light and heavy flavored jets and is typically (2.5--3.0)\%  for single-tagged 
heavy flavor jets and in the range (1--4)\% for single-tagged light jets (the light-quark jet 
mistag probability uncertainty is of order 10\%). The RF uncertainty is approximately
doubled in the samples requiring two $b$-tagged jets. 

\begin{figure*}[t]
\includegraphics[width=5.2in]{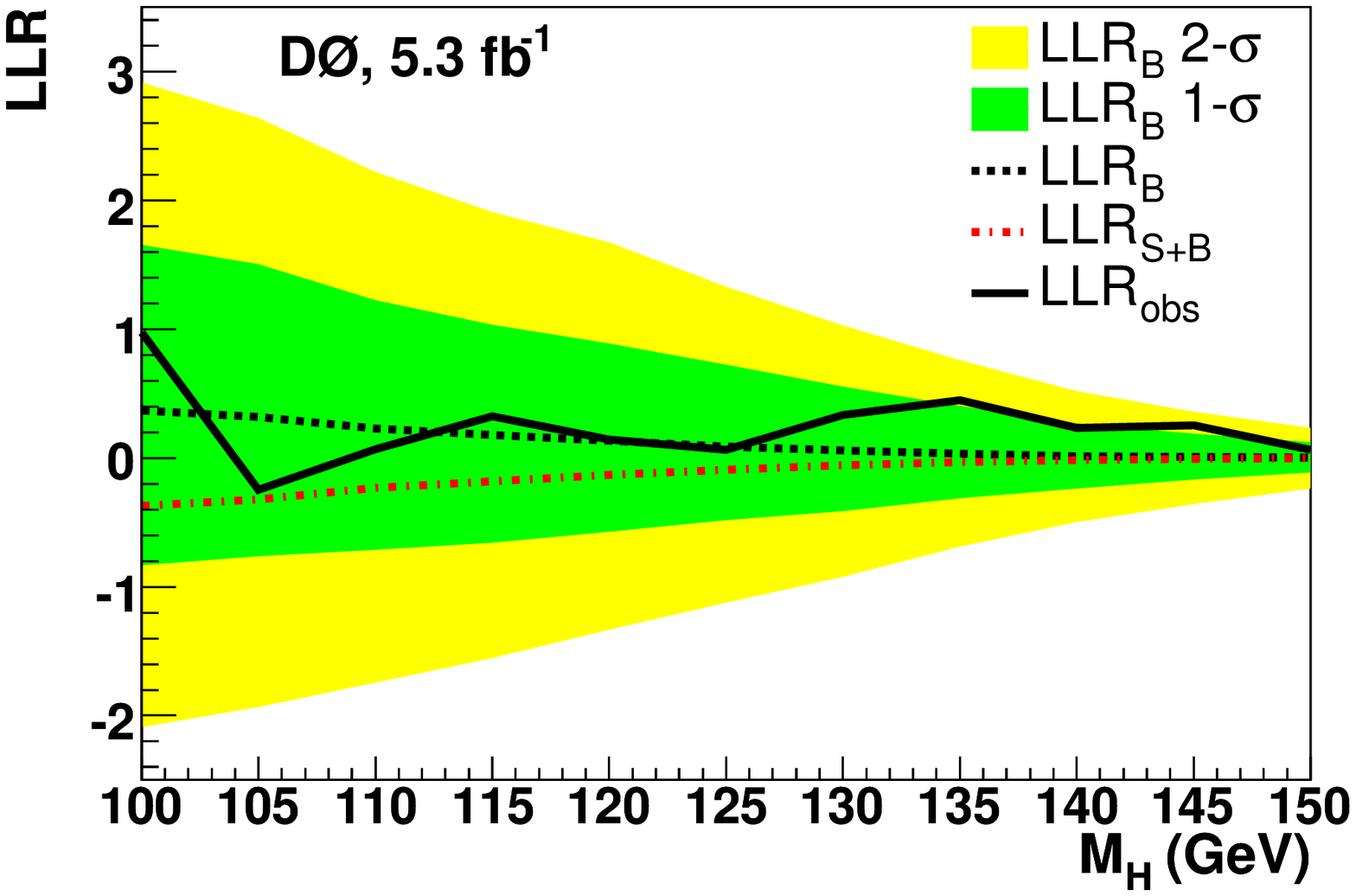}
\vskip -0.3cm
\caption{[color online] The observed LLR as a function of hypothetical Higgs boson mass. Also shown
are the medians of the resulting LLR 
distributions for the background-only hypothesis (dashed line), 
along with the  $\pm 1\sigma$ and $\pm 2\sigma$ values (shaded bands), after generating 
multiple pseudoexperiments at each test mass point. The medians of the 
signal-plus-background hypothesis are shown by the dash-dotted line.}
\label{fig:llr} 
\end{figure*}

Uncertainties in the predicted $t\bar{t}$, single top quark, and diboson cross sections are taken from \cite{ttbar_xsecs,stop_xsecs,mcfm} and 
affect the normalizations of the backgrounds. 
The uncertainty on the CTEQ6L parton density function is estimated 
following the prescription of Ref.\ \cite{CTEQ}.  The 
{{\sc alpgen}}-generated samples include additional normalization factors that change their visible 
cross sections, and their uncertainties are determined separately.   The uncertainty in the reweighting procedure 
applied to the {{\sc alpgen}}-generated event samples affects the shape of the 
{{\sc alpgen}} RF output distributions 
and are typically of the order 2\%. The uncertainty on the {{\sc alpgen}} scale factor
$K^{W\text{+jets}}$ is 6\% and the uncertainty on $S_{Wb\bar{b}}$ is 20\%. The renormalization and factorization scales 
used in {{\sc alpgen}} are 
varied by adjusting each scale simultaneously, by factors of 0.5 and 2.0. This affects the 
shapes of the {{\sc alpgen}} RF output distributions, and the resulting uncertainty is 
of the order 2\%, as is the uncertainty 
arising from the choice of value for the strong coupling constant $\alpha_{S}$.
The uncertainty on the MLM factorization scheme used to match {{\sc alpgen}} 
partons to cone jets is propagated to the RF distribution and results in a 
systematic uncertainty of about 2\%.

The uncertainty in the MJ background modeling is obtained from the data. 
It is determined by varying the parameterization
of the efficiency for loosely selected leptons to enter the final selected sample and 
by also varying the misidentified jet probabilities. The MJ 
uncertainties are anticorrelated with the normalization of the {{\sc alpgen}} samples, and this
is taken into account in the limit setting procedure. The overall experimental systematic uncertainty assigned to the $WH$ distributions 
is about 6\%. The uncertainty of the experimentally measured 
integrated luminosity is treated separately. The uncertainty is $6.1$\% \cite{lumi_ID} and is fully correlated between all of the 
simulated background samples.

\section{Upper Limits on the {\boldmath $WH$} Cross Section}

No excess of events is observed with respect to the background estimation and upper limits 
are therefore derived for the $WH$ production cross section multiplied by the 
corresponding $H\rightarrow b\bar{b}$ branching ratio in units of the SM prediction. 
The limits are
calculated using the modified frequentist  $CL_{s}$ approach \cite{modfreq,modfreq1}, and the procedure is
repeated for each assumed value of $M_{H}$. 

Two hypotheses are considered: the background-only hypothesis 
(B), in which only background contributions are present, and the signal-plus-background (S+B) 
hypothesis in which both signal and background contributions are present. 

The limits are determined using the RF output distributions, 
together with their associated uncertainties, as inputs 
to the limit setting procedure. To preserve the stability of the limit derivation procedure in regions of small background, the width of the 
bin at the largest 
RF output value is adjusted by comparing the total B and 
S+B expectations until the statistical significance for B and S+B is, respectively, greater than $\approx$ 3.6 and 5.0 standard deviations 
from zero. The remaining part of the 
distribution is then divided into 23 equally-sized bins. 
The rebinning procedure is checked for potential biases in the
determination of the final limits, and no such bias is found.

\begin{figure*}[t]
\includegraphics[width=3.5in]{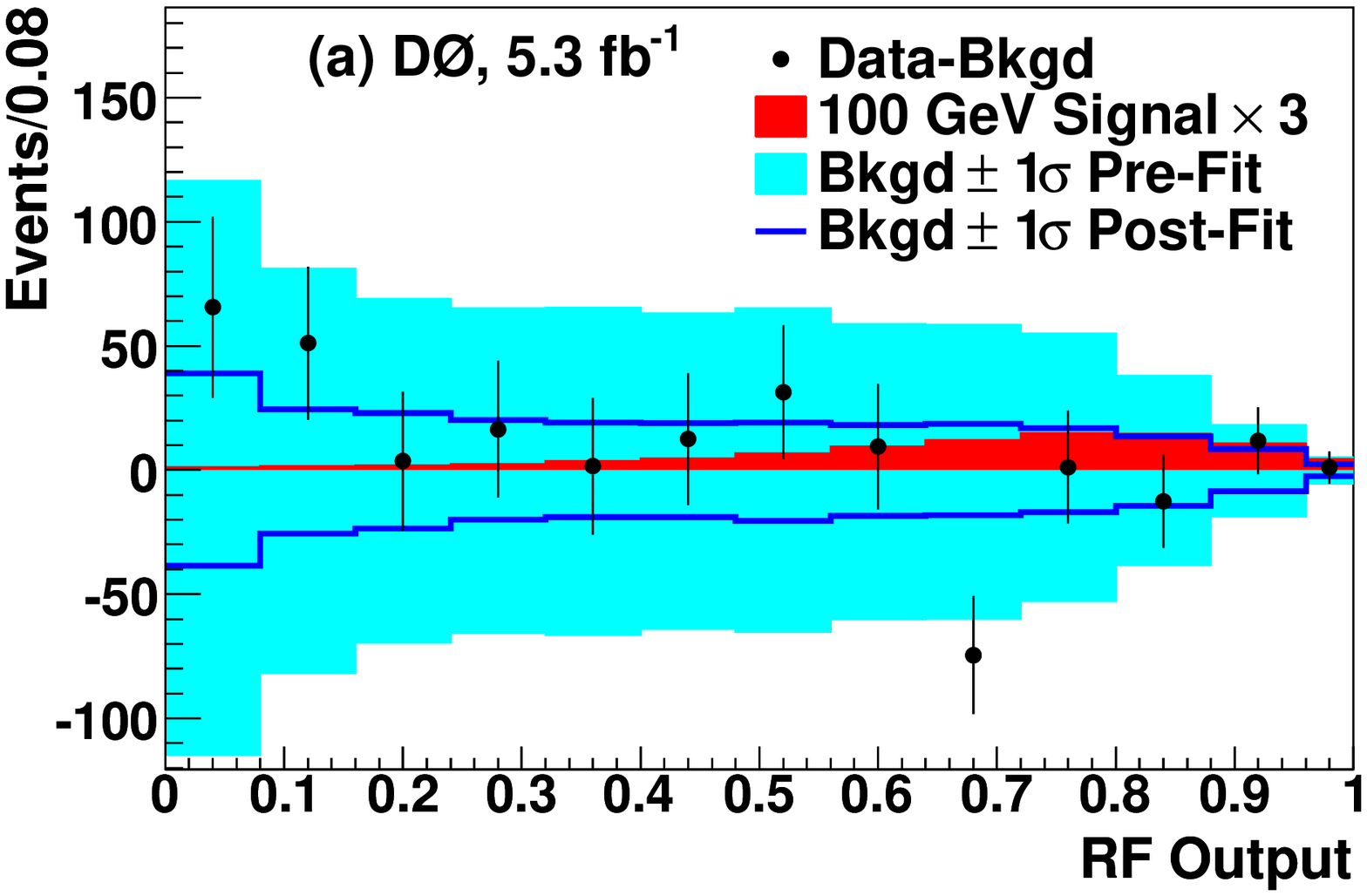}
\includegraphics[width=3.5in]{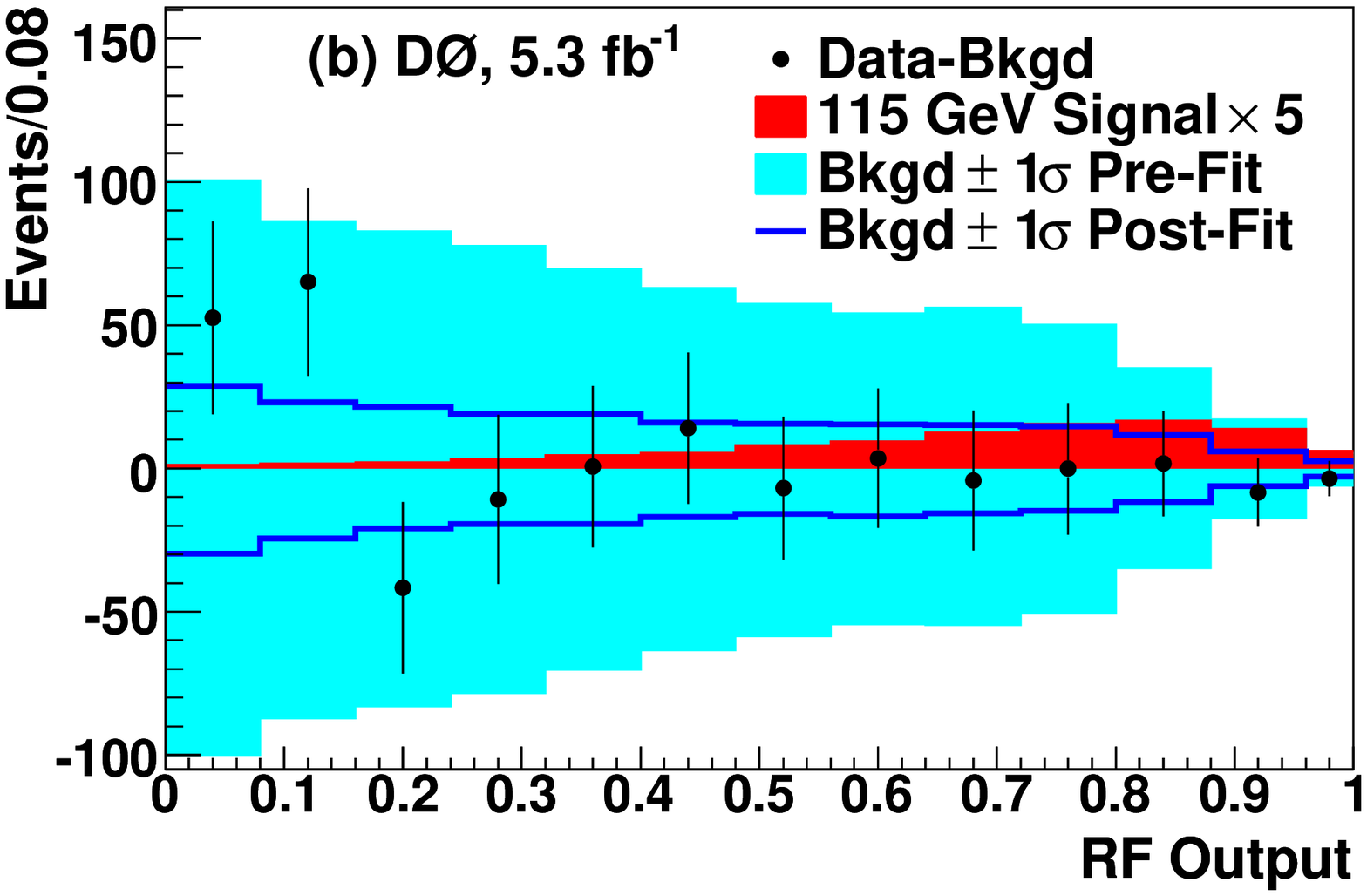}
\includegraphics[width=3.5in]{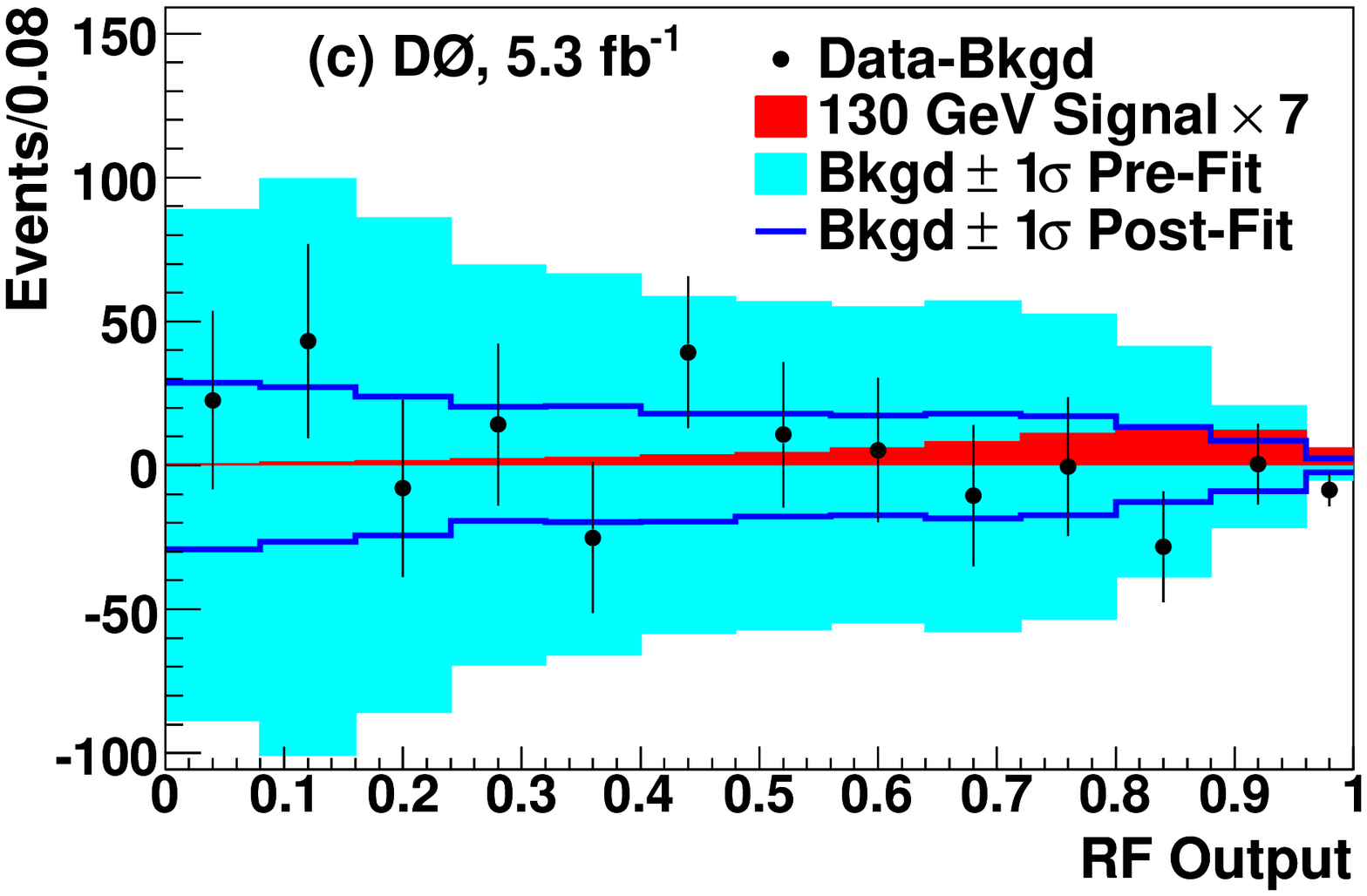}
\includegraphics[width=3.5in]{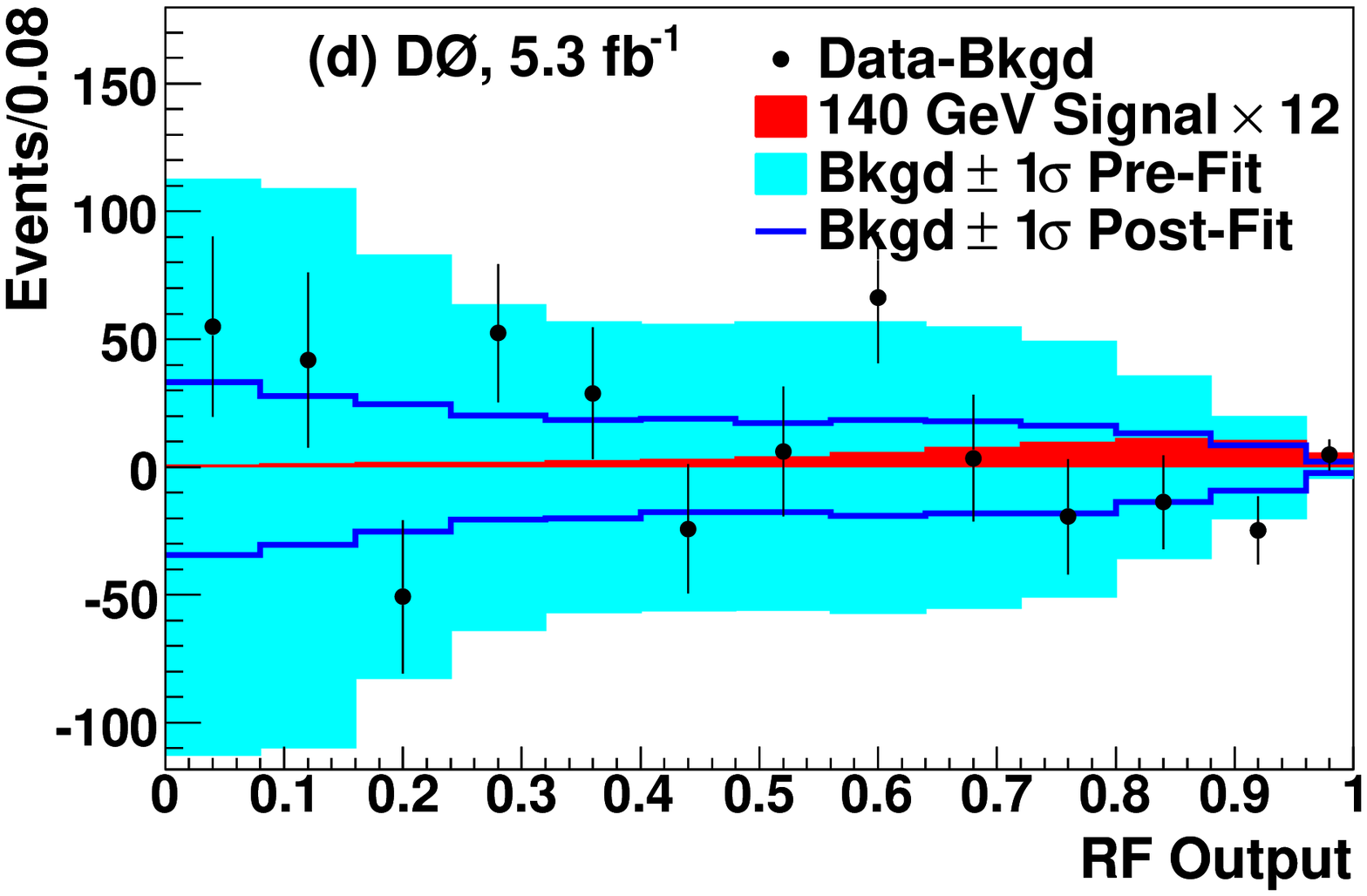}
\caption{[color online] The RF discriminant output distribution minus the total background expectation for 
(a) $M_H=100\rm ~GeV$, (b) $M_H=115\rm ~GeV$, (c) $M_H=130\rm ~GeV$, and (d) $M_H=140\rm ~GeV$. The prefit uncertainties are shown by the shaded bands and 
the post-fit uncertainties are represented by the solid lines. The signal
expectation is shown scaled to the obtained observed upper limit at each test mass point.  }
\label{fig:syst_plot}
\end{figure*}

The result for each hypothesis is obtained by testing the outcome of a large number of simulated pseudoexperiments. For each 
pseudoexperiment, pseudodata are drawn from the RF distributions, by randomly generating the pseudodata 
according to a Poisson statistical parent distribution 
for which the mean is either taken from the background-only or signal-plus-background hypothesis.  A negative Poisson log likelihood 
ratio (LLR) test statistic is used to evaluate the statistical significance of each
experiment,  with the outcomes ordered in terms of their 
statistical significance. The frequency of each outcome defines 
the shapes of the resulting LLR distribution, for both
the background-only and signal-plus-background hypotheses, at each mass point. 

Systematic uncertainties are defined through nuisance parameters that
are assigned Gaussian probability distributions (priors). 
The signal and background predictions are taken to be 
functions of the nuisance parameters and each nuisance parameter 
is sampled from a Gaussian probability distribution in each 
pseudoexperiment. The correlated systematic uncertainties across 
channels (such as the uncertainties on predicted SM cross sections, identification efficiencies, 
and energy calibration, as described in Sec.\ X) are also taken
into account in the limit setting procedure \cite{collie}.

The inclusion of systematic uncertainties in the generation of
pseudoexperiments has the effect of broadening the LLR
distributions and, thus, reducing the ability to resolve
signal-like excesses.  This degradation can be partially reduced
by performing a maximum likelihood fit to each
pseudoexperiment (and data), once each for the
S+B and the background-only hypotheses.  The
maximization is performed over the systematic uncertainties.  The
LLR is evaluated for each outcome using the ratio of maximum
likelihoods for the fit to each hypothesis.
The resulting degradation of the limits due to systematic uncertainties
is of the order of 30\%.

The medians of the obtained background-only LLR distributions  for each tested mass point 
are summarized in Fig.\ \ref{fig:llr}. The resulting medians of the signal-plus-background 
hypothesis LLR distributions are also shown. The corresponding $\pm 1\sigma$ 
and $\pm 2\sigma$ values for the background-only hypothesis at each mass point are represented
by the shaded regions in the figure. The LLR values obtained from the data are also summarized in the figure.

\begin{figure*}[t]
\includegraphics[width=6.3in]{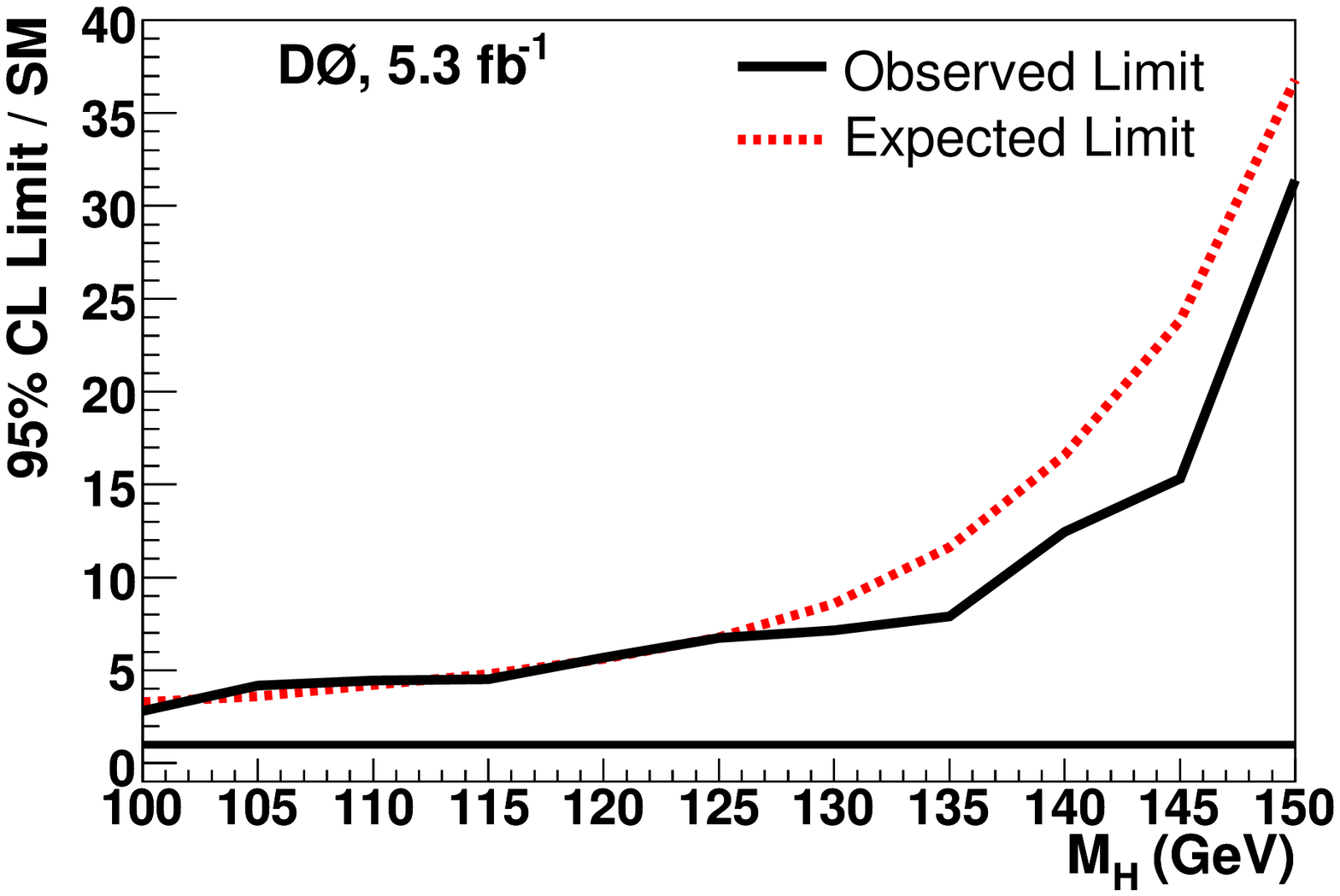}
\caption{[color online] The expected and observed 95\% C.L upper limits from thelikelihood fit, maximized over systematics, as a function of the hypothetical Higgs mass $M_H$. The limits are presented as ratios of 
$\sigma(p\bar{p} \rightarrow WH) \mathcal{BR}(H \rightarrow b \bar{b})$
to the expected SM prediction.}
\label{fig:limits} 
\end{figure*}

The RF discriminant distributions after the background-only profile fit are shown in Fig.\ \ref{fig:syst_plot} after 
subtracting the total background expectation, for the Higgs boson mass 
points 
$M_H=100,115,130,$ and $140\rm ~GeV$. The signal
expectations are shown scaled to the final observed upper limits (rounded to 
the nearest integer) in each case, and the uncertainties in the background before and after the constrained fit are shown by the 
shaded bands and solid lines, respectively.

\begin{table}[h]
\vspace{-0.3cm}
\caption{ The expected and observed 95\% C.L. limits from the likelihood fit,
maximized over systematics, as a function of the hypothetical Higgs mass $M_H$. The limits are presented as ratios of
$\sigma(p\bar{p} \rightarrow WH) \times \mathcal{BR}(H \rightarrow b \bar{b})$
to the expected SM prediction.\\ }
\label{limitvalues}
\begin{tabular}{ccc}
\hline \hline
 &    \multicolumn{2}{c}{Combined 95\% C.L. Limit  $ /  \sigma_{SM}$ }\\
 \hline
 Higgs Mass [GeV] & Expected & Observed \\
\hline
100  &  3.3   & 2.7 \\
105  &  3.6   & 4.0 \\
110  &  4.2   & 4.3 \\
115  &  4.8   & 4.5 \\
120  &  5.6   & 5.8 \\
125  &  6.8   & 6.6 \\
130  &  8.5   & 7.0 \\
135  &  11.5  & 7.6 \\
140  &  16.5  & 12.2 \\
145  &  23.6  & 15.0 \\
150  &  36.8  & 30.4 \\
\hline \hline
\end{tabular}
\end{table}

Upper limits are calculated at 11 discrete values of the Higgs boson mass, spanning the range 100--150 GeV and spaced in units of 
5 GeV, by scaling the expected
signal contribution to the value at which it can be excluded at the 95\% 
C.L. The expected limits are calculated from the background-only LLR distribution whereas the observed limits are quoted with respect 
to the LLR values measured in data. The expected and observed 95\% C.L. upper limits 
results for the  $WH$ cross section multiplied by the branching ratio 
$H\rightarrow b\bar{b}$ are shown, as a function of the Higgs boson mass $M_{H}$, 
in units of the SM prediction in Fig.\ \ref{fig:limits}. The values obtained for the expected and observed limit to SM ratios at each 
mass point are listed in Table \ref{limitvalues} (the uncertainty in the predicted  $WH$ cross section is available in Ref.\ \cite{signal}).

\section{Summary}

A search for the SM associated $WH$ production in data corresponding
to an integrated luminosity of $\cal{L}$ $\approx 5.3 ~\rm fb^{-1}$ collected
with the D0 detector at the Fermilab Tevatron $p\bar{p}$ Collider shows no 
excess beyond the expected contributions from SM backgrounds. 
Statistically independent data samples containing 
$W\rightarrow e \nu$ and $W\rightarrow \mu \nu$ candidates with either two 
or three reconstructed jets in the event and subdivided into two $b$-tagged jets or a single $b$-tagged jet are analyzed 
using a multivariate technique to provide separation of signal and background.
Upper limits are calculated at the 95\% C.L. 
for the $WH$ cross section multiplied by the branching ratio $H\rightarrow b\bar{b}$ for the region $100 < M_{H}<150~\rm GeV$.
The 
observed (expected) upper limits at 95\% C.L. are a factor 4.5 (4.8) 
larger than the SM expectation for a Higgs mass $M_{H}=115~\rm GeV$.
These results, combined with those of Ref.\ \cite{Aaltonen:2011rt} and with other searches
 in this mass region at the Tevatron, provide crucial constraints on the
 Higgs coupling to $b\bar{b}$, complementary to the information obtained,
 for other decay modes, by the LHC experiments.\\

\section{Acknowledgments}
%
We thank the staffs at Fermilab and collaborating institutions,
and acknowledge support from the
DOE and NSF (USA);
CEA and CNRS/IN2P3 (France);
MON, Rosatom and RFBR (Russia);
CNPq, FAPERJ, FAPESP and FUNDUNESP (Brazil);
DAE and DST (India);
Colciencias (Colombia);
CONACyT (Mexico);
NRF (Korea);
FOM (The Netherlands);
STFC and the Royal Society (United Kingdom);
MSMT and GACR (Czech Republic);
BMBF and DFG (Germany);
SFI (Ireland);
The Swedish Research Council (Sweden);
and
CAS and CNSF (China).
%



\begin{thebibliography}{99}

\bibitem{combTEV} The TEVNPH Working Group, arXiv:1107.5518 (2011), http://tevnphwg.fnal.gov.

\bibitem{combD0}  V.~M.~Abazov {\sl et al.} [D0 Collaboration], Phys.\ Lett.\ B {\bf 663}, 26 (2008).


\bibitem{WH} V.~M.~Abazov {\sl et al.} [D0 Collaboration], 
Phys.\ Lett.\ B {\bf 698}, 6 (2011).

\bibitem{RF1}  L.~Breiman, Machine Learning {\bf 45}, 5 (2001).

\bibitem{RF2}   I.~Narsky, arXiv:physics/0507143 (2005); I. Narsky, arXiv:physics/0507157 (2005).

\bibitem{CERN} ALEPH, DELPHI, L3, and OPAL Collaborations, The
LEP Working Group for Higgs Boson Searches, Phys.\ Lett.\ B {\bf 565}, 61 (2003).

\bibitem{EWFIT} LEP, Tevatron, and SLD Electroweak Working Groups, arXiv: 0911.2604, http://lepewwg.web.cern.ch/LEPEWWG/.

\bibitem{D0WH1} V.~M.~Abazov {\sl et al.} [D0 Collaboration], 
Phys.\ Rev.\ Lett.
 {\bf 94}, 091802 (2005).
\bibitem{D0WH2} V.~M.~Abazov {\sl et al.} [D0 Collaboration], Phys.\ Lett.\ B
 {\bf 663}, 26 (2008).
\bibitem{D0WH3} V.~M.~Abazov {\sl et al.} [D0 Collaboration], Phys.\ Rev.\ Lett.
 {\bf 102}, 051803 (2009).

\bibitem{CDFWH1}  D.~Acosta {\sl et al.} [CDF Collaboration], Phys.\ Rev.\ Lett.\ {\bf 95}, 051801 (2005).

\bibitem{CDFWH2} T.~Aaltonen {\sl  et al.} [CDF Collaboration], Phys.\ Rev.\ Lett.
 {\bf 100}, 041801 (2008).
\bibitem{CDFWH3} T.~Aaltonen {\sl et al.} [CDF Collaboration], Phys.\ Rev.\ Lett.
 {\bf 103}, 101802 (2009).

\bibitem{Exclude} T.~Aaltonen {\sl et al.} [CDF and D0 Collaborations], Phys.\ Rev.\ Lett.\ {\bf 104}, 061802 (2010).


\bibitem{ref:ATLAS} G.~Aad {\sl et. al.} [ATLAS Collaboration],
  arXiv:hep-ex/1202.1408 (2012), submitted to Phys.\ Lett.\ B, and
  references therein.
 
\bibitem{ref:CMS} S.~Chatrchyan {\sl et. al.} [CMS Collaboration],
  arXiv:hep-ex/1202.1488 (2012), submitted to Phys.\ Lett.\ B, and 
references therein.

\bibitem{d0det_run2} V.~M.~Abazov {\sl et al.} [D0 Collaboration], 
Nucl.\ Instrum.\ Methods in Phys.\ Res.\ A {\bf 565}, 463 (2006).

\bibitem{d0det_run1} S.~Abachi {\sl et al.} [D0 Collaboration],
Nucl.\ Instrum.\ Methods in Phys.\ Res.\ A {\bf 338}, 185 (1994).

\bibitem{smt_layer} R.~Angstadt {\sl et al.}, 
Nucl.\ Instrum.\ Methods in Phys.\ Res.\ A {\bf 622}, 298 (2010).

\bibitem{d0muo_run2} V.~M.~Abazov {\sl et al.} [D0 Collaboration],
Nucl.\ Instrum.\ Methods in Phys.\ Res.\ A {\bf 552}, 372 (2005).

\bibitem{inelas} S.~Klimenko, J.~Konigsberg, T.~M.~Liss, Fermilab-FN-
0741 (2003).

\bibitem{lumi_ID} T.~Andeen, {\sl et al.}, FERMILAB-TM-2365 (2007).

\bibitem{Boole} G.~Boole, {\it An Investigation of the Laws of Thought} (Walton and Maberly, London, 1854)
[www.gutenberg.org/ebooks/15114].


\bibitem{caltrig} M.~Abolins {\sl et al.}, Nucl.\ Instrum.\ Methods in Phys.\ Res.\ A {\bf 584}, 75 (2008).

\bibitem{ttbar-prd} V.~M.~Abazov {\sl et al.} [D0 Collaboration], Phys.\ Rev.\ D {\bf 76}, 092007 (2007).

\bibitem{d0jet} G.~C.~Blazey {\sl et al.} arXiv:hep-ex/0005012 (2000).

\bibitem{d0jes} V.~M.~Abazov {\sl et al.} [D0 Collaboration], accepted by Phys.\ Rev.\ D. arXiv:hep-ex/1110.3771 (2011).

\bibitem{d0btag} V.~M.~Abazov {\sl et al.} [D0 Collaboration], Nucl.\ Instrum.\ Methods in Phys.\ Res.\ A {\bf 620}, 490 (2010).

\bibitem{geant} R.~Brun and F.~Carminati, CERN Program Library Long
Writeup, Report W5013 (1993);\\  
M. Goossens {\sl et al.}, {{\sc geant}} User's Guide CERN, Geneva, 1994.

\bibitem{alpgen} M.~Mangano {\sl et al.}, J.\ High Energy Phys.\ {\bf 07}, 001 (2003). Version  2.05 was used.

\bibitem{TuneA} R.~Field, TeV4LHC Report of the QCD Working Group edited by M.~G.~Albrow {\sl et al.}, arXiv:hep-ph/0610012, 74 (2006). The results of Tune A were used.

\bibitem{pythia63} T.~Sj\"ostrand {\sl et al.}, {{\sc pythia} 6.3:
  Physics and Manual} hep-ph/0308153 (2003).


\bibitem{signal1} K.~A.~Assamagan {\sl et al.}, arXiv:hep-ph/0406152.

\bibitem{signal2} O.~Brein, A.~Djouadi, and R.~Harlander, Phys.\ Lett.\ B {\bf 579}, 149 (2004).

\bibitem{signal3} M.~L.~Ciccolini, S.~Dittmaier, and M.~Kramer, Phys.\ Rev.\ D {\bf 68}, 073003 (2003).

\bibitem{signal4} J.~Baglio and A.~Djouadi, J.\ High Energy Phys.\ {\bf 10}, 064 (2010).

\bibitem{signal5} A.~Djouadi, J.~Kalinowski, and M.~Spira, Comput.\ Phys.\ Commun.\ {\bf 108}, 56 (1998).

\bibitem{signal} \rm T.~Hahn {\sl et al.}, 
arXiv:hep-ph/0607308.

\bibitem{CTEQ} H.~L.~Lai {\sl et al.}, 
Phys.\ Rev.\ D {\bf 55} 280 (1997); 
J.~Pumplin {\sl et al.}, J.\ High Energy Phys.\ {\bf 07}, 012 (2002).

\bibitem{alw} J.~Alwall {\sl et al.}, Eur.\ Phys.\ C {\bf 53}, 473 (2008).

\bibitem{alpgen_xs} M.~Mangano, M.~Moretti and R.~Pittau, Nucl.\ Phys.\ B {\bf 632} 343 (2002). 



\bibitem{ttbar_xsecs} N.~Kidonakis {\sl et al.}, Phys.\ Rev.\ D {\bf 78}, 074005 (2008).

\bibitem{COMPHEP} A.~Pukhov {\sl et al.}, hep-ph/9908288, (1999).

\bibitem{COMPHEP1} E.~Boos {\sl et al.},  Nucl.\ Instrum.\ Methods in Phys.\ Res.\ A {\bf 534}, 250  (2004).

\bibitem{stop_xsecs} N.~Kidonakis, Phys.\ Rev.\ D {\bf 74}, 114012 (2006).

\bibitem{mcfm} J.~M.~Campbell and R.~K.~Ellis, Phys.\ Rev.\ D {\bf 60}, 113006 (1999). 

\bibitem{alw_data}  V.~M.~Abazov {\sl et al.} [D0 Collaboration], Phys.\ Lett.\ B {\bf 669}, 278 (2008).

\bibitem{tauid}  V.~M.~Abazov {\sl et al.} [D0 Collaboration], Phys.\ Lett.\ B {\bf 670}, 292 (2009).

\bibitem{spincorr} S.~Parke and S.~Veseli, Phys.\ Rev.\ D {\bf 60}, 093003 (1999). 

\bibitem{diboson} V.~M.~Abazov {\sl et al.} [D0 Collaboration] Phys.\ Rev.\ Lett.\ {\bf 102}, 161801 (2009).

\bibitem{modfreq} T.~Junk,  Nucl.\ Instrum.\ Methods in Phys.\ Res.\ A {\bf 434}, 435 (1999). 

\bibitem{modfreq1} A.~Read, J.\ Phys.\ G {\bf 28}, 2693 (2002).

\bibitem{collie} W.~Fisher, FERMILAB-TM-2386-E (2007).

\bibitem{Aaltonen:2011rt} T.~Aaltonen {\sl et al.} [CDF Collaboration],  Phys.\ Rev.\ D {\bf 85}, 072001 (2012).



\end{thebibliography}
\end{document}